\documentclass[preprint]{aastex}
\usepackage[usenames]{color}

\begin{document}

\shortauthors{Welty et al.}
\shorttitle{HD 62542:  Probing an Interstellar Cloud Core}  


\title{HD 62542:  Probing the Bare, Dense Core of a Translucent Interstellar Cloud\footnotemark}
\footnotetext{Based on observations made with the NASA/ESA {\it Hubble Space Telescope}, obtained at the Space Telescope Science Institute, which is operated by the Association of Universities for Research in Astronomy, Inc., under NASA contract NAS 5-26555.  These observations are associated with program 12277. 
Based on observations made with ESO telescopes at the La Silla Paranal Observatory, under program IDs 064.I-0475, 072.C-0682, and 099.C-0637.}

\author{Daniel E. Welty\altaffilmark{1}, Paule Sonnentrucker \altaffilmark{1}, Theodore P. Snow\altaffilmark{2}, Donald G. York\altaffilmark{3,4}}

\altaffiltext{1}{Space Telescope Science Institute, 3700 San Martin Dr., Baltimore, MD 21218; dwelty@stsci.edu}
\altaffiltext{2}{University of Colorado, CASA -- Campus Box 389, Boulder, CO 80309-0001}
\altaffiltext{3}{University of Chicago, Department of Astronomy and Astrophysics, 5640 S. Ellis Ave., Chicago, IL 60637}
\altaffiltext{4}{also, Enrico Fermi Institute}

\begin{abstract}
We discuss the interstellar absorption from many atomic and molecular species seen in high-resolution {\it HST}/STIS UV and high-S/N optical spectra of the moderately reddened B3-5~V star HD~62542.
This remarkable sight line exhibits both very steep far-UV extinction and a high fraction of hydrogen in molecular form -- with strong absorption from CH, C$_2$, CN, and CO, but weak absorption from CH$^+$ and most of the commonly observed diffuse interstellar bands.
Most of the material resides in a single narrow velocity component -- offering a rare opportunity to probe the primarily molecular core of a single interstellar cloud with little associated diffuse atomic gas.
Detailed analyses of the spectra indicate that
(1) the molecular fraction in the main cloud is high [$f$(H$_2$) $\ga$ 0.8];
(2) the gas is fairly cold ($T_{\rm k}$ = 40--43 K; from the rotational excitation of H$_2$ and C$_2$);
(3) the local hydrogen density $n_{\rm H}$ $\sim$ 1500 cm$^{-3}$ (from the C$_2$ excitation, the fine-structure excitation of C$^0$, and simple chemical models);
(4) the unusually high excitation temperatures for $^{12}$CO and $^{13}$CO may be largely due to radiative excitation;
(5) $N$(C$^+$):$N$(CO):$N$(C) $\sim$ 100:10:1;
(6) the depletions of many elements are more severe than those seen in any other sight line, and the detailed pattern of depletions differs from those derived from larger samples of Galactic sight lines; and
(7) the various neutral/first ion ratios do not yield consistent estimates for the electron density, even when the effects of grain-assisted recombination and low-temperature dielectronic recombination are considered.
\end{abstract}


\section{INTRODUCTION}
\label{sec-intro}

While interstellar clouds have long been idealized as uniform spheres or slabs, high-resolution maps of emission due to dust, neutral hydrogen, CO, and other species indicate that both the overall and the internal structure of the clouds are typically more complicated.
Filamentary, sheet-like, and fractal-like structures are apparently quite common (e.g., Chappell \& Scalo 2001; Heiles \& Troland 2003; Kalberla et al. 2016), and clouds containing significant amounts of H$_2$ generally have colder, denser, predominantly molecular cores surrounded by more diffuse atomic gas (e.g., Andersson et al. 1991; Wannier et al. 1999).
Different atomic and molecular species can be concentrated in different (but overlapping) regions within the clouds, based on where the conditions for their survival are most favorable (e.g., Pan et al. 2005).
Moreover, many lines of sight exhibit multiple, closely spaced velocity components (assumed to correspond to individual clouds), which can be characterized by rather different properties (e.g., Welty et al. 1994, 1996).
While higher spectral resolution can enable us to distinguish predominantly diffuse components from other velocity components containing significant amounts of denser gas, we still cannot separate the diffuse and dense gas contributions that are associated with a single cloud at the same velocity.
That inability is particularly vexing for attempts to identify and characterize the so-called ``translucent'' clouds (with A$_{\rm V}$ $\sim$ 1--5 mag), thought to be transitional objects between more diffuse, primarily atomic clouds and the denser molecular clouds involved in forming stars (e.g., Snow \& McCall 2006).
In the cores of such clouds, we expect a molecular fraction $f$(H$_2$) = 2$N$(H$_2$)/($N$(H)+2$N$(H$_2$)) approaching unity, fairly high densities, and enhanced depletions of refractory (and perhaps even nonrefractory) elements into dust grains.
Those characteristics have not yet been observed, however, as most sight lines also include appreciable amounts of more diffuse material (Rachford et al. 2002, 2009; Snow et al. 2002a; Sonnentrucker et al. 2002, 2003).

The sight line toward the moderately reddened B3-5~V star HD~62542 [$E(B-V)$ $\sim$ 0.35--0.37; $A_V$ $\sim$ 1.0--1.2; $R_V$ $\sim$ 2.7--3.1 (Valencic et al. 2004; Fitzpatrick \& Massa 2007; Rachford et al. 2009; Gordon et al. 2009)] offers a rare opportunity to probe the relatively dense molecular core of an interstellar cloud, with little associated diffuse atomic gas.
This remarkable line of sight intersects a ridge of material in the western part of the {\it IRAS} Vela shell, foreground to the Gum Nebula (Sahu 1992; Pereyra \& Magalhaes 2002), apparently swept up and sculpted by stellar winds, radiation pressure, and/or shocks (Cardelli \& Savage 1988; Cardelli et al. 1990; O'Donnell et al. 1992).
A portion of that ridge near HD~62542 has been identified as a {\it Planck} Galactic cold clump (Planck Collaboration 2016; Dirks \& Meyer 2019).
At a distance of about 390 pc (Gaia 2), HD~62542 apparently lies somewhat beyond that structure, however, as there are no signs of interaction between the star and the intervening interstellar material (Whittet et al. 1993).
The far-UV extinction toward HD~62542 is among the steepest known, and the weak, broad 2175\AA\ bump may be shifted to lower wavelengths (Cardelli \& Savage 1988; Fitzpatrick \& Massa 2007; but see also Valencic et al. 2004; Gordon et al. 2009).
The column densities of CN and CH are the highest known for any sight line with A$_V$ $<$ 2 mag; the $N$(CN)/$N$(CH) and $N$(C$_3$)/$N$(C$_2$) ratios are quite high, but CH$^+$ is very weak (Cardelli et al. 1990; Gredel et al. 1991, 1993; \'{A}d\'{a}mkovics et al. 2003).
In addition to fairly strong emission from $^{12}$CO (1-0), emission from $^{12}$CO (3-2) and from the rare CO isotopolog C$^{18}$O (the latter usually seen in emission only for A$_V$ $>$ 3 mag) are also detected toward HD~62542 (van Dishoeck et al. 1991; Gredel et al. 1994).
The typically strongest of the diffuse interstellar bands (DIBs; e.g., those at 5780.6, 5797.2, and 6283.8\AA), which trace predominantly atomic gas, are exceptionally weak toward HD~62542, but the typically much weaker ``C$_2$ DIBs'' (which are correlated with molecular gas) are still detected (Snow et al. 2002b; \'{A}d\'{a}mkovics et al. 2005).
While the available optical and mm-wave spectra had suggested that most of the interstellar material toward HD~62542 resides in a single narrow component at $v_{\odot}$ $\sim$ 14 km~s$^{-1}$ (Cardelli et al. 1990; Gredel et al. 1991, 1993, 1994), the UV spectra discussed in this paper have revealed a number of other components, with much lower total hydrogen column densities, at velocities from 4 to 32 km~s$^{-1}$.
Various diagnostics have yielded estimates for the total hydrogen density $n_{\rm H}$ ranging from several hundred per cm$^3$ (Black \& van Dishoeck 1991) to 10$^4$ cm$^{-3}$ (Cardelli et al. 1990) in the main cloud.
The C$_2$ rotational excitation and the relative weakness of the mm-wave CN emission, however, have indicated that the local kinetic temperature $T_{\rm k}$ $\sim$ 30--45 K and $n_{\rm H}$ $\sim$ 500--1000 cm$^{-3}$ (Gredel et al. 1991, 1993; Sonnentrucker et al. 2007); that density is still rather high for a cloud with $A_{\rm V}$ $\sim$ 1.
Cardelli et al. (1990) argued that the cloud is a small, dense knot whose more diffuse outer layers have been stripped away by the winds and radiation pressure from $\zeta$~Pup and $\gamma^2$~Vel.
While the overall $f$(H$_2$) toward HD~62542 is uncertain, it may be as high as 0.88 (\'{A}d\'{a}mkovics et al. 2005).
The main cloud toward HD~62542 thus may have $f$(H$_2$) $\sim$ 1 -- making it perhaps the best current candidate for a bona fide single, isolated translucent cloud.

In this paper, we discuss high-resolution UV spectra of HD~62542, obtained with the Space Telescope Imaging Spectrograph (STIS) on board the {\it Hubble Space Telescope} ({\it HST}), aimed primarily at characterizing the abundances and physical conditions in the main, largely molecular cloud in that sight line.
Section~\ref{sec-obs} describes the acquisition, processing, and analysis of the spectra.
Section~\ref{sec-res} presents the derived atomic and molecular column densities and explores the excitation of some of those species.
Section~\ref{sec-disc} discusses the local physical conditions and depletions characterizing the clouds toward HD~62542 (as inferred from those atomic and molecular species) and possible implications for our understanding of interstellar clouds.
Section~\ref{sec-sum} provides a summary of our results and conclusions.
Detailed physical/chemical models of the main cloud, based on the data presented here, will be explored in a subsequent paper.

\begin{deluxetable}{llcccc}
\tablecolumns{6}
\tabletypesize{\scriptsize}
\tablecaption{Optical and UV Spectra \label{tab:spec}}
\tablewidth{0pt}

\tablehead{
\multicolumn{1}{c}{Date}&
\multicolumn{1}{c}{Facility}&
\multicolumn{1}{c}{$\lambda$ Range}&
\multicolumn{1}{c}{FWHM}&
\multicolumn{1}{c}{S/N\tablenotemark{a}}&
\multicolumn{1}{c}{Code}}

\startdata
1998 Sep & AAT / UCLES            & 5075--10295 & 5.0        &  70--100 & a \\
1999 Dec & ESO 3.6m / CES+VLC     &  3933, 5893\tablenotemark{b} & 2.00, 1.26 &  65, 85 & c \\
2002 Dec & Keck I / HIRES         &  3985--6420 & 4.5        & 700--1100 & h \\
2003 Nov & ESO VLT / UVES         &  3260--6680 & 4.5--4.9   &  80--250 & u \\
2013 May & Magellan Clay / MIKE   &  3350--9500 & 6.6--8.3   & 300--900 & m \\
2017 Apr & ESO VLT / UVES         &  3050--10422& 6.5--6.8\tablenotemark{c}   & 165--625 & U \\
\hline
2011 Apr & {\it HST} / STIS E140H &  1206--1408 & 2.7        & 15--60 & s \\
2011 Jun & {\it HST} / STIS E140H &  1242--1444 & 2.7        & 15--60 & s \\
2011 Apr & {\it HST} / STIS E230H &  1774--2051 & 2.7        & 25--50 & s \\
2011 Apr & {\it HST} / STIS E230H &  2124--2401 & 2.7        & 55--85 & s \\
\enddata
\tablecomments{The Keck I / HIRES spectra were obtained by \'{A}d\'{a}mkovics et al. (2003); the 2017 UVES spectra were obtained under program 099.C-0637 (M.-F. Nieva, PI); the other optical spectra were obtained by DEW.
The {\it HST} spectra were obtained under guest observer program 12277 (D. Welty, PI).}
\tablenotetext{a}{The S/N values are derived from the scatter in the fitted continua near the various detected absorption lines.}
\tablenotetext{b}{Small intervals around the \ion{Ca}{2} $\lambda$3933 and \ion{Na}{1} $\lambda\lambda$5889,5895 lines were observed.}
\tablenotetext{c}{Profile fits suggest that the actual resolution is closer to 5.1--5.3 km~s$^{-1}$.}
\end{deluxetable}

\begin{figure}
\epsscale{0.9}
\plotone{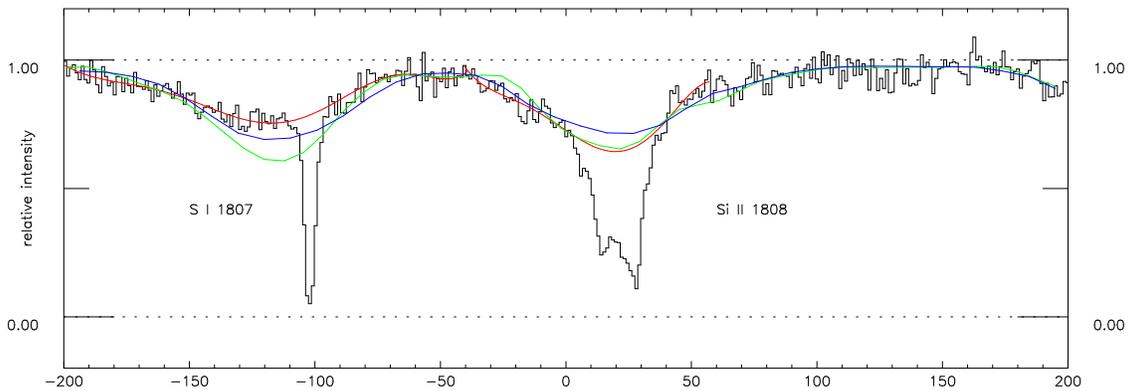}
\caption{Model stellar spectra and continuum fits near the interstellar \ion{S}{1} $\lambda$1807 and \ion{Si}{2} $\lambda$1808 absorption lines toward HD~62542.
The black histogram gives the observed spectrum; the blue and green curves give the predicted spectra from Tlusty stellar models with $T_{\rm eff}$ = 18,000 K [log $g$  = 3.75, $v sin i$ = 30 km~s$^{-1}$ in green; log $g$  = 4.25, $v sin i$ = 40 km~s$^{-1}$ in blue], approximately matched in the stellar continuum regions; the red curves give the adopted local continuum fits around the interstellar absorption lines.
As the predicted strengths of the stellar absorption lines do not perfectly match the observed line strengths, the model spectra are just taken to constrain the stellar $v sin i$ ($\sim$ 35 km~s$^{-1}$) and the curvature for ''reasonable'' fits to the local continua around the interstellar absorption features.}
\label{fig:stel}
\end{figure}


\section{OBSERVATIONS AND DATA ANALYSIS}
\label{sec-obs}

\subsection{Optical and UV Spectra}
\label{sec-opt}

The optical and UV spectra employed in this study of the sight line toward HD~62542 are listed in Table~\ref{tab:spec}.
All of the optical spectra were obtained with echelle spectrographs, at high to moderately high spectral resolution (FWHM = 1.26--8.3 km~s$^{-1}$).
In each case, standard routines within {\sc iraf} were used for the initial processing of the raw data:  bias correction, removal of small-scale variations in instrumental response via a normalized flat field, extraction of one-dimensional spectra from the two-dimensional CCD images, and wavelength calibration (typically using exposures of a Th-Ar lamp).
More detailed descriptions of the characteristics and processing of the spectra obtained with the University College London Echelle Spectrograph (UCLES; Walker \& Diego 1985), the Coud\'{e} Echelle Spectrograph (CES; Enard 1982) and the Ultraviolet and Visual Echelle Spectrograph (UVES; Dekker et al. 2000) can be found in Welty et al. (2006) and D. E. Welty \& P. A. Crowther (2020, in preparation); the 2017 UVES spectra are the standard pipeline data products.  
The High Resolution Echelle Spectrometer (HIRES; Vogt et al. 1994) spectra have been described by \'{A}d\'{a}mkovics et al. (2003).
For this study, the highest resolution CES spectra of the \ion{Na}{1} $\lambda\lambda$5889,5895 and \ion{Ca}{2} $\lambda$3933 lines were used to identify the major interstellar velocity components in the sight line, while the higher signal-to-noise ratio (S/N) HIRES and UVES spectra (primarily) were used to obtain accurate equivalent widths of the interstellar absorption lines from various atomic and molecular species.
The last column of the table gives a one-letter code used to identify the sources of the data in Appendix Tables~\ref{tab:ewatom}, \ref{tab:ewmol}, and \ref{tab:ewc2}.

The high-resolution UV spectra were obtained with {\it HST}/STIS, using the E140H and E230H echelle gratings, during four visits (18 total orbits) in 2011.
The four settings each covered about 200--275 \AA, centered at the nominal wavelength values (1307, 1343, 1913, and 2263 \AA); use of the 0.2$\times$0.09 arcsec slit yielded a resolution of about 2.7 km~s$^{-1}$ in each case.
The standard pipeline-processed spectral data (extracted, wavelength calibrated) were retrieved from the MAST archive\footnotemark, and the multiple individual spectra for each setting were checked for any systematic wavelength offsets via fits to relatively strong interstellar absorption features.
\footnotetext{http://archive.stsci.edu/hst; all of the STIS spectra are available in the MAST archive at https://doi.org/10.17909/t9-20tr-2053.}
After adjustment for those offsets (which in all cases were less than 0.3 km~s$^{-1}$), spectra for the individual orders were combined via sinc interpolation to a common, slightly oversampled heliocentric wavelength grid.
Spectral regions of overlap between adjacent orders were also combined, after checking for mutual consistency.
In order to examine the broad Ly$\alpha$ absorption line of atomic hydrogen, the 15 spectral orders within the range 1200--1255 \AA\ were rebinned to a coarser wavelength scale, matched in the regions of overlap, and merged into a single spectrum.

\begin{figure}
\epsscale{0.9}
\plotone{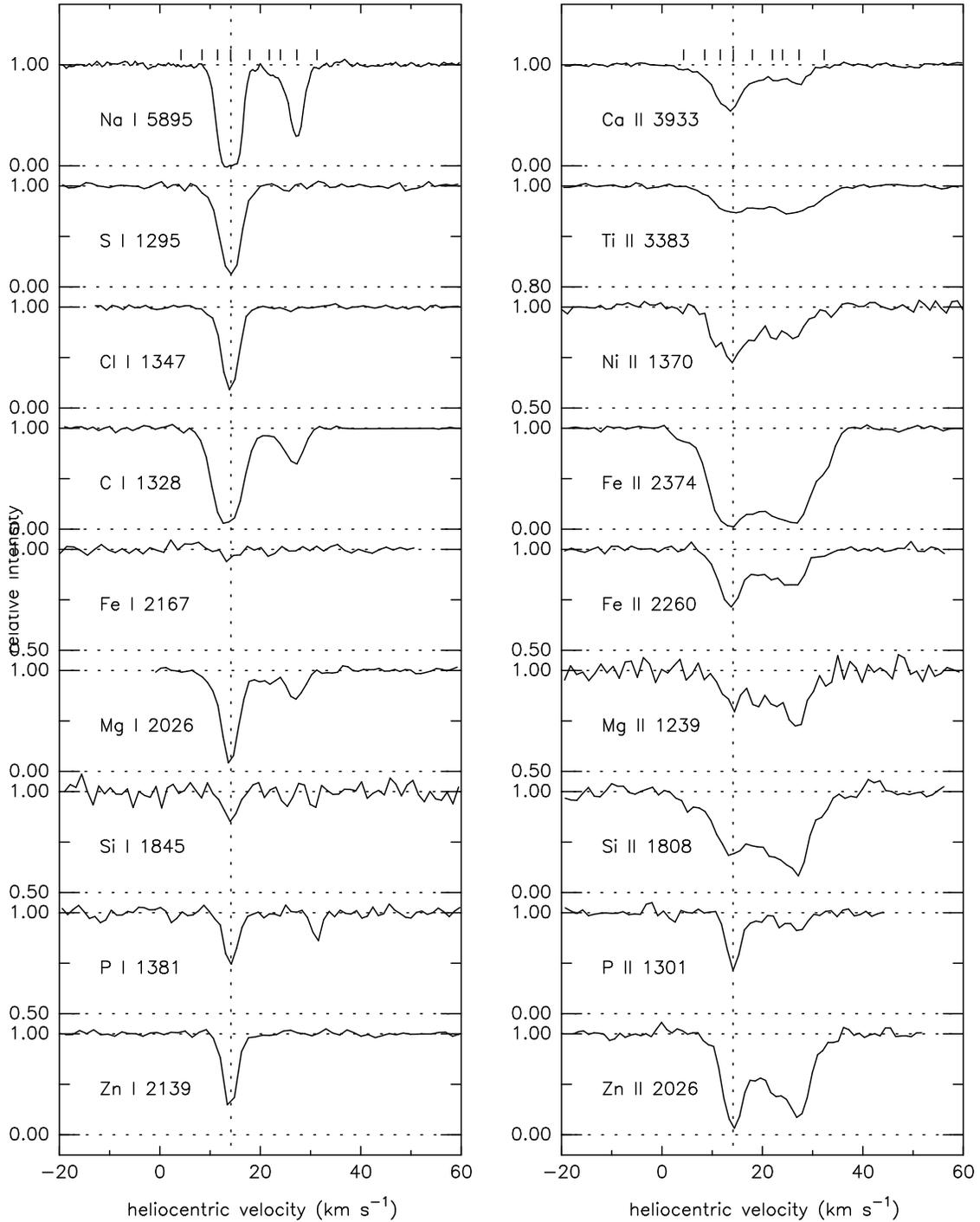}
\caption{Normalized spectra of selected atomic species toward HD~62542.
The UV spectra are from STIS (FWHM $\sim$ 2.7 km~s$^{-1}$); the optical spectra are from ESO/CES (FWHM $\sim$ 1.26--2.0 km~s$^{-1}$) or ESO/UVES (FWHM $\sim$ 6.8 km~s$^{-1}$).
On the left are trace neutral species, concentrated in the main narrow component near 14 km~s$^{-1}$ (vertical dotted line).
On the right are singly ionized species (dominant ions in predominantly neutral gas), with contributions from components spread over velocities from about 4 to 32 km~s$^{-1}$.
Tick marks above the \ion{Na}{1} $\lambda$5895 and \ion{Ca}{2} $\lambda$3933 profiles indicate the individual components determined in detailed fits to those profiles.
The differences in profile for the dominant species reflect both element-to-element and component-to-component differences in depletions.}
\label{fig:atom}
\end{figure}

\begin{figure}
\epsscale{0.45}
\plotone{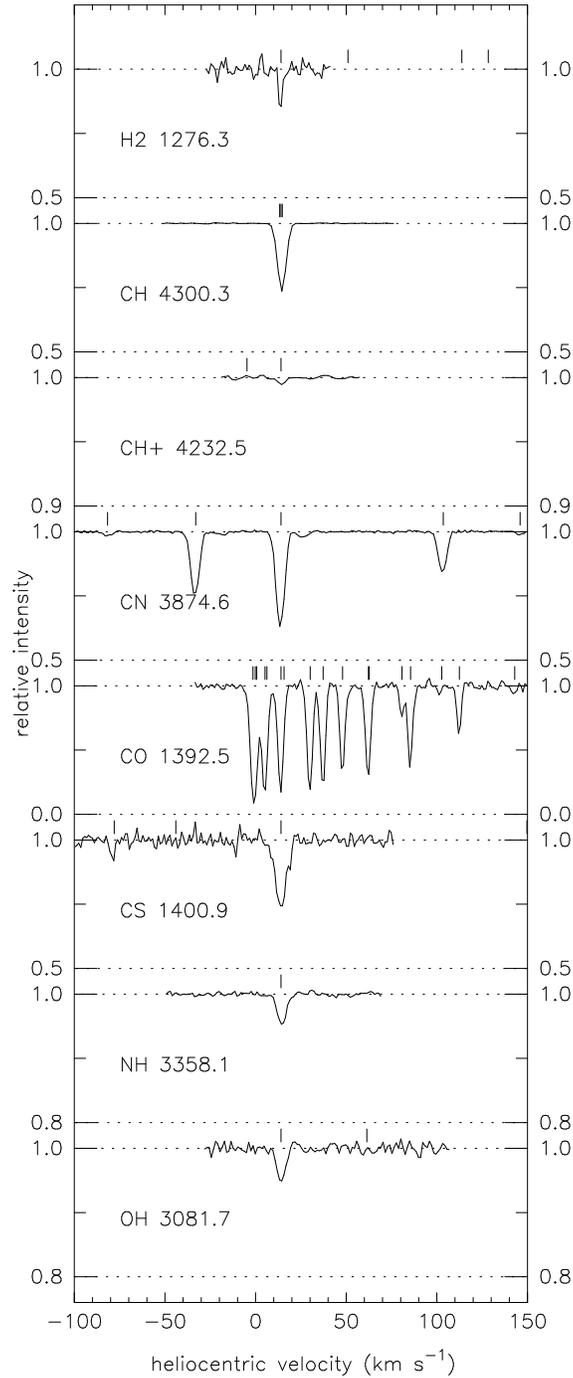}
\caption{Spectra of selected molecular species toward HD~62542.
The UV spectra are from STIS (FWHM $\sim$ 2.7 km~s$^{-1}$); the optical spectra are from Keck/HIRES or VLT/UVES (FWHM $\sim$ 4.5--6.8 km~s$^{-1}$).
The tick marks above the spectra indicate absorption from the main component at 14 km~s$^{-1}$.
For CN $\lambda$3874.6, the R2, R1, P1, and P2 lines flank the main R0 line at that velocity; the weak unmarked features just to the right of the R1 and R0 lines are due to $^{13}$CN.
For CO $\lambda$1392.5, lines from rotational levels $J$=0--6 (in the R, Q, and P branches) are detected; see also Appendix Figures 21 and 22.
For CS $\lambda$1400.9, the weak line near $-$80 km~s$^{-1}$ is due to \ion{Sn}{2} in the main 14 km~s$^{-1}$ component.
For several other cases, the ''extra'' tick marks give the locations of nearby lines of other species (whether detected or not).}
\label{fig:mol}
\end{figure}

\subsection{Equivalent Widths of Absorption Features}
\label{sec-ew}

Small spectral regions around atomic and molecular lines of interest were extracted from the processed optical and UV spectra of HD~62542, in order to measure the strengths of those features and to perform detailed fits to the line profiles (where complex velocity component structure could be discerned).
Those spectral segments were normalized via low-order polynomial fits to the continuum regions adjacent to the absorption features.
The continuum fits required some care -- particularly for the broader absorption features due to the stronger lines from some dominant species -- given the relatively low projected rotational velocity ($v sin i$ $\sim$ 35 km~s$^{-1}$) of HD~62542\footnotemark and, in a number of cases, the presence of stellar absorption features adjacent to or overlapping the interstellar lines.
\footnotetext{The $v sin i$ = 113 km~s$^{-1}$ found by Balona (1975) is too high.}
For some of those cases, the continuum fits were checked for reasonableness via comparisons with synthetic stellar spectra generated with the Tlusty code (Hubeny \& Lanz 1995; Lanz \& Hubeny 2007), for models with $T_{\rm eff}$ = 18000 K, log $g$ = 3.75 or 4.25, and $v sin i$ = 30 or 40 km~s$^{-1}$ (Figure~\ref{fig:stel}).
As the predicted strengths of the stellar absorption lines generally do not perfectly match the observed line strengths, the model spectra are just taken to constrain the stellar $v sin i$ and the curvature for ''reasonable'' fits to the local continua around the interstellar absorption features.
For the \ion{Na}{1} $\lambda\lambda$5889, 5895 and \ion{Ca}{2} $\lambda$3933 line profiles, stellar contributions from \ion{Na}{1}, \ion{C}{2} ($\lambda$5889), \ion{Ca}{2}, and \ion{S}{2} ($\lambda$3933) -- consistent with those found by Crawford (1990) around the \ion{Ca}{2} lines but stronger than that around the \ion{Na}{1} lines -- were fitted and removed.

Figures~\ref{fig:atom} and \ref{fig:mol} show the normalized profiles for interstellar absorption due to some of the atomic and molecular species detected toward HD~62542.
The absorption from both trace neutral atomic species (left-hand column of Fig.~\ref{fig:atom}) and molecules (Fig.~\ref{fig:mol}) is dominated by an apparently single strong, narrow component near a heliocentric velocity of 14 km~s$^{-1}$; several of the stronger lines from the trace neutral species show additional weaker absorption components at somewhat higher velocities.
Those additional components are relatively more prominent for the various singly ionized species (which are dominant ionization states in predominantly neutral interstellar clouds), whose profiles are shown in the right-hand column of Fig.~\ref{fig:atom}; in some cases (e.g., \ion{Mg}{2}, \ion{Si}{2}), the ''main'' component near 14 km~s$^{-1}$ appears not to have the highest column density.

Both the equivalent widths of detected absorption features and apparent optical depth (AOD) estimates for the corresponding column densities (e.g., Hobbs 1969; Sembach \& Savage 1992) were obtained by direct integration over the line profiles in the normalized spectra; values for lines due to various atomic and molecular species are listed in Appendix Tables~\ref{tab:ewatom}, \ref{tab:ewmol}, and \ref{tab:ewc2}.
In general, the equivalent widths measured for the CH, C$_2$, and CN lines found in the optical spectra agree quite well with previously reported values (Cardelli et al. 1990; Gredel et al. 1991, 1993).
The tables include some lines which are present in the spectra, but which could not be measured due to blending with either stellar or other interstellar absorption lines.
While transition $f$ values were taken from Morton (2000, 2003) for most of the atomic species, the $f$ values used for \ion{C}{1} are those adopted by Jenkins \& Tripp (2001, 2011); other exceptions are noted in Table~\ref{tab:ewatom}.
For species with measurements for multiple lines, differences in the AOD column density estimates may reflect problems with the $f$ values (generally for weaker lines) and/or saturation effects (for stronger lines).
In a few cases where the AOD column density estimates obtained from relatively weak lines seemed inconsistent, we have adopted revised $f$ values, as noted in Table~\ref{tab:ewatom}.
In general, the AOD column densities are lower limits to the true values, and they are consistent with the total sight-line column densities obtained from detailed fits to the line profiles (see next section).
The rms deviation of the scatter in the continuum regions yielded empirical estimates for the local S/N ratios (see the ranges for each data set given in Table~\ref{tab:spec}), which were used to estimate both the uncertainties in the equivalent widths of detected lines and upper limits for undetected lines.
The uncertainties (1$\sigma$) and upper limits (3$\sigma$) include contributions from both photon noise (Jenkins et al. 1973) and continuum placement (Sembach \& Savage 1992), added in quadrature.
The two sources of uncertainty are generally comparable for the narrower lines (e.g., from most trace neutral species), but the continuum uncertainties are larger (often by of order 50\%) for the broader lines from dominant singly ionized species.
Blending with stellar lines can also increase the effective uncertainties for some of the interstellar absorption features. 

\begin{figure}
\epsscale{1.0}
\plotone{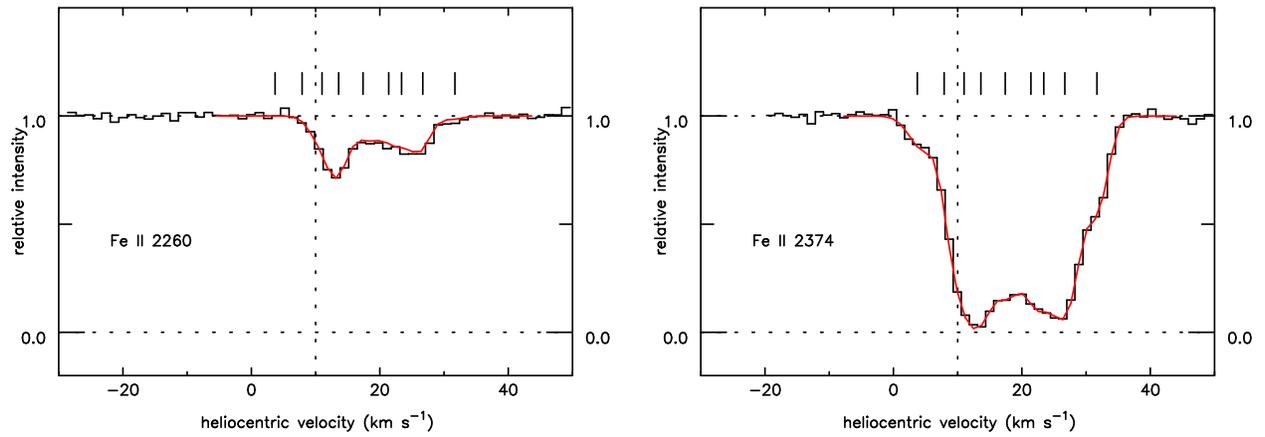}
\caption{Normalized spectra of the weak $\lambda$2260 and strong $\lambda$2374 lines of \ion{Fe}{2} toward HD~62542.
The black histogram is the observed spectrum; the red smooth curve is the adopted fit.
The tick marks give the positions of the nine components adopted in the fits (see Table~\ref{tab:acmp}).}
\label{fig:fe2fit}
\end{figure}

\begin{deluxetable}{rcccccccccc}
\rotate
\tablecolumns{11}
\tabletypesize{\scriptsize}
\tablecaption{Component Structures (Atomic Species) \label{tab:acmp}}
\tablewidth{0pt}

\tablehead{
\multicolumn{1}{c}{Comp}&
\multicolumn{1}{c}{$v$}&
\multicolumn{1}{c}{$b$}&
\multicolumn{1}{c}{Na I}&
\multicolumn{1}{c}{C I}&
\multicolumn{1}{c}{C I*}&
\multicolumn{1}{c}{C I**}&
\multicolumn{1}{c}{C I$_{\rm tot}$}&
\multicolumn{1}{c}{Mg I}&
\multicolumn{1}{c}{S I}&
\multicolumn{1}{c}{Zn I}\\
\multicolumn{1}{c}{ }&
\multicolumn{1}{c}{(km~s$^{-1}$)}&
\multicolumn{1}{c}{(km~s$^{-1}$)}&
\multicolumn{1}{c}{($\times$ 10$^{10}$)}&
\multicolumn{1}{c}{($\times$ 10$^{12}$)}&
\multicolumn{1}{c}{($\times$ 10$^{12}$)}&
\multicolumn{1}{c}{($\times$ 10$^{12}$)}&
\multicolumn{1}{c}{($\times$ 10$^{12}$)}&
\multicolumn{1}{c}{($\times$ 10$^{11}$)}&
\multicolumn{1}{c}{($\times$ 10$^{11}$)}&
\multicolumn{1}{c}{($\times$ 10$^{11}$)}}

\startdata
 1& [4.2]       &  [1.5]      &    $<$6.0    &  \nodata   &   \nodata   &  \nodata   &   \nodata   &  \nodata   &   \nodata   &   \nodata   \\
 2& [8.4]       &  [2.0]      &    $<$6.0    & 1.8$\pm$0.3&  2.3$\pm$0.5& 0.9$\pm$0.4&  5.0$\pm$0.7& 4.3$\pm$1.3&   \nodata   &   \nodata   \\
 3& 11.5$\pm$0.2&  [1.0]      & 24.0$\pm$6.0 &11.2$\pm$1.3& 13.5$\pm$1.4&14.8$\pm$1.4& 39.5$\pm$2.4& 3.9$\pm$2.3& 16.0$\pm$4.0&   \nodata   \\
 {\bf 4}& {\bf 14.1$\pm$0.1}&{\bf 0.85$\pm$0.10}&{\bf 11000$\pm$2000}& {\bf 800$\pm$80} & {\bf 1260$\pm$100}& {\bf 490$\pm$30} & {\bf 2550$\pm$132}& {\bf 650$\pm$60} & {\bf 4400$\pm$200}& {\bf 13.5$\pm$3.5}\\
  &             &    b        &    0.85      &   [0.9]    &    [0.85]   &   [0.85]   &   \nodata   &   [0.95]   &    [0.9]    &    [0.8]    \\   
 5& 17.9$\pm$0.3&  [1.0]      &  3.5$\pm$1.2 & 2.4$\pm$0.5&  2.0$\pm$0.6& 1.6$\pm$0.5&  6.0$\pm$0.9& 4.3$\pm$1.3&  6.2$\pm$3.6&   \nodata   \\
 6& 21.8$\pm$0.2&  [1.0]      &  5.3$\pm$0.9 & 0.4$\pm$0.2& $<$1.5      & $<$1.2     & $<$3.1      & 7.5$\pm$1.4& $<$5.4      &   \nodata   \\
 7& 24.0$\pm$0.2&  [1.0]      &  8.1$\pm$0.9 & 1.2$\pm$0.3&  0.6$\pm$0.5& $<$1.2     &  1.8$\pm$0.6& $<$3.9     & $<$4.8      &   \nodata   \\
 8& 27.3$\pm$0.1& 1.4$\pm$0.1 & 76.0$\pm$2.0 & 5.3$\pm$0.3&  1.2$\pm$0.5& 0.9$\pm$0.4&  7.4$\pm$0.7&22.9$\pm$1.7&  2.6$\pm$1.5&   \nodata   \\
 9& 31.3$\pm$0.3&  [1.0]      &  1.9$\pm$0.4 & $<$0.6     & $<$1.5      & $<$1.2     & $<$3.3      & $<$3.0     & $<$3.9      &   \nodata   \\
\hline
\multicolumn{1}{c}{Comp}&
\multicolumn{1}{c}{$v$}&
\multicolumn{1}{c}{$b$}&
\multicolumn{1}{c}{Ca II}&
\multicolumn{1}{c}{Mg II}&
\multicolumn{1}{c}{Si II}&
\multicolumn{1}{c}{P II}&
\multicolumn{1}{c}{Ti II}&
\multicolumn{1}{c}{Fe II}&
\multicolumn{1}{c}{Ni II}&
\multicolumn{1}{c}{Zn II}\\
\multicolumn{1}{c}{ }&
\multicolumn{1}{c}{(km~s$^{-1}$)}&
\multicolumn{1}{c}{(km~s$^{-1}$)}&
\multicolumn{1}{c}{($\times$ 10$^{10}$)}&
\multicolumn{1}{c}{($\times$ 10$^{13}$)}&
\multicolumn{1}{c}{($\times$ 10$^{13}$)}&
\multicolumn{1}{c}{($\times$ 10$^{12}$)}&
\multicolumn{1}{c}{($\times$ 10$^{10}$)}&
\multicolumn{1}{c}{($\times$ 10$^{12}$)}&
\multicolumn{1}{c}{($\times$ 10$^{11}$)}&
\multicolumn{1}{c}{($\times$ 10$^{11}$)}\\
\hline
 1 &  4.3$\pm$0.4 &[1.5]&   2.4$\pm$0.5 &  $<$14.3    &  6.5$\pm$1.7 & $<$5.4      & $<$0.9      &  4.0$\pm$1.0& $<$10.0      & $<$0.8        \\
 2 &  8.5$\pm$0.4 &[2.0]&   6.8$\pm$1.0 &  4.9$\pm$5.9&  8.7$\pm$2.3 & $<$6.5      & 1.1$\pm$0.6 &  9.0$\pm$1.2&  4.3$\pm$4.6 &  0.6$\pm$0.4  \\
 3 & 11.6$\pm$0.2 &[1.5]&  19.2$\pm$1.5 &  6.0$\pm$6.6& 19.4$\pm$3.6 & $<$9.3      & 5.1$\pm$1.0 & 45.6$\pm$6.4& 32.1$\pm$8.6 &  1.9$\pm$0.8  \\
 {\bf 4} & {\bf 14.2$\pm$0.1} &{\bf [1.5]}&  {\bf 25.8$\pm$1.7} & {\bf 25.9$\pm$6.5}& {\bf 30.0$\pm$4.5} &{\bf 73.0$\pm$5.4} & {\bf 4.9$\pm$0.9} & {\bf 65.3$\pm$8.4}& {\bf 39.0$\pm$6.8} & {\bf 74.0$\pm$14.3} \\
   &              &  b  &     [1.5]     &    [1.2]    &    [1.2]     &   [1.2]     &   [1.5]     &    [1.2]    &    [1.2]     &    [1.2]      \\   
 5 & 18.0$\pm$0.2 &[2.5]&  14.4$\pm$0.9 & 36.8$\pm$7.0& 41.4$\pm$3.4 & 8.9$\pm$2.7 & 6.9$\pm$0.7 & 47.5$\pm$3.1& 41.8$\pm$5.8 &  9.3$\pm$0.7  \\
 6 &[22.0]        &[1.5]&   5.0$\pm$0.8 & 23.8$\pm$8.1& 28.9$\pm$4.9 & $<$7.7      & 3.2$\pm$1.0 & 24.3$\pm$3.8& 11.9$\pm$5.4 &  9.5$\pm$0.9  \\
 7 &[24.0]        &[1.5]&   5.5$\pm$0.8 & 13.5$\pm$8.5& 21.6$\pm$5.4 & 7.4$\pm$2.7 & 3.3$\pm$0.9 & 26.7$\pm$4.6& 17.9$\pm$5.7 &  5.5$\pm$1.0  \\
 8 & 27.3$\pm$0.1 &[1.8]&  12.5$\pm$0.6 & 61.6$\pm$7.5& 81.4$\pm$9.6 &16.7$\pm$2.3 & 8.6$\pm$0.6 & 64.4$\pm$5.4& 28.4$\pm$5.3 & 31.6$\pm$2.0  \\
 9 & 32.3$\pm$0.6 &[1.5]&   1.3$\pm$0.4 &  5.3$\pm$4.9&  9.7$\pm$1.7 & $<$4.9      & 3.1$\pm$0.5 & 12.7$\pm$0.9&  5.2$\pm$3.5 &  1.9$\pm$0.3  \\
\enddata
\tablecomments{Velocites and $b$ values for the neutral and singly ionized species were determined from fits to the higher resolution Na~I and Ca~II profiles, respectively.
Values in square braces were fixed in the fits.
Only the $b$ value for the main component (4, in bold) was varied for different species.}
\end{deluxetable}

\subsection{Fits to Absorption-line Profiles}
\label{sec-fits}

Multicomponent fits to the absorption-line profiles were performed in order to determine column densities for the ''individual'' components discernible for the various neutral and singly ionized atomic and molecular species that trace the predominantly neutral gas toward HD~62542, using the program {\sc fits6p} (e.g., Welty et al. 2003) and a variant of that program which can simultaneously fit multiple, widely separated lines (from the same or different species) having a common component structure.
Initial, independent fits to the highest resolution CES spectra of \ion{Na}{1} and \ion{Ca}{2} yielded very similar component velocities and line widths ($b$ values) for those two species (columns 2 and 3 of Table~\ref{tab:acmp}).
For the main component near 14 km~s$^{-1}$ -- which strongly dominates the absorption for the molecular and trace neutral atomic species -- the $b$ value for each species was determined by requiring consistent fits to both strong and weak lines (where possible -- e.g., for \ion{Na}{1}, the strong D lines at 5890 and 5896 \AA\ and the much weaker $\lambda$3302 doublet).
The resulting main-component $b$ values range from 0.8 to 0.95 km~s$^{-1}$ for the trace neutral species, and from 1.0 to 1.5 km~s$^{-1}$ for the dominant singly ionized species.
For neutral carbon, separate simultaneous fits were performed for the unblended lines of \ion{C}{1}, \ion{C}{1}*, and \ion{C}{1}**; the adopted $b$ = 0.85--0.90 km~s$^{-1}$ and the resulting total log[$N$(\ion{C}{1})] = 15.41 for the main component are in good agreement with the values (1.0 km~s$^{-1}$, 15.46) obtained by Dirks \& Meyer (2019) from the same spectra.\footnotemark
\footnotetext{Contributions from $^{13}$C, assuming the average interstellar $^{12}$C/$^{13}$C ratio $\sim$ 70 (see Sec.~\ref{sec-iso}), were considered for the \ion{C}{1} $\lambda$1328 multiplet, for which the difference in velocity (relative to the $^{12}$C lines) is of order $-$2 km~s$^{-1}$ (Morton 2003; Berengut et al. 2006).
Those contributions produced only slight changes in the line profiles, however.
Contributions from $^{13}$C were not considered for the other \ion{C}{1} multiplets, where the velocity differences are even smaller ($\la$ 0.25 km~s$^{-1}$; Berengut et al. 2006).}
For these small $b$ values, even relatively weak absorption lines will yield noticeable differences in the column densities derived from the equivalent widths (assumed optically thin), from AOD integrations over the line profiles, and from fits to the line profiles.
As an example, for $b$ = 1.0 km~s$^{-1}$, the $N$(\ion{O}{1}) obtained for the main component from the weak $\lambda$1355 line ($W$ = 6.8 m\AA) are about 3.5, 4.5, and 5.8 $\times$ 10$^{17}$ cm$^{-2}$, respectively, for the three methods.
Given the relatively low temperatures ($T$ $\sim$ 40--43 K) obtained for that component from analyses of the excitation of H$_2$ and C$_2$ (Sec.~\ref{sec-tk}), the line widths appear to be dominated by turbulence (or unresolved velocity structure).
For all other components, both the relative velocities and the $b$ values determined in the \ion{Na}{1} and \ion{Ca}{2} fits were used (with slight adjustments in some cases) in fitting the lines from other trace neutral and singly ionized species, respectively.

The individual component column densities derived from those detailed profile fits are given in the upper and lower parts of Table~\ref{tab:acmp}, respectively.
As examples, the adopted fits to the absorption from the weak $\lambda$2260 and strong $\lambda$2374 lines of \ion{Fe}{2} are shown in Figure~\ref{fig:fe2fit}, and the fits to the seven observed \ion{C}{1} multiplets are shown in Appendix Figure~\ref{fig:c1}.
Additional very weak components between about $-$7 and 2 km~s$^{-1}$ and near 37 km~s$^{-1}$ may be seen in the strongest lines of some dominant species (e.g., \ion{Si}{2} $\lambda$1304, \ion{Fe}{2} $\lambda$2344), but the column densities are rather uncertain due to the difficulty in accounting for stellar lines in fitting the local continua around those weak interstellar features.
Estimated column densities for those weak, outlying components are included in the total sight line and ''other components'' column densities in Table~\ref{tab:cdatom}, but the individual values are not listed in Table~\ref{tab:acmp}.
The column densities for the main component can be sensitive to $b$ if only single, moderately strong lines are available (e.g., for \ion{Zn}{1} and \ion{Zn}{2}). 
For \ion{Zn}{2}, decreasing the main-component $b$ from the adopted 1.2 km~s$^{-1}$ to 1.0 km~s$^{-1}$ (which also yields acceptable fits to the line profile) increases the derived $N_{\rm m}$ by a factor of 2.
The column densities for the other, generally weaker components can be uncertain if only relatively weak or very strong lines are available (e.g., for \ion{C}{2}, \ion{O}{1}, some of the trace neutral species, and the dominant ions of many of the heaviest elements) -- due to blending with the main component and/or stellar lines and/or to uncertainties in continuum placement.

Absorption from molecular species is detected only for the main component near 14 km~s$^{-1}$.
As for the neutral atomic species, the column densities and $b$ values listed in Table~\ref{tab:cdmol} for that component were determined by requiring consistent fits for both weak and strong lines.
For CH (with three multiplets in the optical spectra), $b$ = 1.0 km~s$^{-1}$ was adopted.
For CN (three multiplets) and C$_2$ (one UV and three optical multiplets), however, the best fits were for $b$ = 0.7 km~s$^{-1}$ -- as found for CN by Gredel et al. (1991).
For CO, $b$ = 0.5 km~s$^{-1}$ gave the most consistent column densities for the many A-X band lines from rotational levels $J$ = 0--6 (Appendix Table~\ref{tab:cobands}); the values obtained from fits assuming $b$ = 0.4 and 0.7 km~s$^{-1}$ were noticeably less consistent.
For $b$ = 0.5 km~s$^{-1}$, the total log[$N$($^{12}$CO)] = 16.42 obtained from the permitted CO A-X bands is consistent with the values determined from the much weaker CO intersystem bands and from the observed CO emission (van Dishoeck et al. 1991); the adopted $b$ and derived total log[$N$($^{12}$CO)] are also in good agreement with the values (0.5 km~s$^{-1}$, 16.49) obtained from the same STIS spectra by Dirks \& Meyer (2019).
The corresponding FWHM = 1.665 $\times$ $b$ $\sim$ 0.8 km~s$^{-1}$ for $^{12}$CO is smaller than the width of the $^{12}$CO (1--0) emission line seen toward HD~62542 (1.5 km~s$^{-1}$ for a 43\arcsec\ beam; Gredel et al. 1994), but is comparable to the widths of the weaker $^{13}$CO and C$^{18}$O emission lines (0.4--1.1 km~s$^{-1}$).
The fits to the absorption from the observed $^{12}$CO and $^{13}$CO bands are shown in Appendix Figures~\ref{fig:co12} and \ref{fig:co13}.

\clearpage

\begin{deluxetable}{lrrrrrrr}
\tablecolumns{8}
\tabletypesize{\scriptsize}
\tablecaption{Column Densities and Depletions -- Atomic Species \label{tab:cdatom}}
\tablewidth{0pt}

\tablehead{
\multicolumn{1}{c}{Species}&
\multicolumn{1}{c}{A$_{\odot}$\tablenotemark{a}}&
\multicolumn{2}{c}{Main component}&
\multicolumn{2}{c}{All other components}&
\multicolumn{2}{c}{Total sight line}\\
\multicolumn{2}{c}{ }&
\multicolumn{1}{c}{log[$N_{\rm m}$]}&
\multicolumn{1}{c}{Depletion}&
\multicolumn{1}{c}{log[$N_{\rm o}$]\tablenotemark{b}}&
\multicolumn{1}{c}{Depletion}&
\multicolumn{1}{c}{log[$N$]}&
\multicolumn{1}{c}{Depletion}}

\startdata
H                     &      &[20.48]         & \nodata &[20.30]         & \nodata & 20.70$\pm$0.18 & \nodata \\
H$_2$                 &      & 20.81$\pm$0.21 & \nodata &[$<$18.8]       & \nodata & 20.81$\pm$0.21 & \nodata \\
H$_{\rm tot}$         &      & 21.20$\pm$0.17 & \nodata &[20.30]         & \nodata & 21.26$\pm$0.18 & \nodata \\
 & \\
B II                  & 2.85 &$<$10.91        &$<-$1.14 &   \nodata      & \nodata &$<$11.04        &$<-$1.07 \\
C I                   & 8.46 & 14.90$\pm$0.04 & \nodata & 13.35$\pm$0.05 & \nodata & 14.92$\pm$0.04 & \nodata \\
C I*                  &      & 15.10$\pm$0.03 & \nodata & 13.29$\pm$0.04 & \nodata & 15.11$\pm$0.03 & \nodata \\
C I**                 &      & 14.69$\pm$0.03 & \nodata & 13.26$\pm$0.04 & \nodata & 14.71$\pm$0.03 & \nodata \\
C I (total)           &      & 15.41$\pm$0.02 & \nodata & 13.78$\pm$0.03 & \nodata & 15.42$\pm$0.02 & \nodata \\
C II                  &      &$<$17.96        & \nodata &   \nodata      & \nodata &$<$18.08        &$<+$0.36 \\
N I                   & 7.90 &    \nodata     & \nodata &   \nodata      & \nodata &$>$15.02        &$>-$2.14 \\
O I\tablenotemark{c}  & 8.76 & 17.76$\pm$0.02 & $-$0.20 &   \nodata      & \nodata &[17.82]         &[$-$0.20]\\
O I*                  &      & 13.60$\pm$0.05 & \nodata &   \nodata      & \nodata &[13.65]         & \nodata \\
O I**                 &      & 13.32$\pm$0.04 & \nodata &   \nodata      & \nodata &[13.37]         & \nodata \\
Na I                  & 6.37 & 14.04$\pm$0.08 & \nodata & 12.07$\pm$0.02 & \nodata & 14.05$\pm$0.08 & \nodata \\
Mg I                  & 7.62 & 13.81$\pm$0.04 & \nodata & 12.63$\pm$0.04 & \nodata & 13.84$\pm$0.04 & \nodata \\
Mg II                 &      & 14.41$\pm$0.11 & $-$2.41 & 15.18$\pm$0.05 & $-$0.74 & 15.25$\pm$0.05 & $-$1.63 \\
Al I                  & 6.54 &$<$10.62        & \nodata &   \nodata      & \nodata &$<$10.74        & \nodata \\
Al III                &      &    \nodata     & \nodata &   \nodata      & \nodata & 12.45$\pm$0.05 & \nodata \\
Si I                  & 7.61 & 11.60$\pm$0.07 & \nodata &   \nodata      & \nodata &[11.61]         & \nodata \\
Si II                 &      & 14.48$\pm$0.07 & $-$2.33 & 15.34$\pm$0.03 & $-$0.57 & 15.39$\pm$0.02 & $-$1.48 \\
Si IV                 &      &    \nodata     & \nodata &   \nodata      & \nodata & 12.83$\pm$0.03 & \nodata \\
P I                   & 5.54 & 12.01$\pm$0.03 & \nodata &   \nodata      & \nodata &[12.02]         & \nodata \\
P II                  &      & 13.86$\pm$0.03 & $-$0.88 & 13.52$\pm$0.06 & $-$0.32 & 14.03$\pm$0.03 & $-$0.77 \\
S I                   & 7.26 & 14.64$\pm$0.02 & \nodata & 12.39$\pm$0.10 & \nodata & 14.65$\pm$0.02 & \nodata \\
S II                  &      &    \nodata     & \nodata &   \nodata      & \nodata &$>$15.62        &$>-$0.90 \\
Cl I                  & 5.33 & 14.30$\pm$0.10 & $-$0.23 & 11.70$\pm$0.16 & \nodata &[14.30]         & \nodata \\
K I                   & 5.18 & 12.08$\pm$0.07 & \nodata & 10.26$\pm$0.12 & \nodata & 12.09$\pm$0.07 & \nodata \\
Ca I                  & 6.41 &  9.10$\pm$0.09 & \nodata &   \nodata      & \nodata & [9.12]         & \nodata \\
Ca II                 &      & 11.41$\pm$0.03 & $-$4.20 & 11.83$\pm$0.02 & \nodata & 11.97$\pm$0.02 &$>-$3.70 \\
Ti I                  & 5.00 &$<$10.03        & \nodata &   \nodata      & \nodata &$<$10.15        & \nodata \\
Ti II                 &      & 10.69$\pm$0.08 & $-$3.50 & 11.50$\pm$0.03 & $-$1.80 & 11.56$\pm$0.03 & $-$2.70 \\
V II                  & 4.07 &$<$11.88        &$<-$1.39 &   \nodata      & \nodata &$<$12.00        &$<-$1.33 \\
Cr I                  & 5.72 & $<$9.68        & \nodata &   \nodata      & \nodata & $<$9.80        & \nodata \\
Mn I                  & 5.58 &$<$10.67        & \nodata &   \nodata      & \nodata &$<$10.79        & \nodata \\
Mn II                 &      &$<$12.72        &$<-$2.06 &   \nodata      & \nodata &$<$12.84        &$<-$2.00 \\
Fe I                  & 7.54 & 11.28$\pm$0.15 & \nodata &   \nodata      & \nodata &[11.30]         & \nodata \\
Fe II\tablenotemark{d}&      & 13.81$\pm$0.06 & $-$2.93 & 14.37$\pm$0.02 & $-$1.47 & 14.48$\pm$0.02 & $-$2.32 \\
Co II                 & 4.98 &$<$12.48        &$<-$1.70 &   \nodata      & \nodata &$<$12.60        &$<-$1.64 \\
Ni I                  & 6.29 &$<$10.68        & \nodata &   \nodata      & \nodata &$<$10.80        & \nodata \\
Ni II                 &      & 12.59$\pm$0.08 & $-$2.90 & 13.15$\pm$0.05 & $-$1.44 & 13.26$\pm$0.04 & $-$2.29 \\
Cu II                 & 4.34 & 11.66$\pm$0.11 & $-$1.88 & 11.52$\pm$0.11 & $-$1.12 & 11.90$\pm$0.14 & $-$1.70 \\
Zn I                  & 4.70 & 12.13$\pm$0.12 & \nodata &   \nodata      & \nodata &[12.14]         & \nodata \\
Zn II                 &      & 12.87$\pm$0.08 & $-$1.03 & 12.78$\pm$0.02 & $-$0.22 & 13.13$\pm$0.05 & $-$0.83 \\
Ga II                 & 3.17 & 10.53$\pm$0.13 & $-$1.84 & 10.70$\pm$0.13 & $-$0.77 & 10.92$\pm$0.15 & $-$1.51 \\
Ge II                 & 3.70 & 11.89$\pm$0.06 & $-$1.01 &   \nodata      & \nodata & 11.94$\pm$0.11 & $-$1.02 \\
As II                 & 2.40 &$<$11.52        &$<-$0.08 &   \nodata      & \nodata &$<$11.64        &$<-$0.02 \\
Kr I                  & 3.36 & 12.36$\pm$0.06 & $-$0.20 &   \nodata      & \nodata & 12.41$\pm$0.11 & $-$0.21 \\
Cd II                 & 1.81 & 10.97$\pm$0.12 & $-$0.03 &   \nodata      & \nodata & 11.18$\pm$0.14 & $+$0.12 \\
Sn II                 & 2.19 & 10.88$\pm$0.10 & $-$0.51 &   \nodata      & \nodata & 10.89$\pm$0.15 & $-$0.55 \\
Pb II                 & 2.13 &$<$11.42        &$<+$0.09 &   \nodata      & \nodata &$<$11.54        &$<+$0.15 \\
\enddata
\tablecomments{Uncertainties are 1$\sigma$; limits are 3$\sigma$.  
Values in square braces are estimated / assumed.  
For H and H$_2$, see Sec.~\ref{sec-isnh}.  
For several trace neutral species, the total is assumed to be 0.01-0.02 dex higher than the main-component value (as found for \ion{C}{1}, \ion{Na}{1}, \ion{S}{1}, and \ion{K}{1}).  
For \ion{O}{1}, the total is assumed to have the same (very mild) depletion as found for the main component.
See Appendix Table~\ref{tab:ewatom} for the absorption lines measured for each species.}
\tablenotetext{a}{Protosolar abundances from Lodders (2003), which are 0.07--0.08 dex higher than the current solar photospheric values.
The protosolar values are adopted here to facilitate comparisons with the surveys of Jenkins (2009) and Ritchey et al. (2018).}
\tablenotetext{b}{Total for all components other than the main component near 14 km~s$^{-1}$.}
\tablenotetext{c}{For $b$ = 1.0$\pm$0.1 km~s$^{-1}$.  
The smaller log[$N$(O I)] = 17.43 obtained by Jensen et al. (2005) was based on equivalent widths of the very strong $\lambda$1039 and $\lambda$1302 lines, with $b$ = 18.1 km~s$^{-1}$ for the effective curve of growth.}
\tablenotetext{d}{The much larger log[$N$(Fe II)] = 15.45 obtained by Snow et al. (2002a) may be due to inclusion of some stellar absorption in the equivalent widths measured in the lower resolution, lower S/N ratio {\it FUSE} spectra.}
\end{deluxetable}

\begin{deluxetable}{lrcl}
\tablecolumns{4}
\tabletypesize{\scriptsize}
\tablecaption{Column Densities -- Molecules \label{tab:cdmol}}
\tablewidth{0pt}

\tablehead{
\multicolumn{1}{c}{Molecule}&
\multicolumn{1}{c}{log[$N_{\rm m}$]}&
\multicolumn{1}{c}{$b$}&
\multicolumn{1}{c}{Excitation}}

\startdata
H$_2$     &    20.81$\pm$0.21 &     & T$_{01}$ = 43$\pm$11        \\
 & \\
CH        &    13.54$\pm$0.04 & 1.0 & \nodata                     \\
CH$^+$    &    11.68$\pm$0.07 &     & \nodata                     \\
CH$_2$    &($<$13.01)         &     & \nodata                     \\
C$_2$     &    13.98$\pm$0.04 & 0.7 & $T_{\rm k}$ = 40$\pm$5      \\
C$_3$     &    13.02$\pm$0.02 &     & $T_{\rm low}$ = 75          \\
          &    \nodata        &     & $T_{\rm high}$ = 137        \\
CN        &    13.55$\pm$0.04 & 0.7 & $T_{01}$ = 2.89$\pm$0.15    \\
          &    \nodata        &     & $T_{12}$ = 2.95$\pm$0.19    \\
$^{13}$CN &    11.65$\pm$0.06 &     & \nodata                     \\
CO        &    16.42$\pm$0.05 & 0.5 & $T_{\rm ex}$ = 11.7$\pm$0.4 \\ 
$^{13}$CO &    14.63$\pm$0.03 & 0.5 & $T_{\rm ex}$ =  7.7$\pm$0.2 \\ 
C$^{18}$O &    12.87$\pm$0.08 &     & \nodata                     \\ 
CO$^+$    &($<$12.59)         &     & \nodata                     \\
CS        &    12.12$\pm$0.03 &     & \nodata                     \\ 
NH        &    12.94$\pm$0.06 &     & \nodata                     \\
NO        &($<$13.86)         &     & \nodata                     \\
NO$^+$    &($<$13.54)         &     & \nodata                     \\
OH        &    14.00$\pm$0.10 &     & $T_{\rm ex}$ $\sim$ 1.7 K   \\
OH$^+$    & $<$12.81          &     & \nodata                     \\
H$_2$O    &($<$12.95)         &     & \nodata                     \\
SH$^+$    &($<$13.14)         &     & \nodata                     \\
HCl       &($<$11.83)         &     & \nodata                     \\
\enddata
\tablecomments{Data for H$_2$ are from Rachford et al. (2002); data for C$_3$ are from \'{A}d\'{a}mkovics et al. (2003).
See Appendix Tables~\ref{tab:ewmol} and \ref{tab:ewc2} for the absorption lines measured for each species.
The total column density for C$_2$ includes small estimated contributions from (unmeasured) levels with $J$ $>$ 18.
The total column density for $^{13}$CN assumes the same excitation temperature as for CN.
Column density limits in parentheses are based on measurements of a single line, with no information on molecular excitation (see Table~\ref{tab:ewmol}).
Absorption from C$_2^-$, C$_2^+$, CN$^+$, SiO, SH, and AlH was not detected (Table~\ref{tab:ewmol}), but reliable $f$ values are not available for those transitions.}
\end{deluxetable}

\clearpage

\section{INTERSTELLAR COLUMN DENSITIES}
\label{sec-res}

\subsection{H, H$_2$, and H$_{\rm tot}$}
\label{sec-isnh}

Fits to the STIS Ly$\alpha$ profile, using the usual continuum reconstruction method (e.g., Savage et al. 1977; Diplas \& Savage 1994), yield a total column density of atomic hydrogen $N$(H) $\sim$ 16.5$\pm$1.5 $\times$ 10$^{20}$ cm$^{-2}$ for the sight line.
Given the spectral type assigned to HD~62542, which has ranged from B3~V to B5~V (Feast et al. 1955; Houk 1978; Cardelli \& Savage 1988), however, that measured $N$(H) likely includes a significant stellar contribution (Savage \& Panek 1974; Diplas \& Savage 1994).
Using the photometry of Kilkenny (1978), the resulting Str\"{o}mgren index [$c_1$] = $c_1$ $-$ 0.2($b-y$) = 0.364 implies a stellar contribution of about 11$\pm$2 $\times$ 10$^{20}$ cm$^{-2}$ (Fig.~2 of Diplas \& Savage 1994), which would then yield a total interstellar $N$(H) $\sim$ 5.5$\pm$2.5 $\times$ 10$^{20}$ cm$^{-2}$ toward HD~62542.

Bearing in mind that this sight line exhibits some unusual properties, additional estimates for the interstellar $N$(H) -- for the sight line as a whole, for the main component, and for the sum of the other components -- may be obtained from other measured quantities and the corresponding mean relationships established for the local Galactic interstellar medium (ISM).
In the following (and throughout the rest of this paper), column densities without subscripts refer to the entire sight line; subscripts ''m'' and ''o'' denote values for the main component and the sum of the other components, respectively: 
\begin{enumerate}
\item{For an average Galactic gas-to-dust ratio $N$(H$_{\rm tot}$)/$E(B-V)$ = 5.6$\pm$1.4 $\times$ 10$^{21}$ cm$^{-2}$ (e.g., Bohlin et al. 1978; Welty et al. 2012), the adopted color excess, $E(B-V)$ = 0.35, would suggest a total hydrogen column density $N$(H$_{\rm tot}$) = $N$(H) + 2$N$(H$_2$) $\sim$ 20$\pm$5 $\times$ 10$^{20}$ cm$^{-2}$.
With the $N$(H$_2$) = 6.5$^{+4.0}_{-2.4}$ $\times$ 10$^{20}$ cm$^{-2}$ derived from {\it FUSE} spectra (Rachford et al. 2002), that $N$(H$_{\rm tot}$) would imply $N$(H) $\sim$ 7 $\times$ 10$^{20}$ cm$^{-2}$ -- slightly higher than the value obtained from Ly$\alpha$.}
\item{The $N$(CH) toward HD~62542 is slightly above (but consistent with) the general good correlation with $N$(H$_2$) observed in the Galactic ISM (e.g., Danks et al. 1984; Rachford et al. 2002; Sheffer et al. 2008; see Fig.~\ref{fig:molvsh2} below).
Fits to the high-S/N ratio Keck CH $\lambda$4300 profile then suggest that the main component near 14 km~s$^{-1}$ contains at least 99 percent of the CH (and H$_2$) toward HD~62542.
The other components in the sight line thus contain primarily atomic hydrogen.}
\item{The equivalent width of the $\lambda$5780.6 diffuse interstellar band is generally fairly tightly correlated with $N$(H) in the local Galactic ISM (Herbig 1993; Friedman et al. 2011), with an average $N$(H)/$W$(5780.6) = 7.0$\pm$1.8 $\times$ 10$^{18}$ cm$^{-2}$ m\AA$^{-1}$ (Welty et al. 2006, 2012).
The observed $W$(5780.6) = 27$\pm$3 m\AA\ (from the average of the UVES spectra) would thus correspond to a somewhat smaller $N$(H) $\sim$ 1.9$\pm$0.5 $\times$ 10$^{20}$ cm$^{-2}$.
[Note, however, that there are sight lines (e.g., in the Sco-Oph and Orion Trapezium regions; Herbig 1993; Welty et al. 2006, 2012) in which the $\lambda$5780.6 DIB is weaker than would be expected from the observed $N$(H) -- so $N$(H) could be higher than that.]}
%
\begin{figure}
\epsscale{1.0}
\plotone{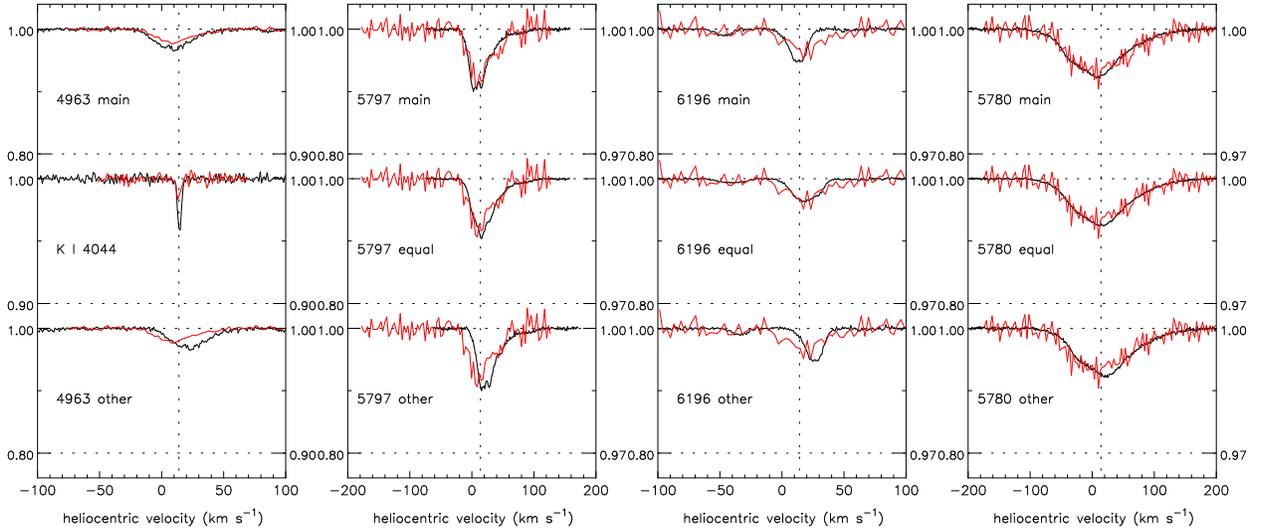}
\caption{Velocity shifts of several DIBs toward HD~62542 (red; UVES spectra), relative to those toward 20~Aql (black; HARPS spectra), scaled to similar central depths [note the different y-axis scales for HD~62542 (right) and 20~Aql (left)].
On the top line of each panel, the 20~Aql profile is shifted to 14 km~s$^{-1}$ (the velocity of the main, predominantly molecular component toward HD~62542); on the bottom line of each panel, the 20~Aql profile is shifted to 27 km~s$^{-1}$ (the velocity of the generally strongest of the predominantly atomic ''other'' components); in the middle is a composite constructed from the 20~Aql profile, with equal contributions at 14 and 27 km~s$^{-1}$ (except in the left-most panel, which compares the aligned \ion{K}{1} $\lambda$4044 profiles).
Toward HD~62542, the $\lambda$4963.9 C$_2$-DIB (which can be present in diffuse molecular gas) appears to be concentrated almost exclusively in the main component, the $\lambda$5797.2 DIB appears to have some contribution from the higher velocity component, and the $\lambda$6196.0 and $\lambda$5780.6 DIBs (which trace primarily atomic gas) appear to have roughly equal contributions from the two components.}
\label{fig:dibs4}
\end{figure}

\item{While most of the well-known DIBs (e.g., $\lambda\lambda$4428.5, 5780.6, 6283.8) appear to trace predominantly atomic gas, the C$_2$-DIBs (and, to a lesser extent, some others such as $\lambda$5797.2) can also be present in diffuse molecular gas (Thorburn et al. 2003; Welty 2014; Welty et al. 2014; Lan et al. 2015; Fan et al. 2017).
Sight lines in which primarily molecular and primarily atomic components are separated in velocity thus may exhibit velocity shifts among those DIBs -- which can then provide some indication of the relative distributions (in velocity) of the molecular and atomic gas (e.g., Cox et al. 2005; Welty et al. 2014). 
Figure~\ref{fig:dibs4} compares the UVES profiles of several DIBs toward HD~62542 with the corresponding profiles seen in HARPS spectra of the similarly reddened 20~Aql (where the DIBs are among the narrowest known -- and thus may exhibit the ''intrinsic'' DIB profiles).
In the top and bottom sections of each panel, the 20~Aql profiles are aligned at 14 and 27 km~s$^{-1}$ (i.e., with the dominant main molecular component and the generally strongest of the predominantly atomic components toward HD~62542, respectively); in the middle is a composite based on equal contributions from a 20~Aql profile at each of those two velocities.
While the DIBs toward HD~62542 are fairly weak, these rather crude comparisons suggest that the $\lambda$4963.9 C$_2$-DIB is nearly all in the main component at 14 km~s$^{-1}$, the $\lambda$5797.2 DIB is largely in that main component, and the $\lambda$6196.0 and $\lambda$5780.6 DIBs may have roughly equal contributions from the two components.
The profiles and velocities of the DIBs toward HD~62542 thus appear to be consistent with the H$_2$ being concentrated in the main 14 km~s$^{-1}$ component and with the total H being divided roughly equally between that main component and the other components (together, which are seen only in the atomic lines).}
\item{In the Galactic ISM, the column density of \ion{Na}{1} exhibits a nearly quadratic dependence on $N$(H$_{\rm tot}$) (Welty \& Hobbs 2001), so that the total $N_{\rm o}$(\ion{Na}{1}) $\sim$ 11.9 $\times$ 10$^{11}$ cm$^{-2}$ for all the other components (Tables~\ref{tab:acmp} and \ref{tab:cdatom}) suggests $N_{\rm o}$(H) $\sim$ 2.5$\pm$0.6 $\times$ 10$^{20}$ cm$^{-2}$ for those components.
The depletion indicators log[$N$(\ion{Ca}{2})/$N$(\ion{Na}{1})] and log[$N$(\ion{Ti}{2})/$N$(H$_{\rm tot}$)] are also fairly well correlated ($r$ = 0.87) in the Galactic ISM (D. E. Welty \& P. A. Crowther, in preparation).
The column densities of \ion{Na}{1}, \ion{Ca}{2}, and \ion{Ti}{2} for all the other components then yield a very similar estimate for $N_{\rm o}$(H) $\sim$ $N_{\rm o}$(H$_{\rm tot}$) $\sim$ 2.4 $\times$ 10$^{20}$ cm$^{-2}$.}
%
\begin{figure}[h!]
\epsscale{1.0}
\plotone{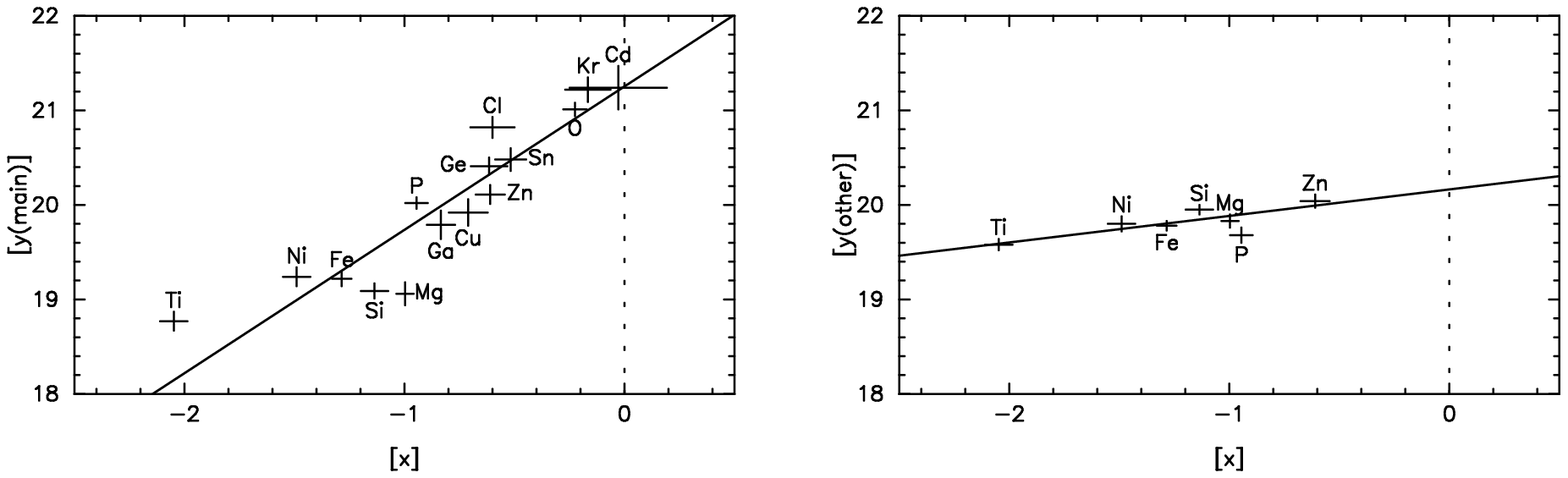}
\caption{Application of equation 24a from Jenkins (2009), used to estimate the depletion strength factor F$_*$ and $N$(H$_{\rm tot}$) from the relative depletions of various species, for the main component toward HD~62542 (left panel) and the other components (right panel).
The y values are derived from the column densities of dominant species listed in Table~\ref{tab:cdatom}, the proto-solar abundances from Lodders (2003), and the coefficients derived by Jenkins (2009) and Ritchey et al. (2018).
The x values are the corresponding $A_{\rm X}$ coefficients from the latter two references.
The depletion factor is given by the slope of the fitted line:  1.52$\pm$0.19 for the main component and 0.28$\pm$0.10 for the other components.
The estimate for log[$N$(H$_{\rm tot}$)] is given by the y value of the line at x = 0.0:  21.25$\pm$0.19 for the main component and 20.16$\pm$0.13 for the other components.}
\label{fig:depl_fh}
\end{figure}

\begin{deluxetable}{lccccccccc}
\tablecolumns{10}
\tabletypesize{\scriptsize}
\tablecaption{Estimates for the Column Densities of Atomic, Molecular, and Total Hydrogen \label{tab:hest}}
\tablewidth{0pt}

\tablehead{
\multicolumn{1}{c}{Indicator}&
\multicolumn{3}{c}{Main component}&
\multicolumn{3}{c}{All other components}&
\multicolumn{3}{c}{Total sight line}\\
\multicolumn{1}{c}{ }&
\multicolumn{1}{c}{$N_{\rm m}$(H)}&
\multicolumn{1}{c}{$N_{\rm m}$(H$_2$)}&
\multicolumn{1}{c}{$N_{\rm m}$(H$_{\rm tot}$)}&
\multicolumn{1}{c}{$N_{\rm o}$(H)}&
\multicolumn{1}{c}{$N_{\rm o}$(H$_2$)}&
\multicolumn{1}{c}{$N_{\rm o}$(H$_{\rm tot}$)}&
\multicolumn{1}{c}{$N$(H)}&
\multicolumn{1}{c}{$N$(H$_2$)}&
\multicolumn{1}{c}{$N$(H$_{\rm tot}$)}}

\startdata
Lyman $\alpha$ (STIS) & \nodata    & \nodata    & \nodata        & \nodata     & \nodata& \nodata    & 5.5$\pm$2.5    & \nodata    &(19$\pm$7)    \\
H$_2$ (FUSE)          & \nodata    & \nodata    & \nodata        & \nodata     & \nodata& \nodata    & \nodata        & 6.5$\pm$3.2& \nodata      \\
 & \\
$E(B-V)$              & \nodata    & \nodata    & \nodata        & \nodata     & \nodata& \nodata    & ($\sim$7)      & \nodata    & 20$\pm$5     \\
$N_{\rm r}$(Na~I)     & \nodata    & \nodata    & \nodata        &(2.5$\pm$0.6)& \nodata& 2.5$\pm$0.6& \nodata        & \nodata    & \nodata      \\
$N_{\rm r}$(Na~I, Ca~II, Ti~II)&\nodata&\nodata & \nodata        &(2.4$\pm$0.6)& \nodata& 2.4$\pm$0.6& \nodata        & \nodata    & \nodata      \\
$W$(5780.6)           & \nodata    & \nodata    & \nodata        & \nodata     & \nodata& \nodata    &$\ga$1.9$\pm$0.5& \nodata    &($\ga$15$\pm$6)\\
Diffuse band profiles &$\sim$0.5$N$(H)& \nodata & \nodata        &$\sim$0.5$N$(H)&\nodata&\nodata    & \nodata        & \nodata    & \nodata      \\
Depletions            & ($\sim$5)  & \nodata    & 18$^{+10}_{-6}$&(1.4$\pm$0.5)& \nodata& 1.4$\pm$0.5& ($\sim$6)      & \nodata    &(19$\pm$8)    \\
$N$(CH)               & \nodata    & 6.5$\pm$3.2& \nodata        & \nodata     & $<$0.06& \nodata    & \nodata        & \nodata    & \nodata      \\
 & \\
Adopt                 &   3$\pm$1  & 6.5$\pm$3.2& 16$\pm$6       & 2.0$\pm$0.5 & $<$0.06& 2.0$\pm$0.5&   5$\pm$2  & 6.5$\pm$3.2&   18$\pm$7   \\
\enddata
\tablecomments{All column densities are $\times$ 10$^{20}$ cm$^{-2}$.  Values in parentheses for atomic or total hydrogen use the observed $N$(H$_2$) and the estimated total or atomic hydrogen column densities, respectively.}
\end{deluxetable}

\item{If the pattern of elemental depletions is similar to that generally found in the Galactic ISM, then an estimate for $N$(H$_{\rm tot}$) may be obtained from the observed column densities of various dominant ions and the average elemental depletion coefficients derived by Jenkins (2009; eqn.~24a; see Sec.~\ref{sec-depl} below).
Figure~\ref{fig:depl_fh} shows the application of this procedure to the main component toward HD~62542 (left panel) and to the other components (right panel).
In each case, the depletion strength factor F$_*$ is given by the slope of the fitted line, and the estimated log[$N$(H$_{\rm tot}$)] is given by the y value at x = 0. 
For the main component toward HD~62542, the pattern of depletions is somewhat unusual (Sec.~\ref{sec-depl}), and the observed column densities yield an estimated $N_{\rm m}$(H$_{\rm tot}$) $\sim$ 18$^{+10}_{-6}$ $\times$ 10$^{20}$ cm$^{-2}$, with F$_*$ = 1.52$\pm$0.19.
On the whole, the depletions corresponding to the sum of all of the other components are slightly more severe than the Galactic warm cloud pattern, and the observed column densities yield an estimated $N_{\rm o}$(H$_{\rm tot}$) $\sim$ $N_{\rm o}$(H) $\sim$ 1.4$\pm$0.5 $\times$ 10$^{20}$ cm$^{-2}$, with F$_*$ = 0.28$\pm$0.10.
We note, however, that application of this procedure to the sum of all the components toward HD~62542 -- combining components with very different depletion properties -- yields F$_*$ = 0.85$\pm$0.11 (slightly less severe than the Galactic cold cloud pattern) and an estimated $N$(H$_{\rm tot}$ $\sim$ 11.5$^{+4.0}_{-3.0}$ $\times$ 10$^{20}$ cm$^{-2}$ -- which is smaller than the sum of $N_{\rm m}$(H$_{\rm tot}$) and $N_{\rm o}$(H$_{\rm tot}$); see the discussion in Sec.~\ref{sec-depl} below.}
\end{enumerate}

In view of these various estimates for the column densities of H, H$_2$, and H$_{\rm tot}$, we have adopted $N$(H) = 5$\pm$2 $\times$ 10$^{20}$ cm$^{-2}$ [from Ly$\alpha$ and $W$(5780.6)], with $\sim$2 $\times$ 10$^{20}$ cm$^{-2}$ in the other components (all together, from \ion{Na}{1} and the depletions) and the remainder ($\sim$3 $\times$ 10$^{20}$ cm$^{-2}$) in the main component (Table~\ref{tab:hest}).
The total hydrogen column density for the sight line is then $N$(H$_{\rm tot}$) = 18$\pm$7 $\times$ 10$^{20}$ cm$^{-2}$, with 16 $\times$ 10$^{20}$ cm$^{-2}$ in the main component.
If $N_{\rm m}$(H) is of order 3 $\times$ 10$^{20}$ cm$^{-2}$ in the main component, then $f$(H$_2$) $\sim$ 0.81 there (consistent with the estimate of \'{A}d\'{a}mkovics et al. 2005, which was based on more limited data).
The molecular fraction in the other components is $\la$ 0.06.

\begin{deluxetable}{rccccccccc}
\tablecolumns{10}
\tabletypesize{\scriptsize}
\tablecaption{Component Ratios (Atomic Species) \label{tab:arat}}
\tablewidth{0pt}

\tablehead{
\multicolumn{1}{c}{Comp}&
\multicolumn{1}{c}{$v$}&
\multicolumn{1}{c}{Na I/Ca II}&
\multicolumn{1}{c}{Mg I/Mg II}&
\multicolumn{1}{c}{Mg II/Si II}&
\multicolumn{1}{c}{Si II/Zn II}&
\multicolumn{1}{c}{P II/Zn II}&
\multicolumn{1}{c}{Ti II/Zn II}&
\multicolumn{1}{c}{Fe II/Zn II}&
\multicolumn{1}{c}{Fe II/Ni II}\\
\multicolumn{2}{c}{ }&
\multicolumn{1}{c}{tm/ds}&
\multicolumn{1}{c}{ts/ds}&
\multicolumn{1}{c}{ds/ds}&
\multicolumn{1}{c}{ds/dm}&
\multicolumn{1}{c}{dm/dm}&
\multicolumn{1}{c}{ds/dm}&
\multicolumn{1}{c}{ds/dm}&
\multicolumn{1}{c}{ds/ds}}

\startdata
 1 &  4.3 & $<$2.5        &  \nodata        &  $<$2.20     &  $>$813       &  \nodata     &  \nodata      &  $>$50      &  $>$4.0       \\
 2 &  8.5 & $<$0.9        &    0.009:       &     0.56:    & 1450$\pm$1040 &  $<$108.3    & 0.18$\pm$0.16 & 150$\pm$102 &    21.0:      \\
 3 & 11.6 &   1.3$\pm$0.3 &    0.007:       &     0.31:    & 1021$\pm$470  &   $<$48.9    & 0.27$\pm$0.13 & 240$\pm$107 & 14.2$\pm$4.3  \\
 {\bf 4} & {\bf 14.2} & {\bf 426.4$\pm$82.5}& {\bf 0.251$\pm$0.067} & {\bf 0.86$\pm$0.25}&   {\bf 41$\pm$10}   &  {\bf 9.9$\pm$2.0} &{\bf 0.007$\pm$0.002}&   {\bf 9$\pm$2}   & {\bf 16.7$\pm$3.6}  \\
 5 & 18.0 &   0.2$\pm$0.1 &0.0012$\pm$0.0004& 0.89$\pm$0.18&  445$\pm$50   &  9.6$\pm$3.0 &0.074$\pm$0.009&  51$\pm$5   & 11.4$\pm$1.7  \\
 6 & 22.0 &   1.1$\pm$0.2 &0.0032$\pm$0.0012& 0.82$\pm$0.31&  304$\pm$59   &    $<$8.1    &0.034$\pm$0.011&  26$\pm$5   & 20.4$\pm$9.8  \\
 7 & 24.0 &   1.5$\pm$0.3 & $<$0.003        & 0.63$\pm$0.42&  393$\pm$121  & 13.5$\pm$5.5 &0.062$\pm$0.021&  49$\pm$12  & 14.9$\pm$5.4  \\
 8 & 27.3 &   6.1$\pm$0.3 &0.0037$\pm$0.0005& 0.76$\pm$0.13&  258$\pm$35   &  5.3$\pm$0.8 &0.027$\pm$0.003&  20$\pm$2   & 22.7$\pm$4.6  \\
 9 & 32.3 &   1.5$\pm$0.5 & $<$0.006        & 0.55$\pm$0.51&  511$\pm$120  &   $<$25.8    & 0.16$\pm$0.03 &  67$\pm$12  & 24.4$\pm$16.5 \\
\enddata
\tablecomments{The two-letter codes in the second line of the header crudely characterize the ionization and depletion behavior of each species: t = trace, d = dominant, m = mild depletion, s = severe depletion.
Pairs with similar behavior generally have similar ratios for all components, while pairs with different behavior generally exhibit different ratios for the main component (in bold) and for the other components.
Colons denote very uncertain values.}
\end{deluxetable}

\clearpage

\begin{figure}
\epsscale{0.7}
\plotone{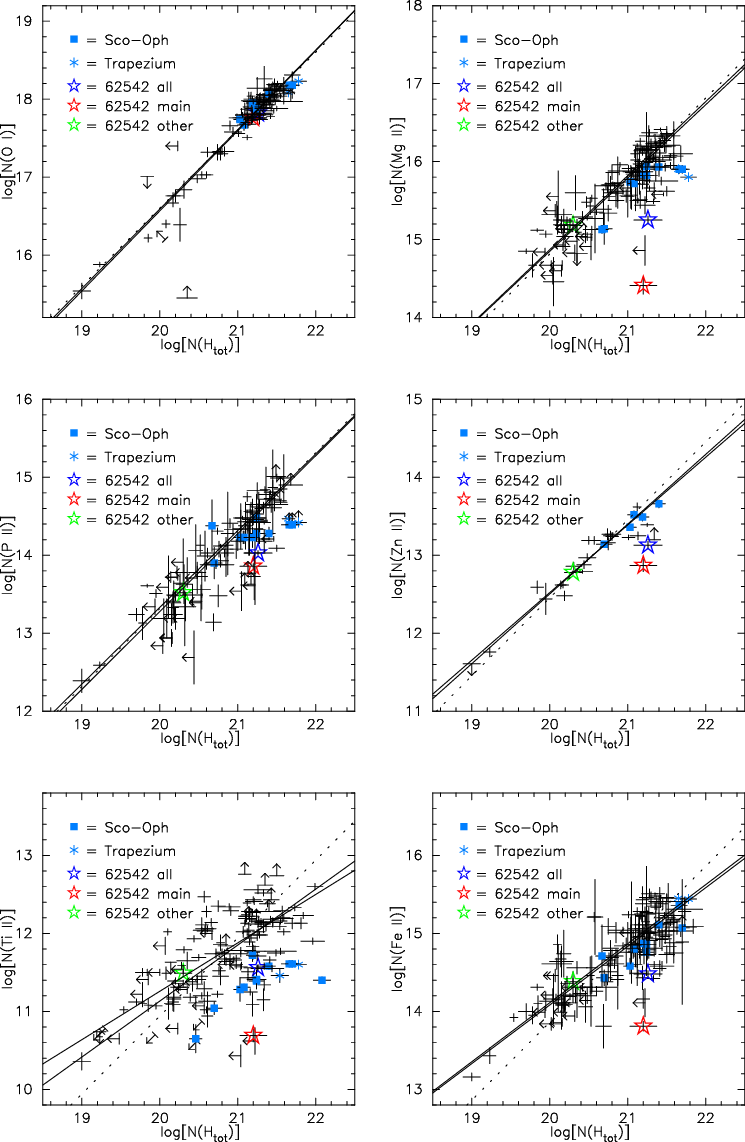}
\caption{Column densities of dominant species \ion{O}{1}, \ion{Mg}{2}, \ion{P}{2}, \ion{Zn}{2}, \ion{Ti}{2}, and \ion{Fe}{2} vs. $N$(H$_{\rm tot}$) for Galactic sight lines.
The stars denote the sight line toward HD~62542: red = main (14 km~s$^{-1}$) component; green = all other components; blue = all components.
The solid lines show weighted and unweighted fits to the data (not including the Sco-Oph and Trapezium sight lines); the dotted lines have slope = 1.
Toward HD~62542, the column densities of all but \ion{O}{1} are clearly deficient, relative to the general Galactic trends -- reflecting the enhanced depletions of those species in this unusual sight line.
The column densities for \ion{Mg}{2} and for the more refractory \ion{Ti}{2} and \ion{Fe}{2} are particulary low in the main component -- indicating very severe depletions there; \ion{P}{2} and \ion{Zn}{2} also lie below the general trends.
Note also the lower abundances of \ion{Ti}{2} in the Sco-Oph sight lines (Welty \& Crowther 2010).}
\label{fig:dom_htot}
\end{figure}

\clearpage

\begin{figure}
\epsscale{0.8}
\plotone{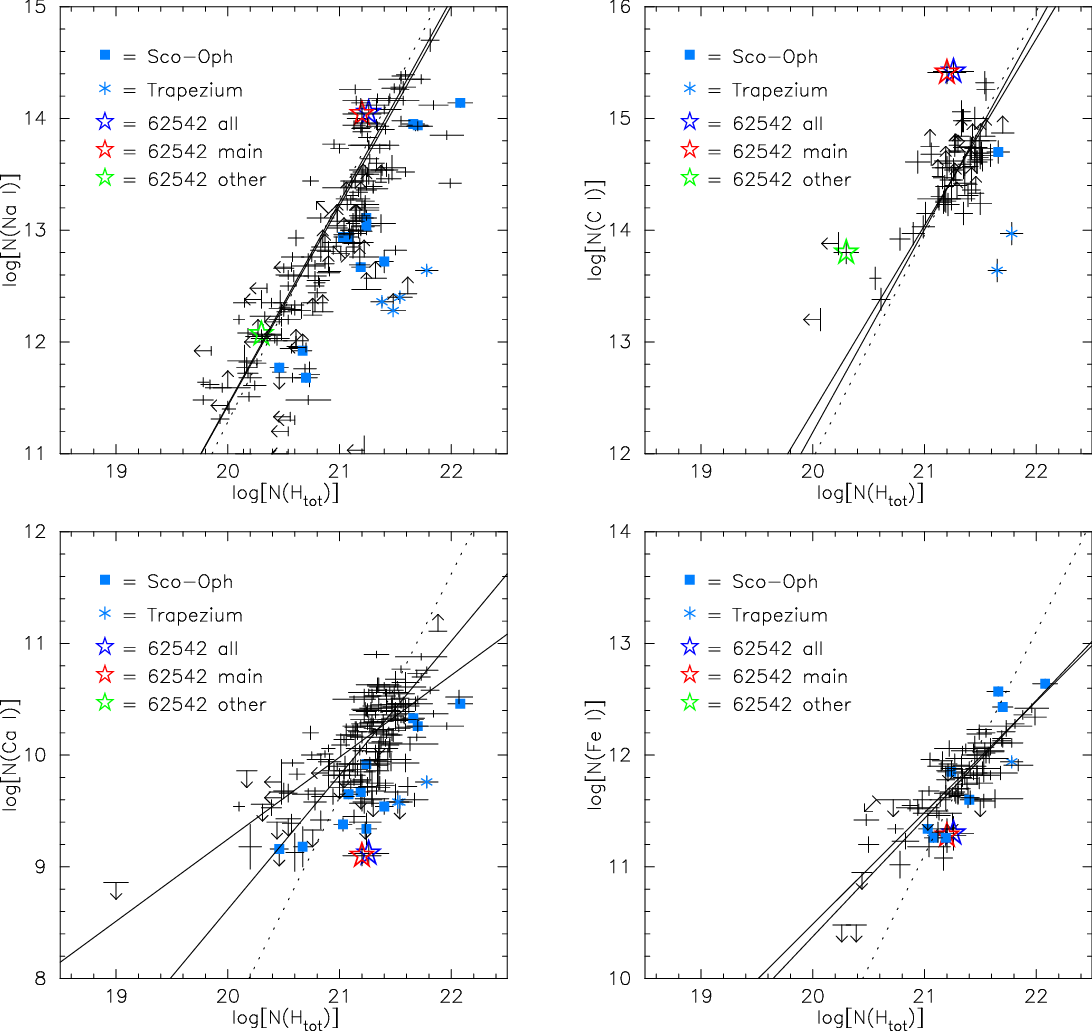}
\caption{Column densities of trace neutral species \ion{Na}{1}, \ion{C}{1}, \ion{Ca}{1}, and \ion{Fe}{1} vs. $N$(H$_{\rm tot}$) for Galactic sight lines.
The stars denote the sight line toward HD~62542: red = main (14 km~s$^{-1}$) component; green = all other components; blue = all components.
The solid lines show weighted and unweighted fits to the data (not including the Sco-Oph and Trapezium sight lines); the dotted lines have slope = 2 (as would be expected from considerations of ionization balance).
Toward HD~62542, $N$(\ion{Na}{1}) is consistent with the general Galactic relationship, $N$(\ion{C}{1}) is somewhat enhanced, and $N$(\ion{Ca}{1}) and $N$(\ion{Fe}{1}) are deficient -- reflecting the specific ionization and depletion behavior of each of those species in this unusual sight line (see Sec.~\ref{sec-depltr}).
The lower abundances of the trace neutral species in most of the Sco-Oph and Trapezium sight lines may be due to stronger than average radiation fields there (e.g., Welty \& Hobbs 2001).}
\label{fig:tr_htot}
\end{figure}

\subsection{Atomic Species}
\label{sec-atom}

The combination of optical and UV spectra obtained for HD~62542 has yielded detections or useful limits for many neutral and singly ionized atomic species.
For the species with potentially detectable lines covered by those spectra, Table~\ref{tab:cdatom} lists both the total column densities and (where discernible) the separate contributions from the main component and from the sum of all the other components, as derived from either the detailed profile fits or the AOD integrations.
The table also lists values for elemental depletions, based on the adopted $N$(H$_{\rm tot}$), the column densities of the dominant ions in predominantly neutral gas, and the protosolar abundances given by Lodders (2003).
We have adopted the recommended protosolar abundances -- which are 0.07--0.08 dex higher than the current solar values -- to facilitate comparisons with the depletion parameters determined in the recent abundance surveys of Jenkins (2009) and Ritchey et al. (2018).
Figures~\ref{fig:dom_htot} and \ref{fig:tr_htot} show the total column densities of a few of the singly ionized and neutral atomic species (respectively), versus $N$(H$_{\rm tot}$), in relation to the general trends for those species found for the local Galactic ISM.
The Galactic data have been drawn from a number of sources:  for \ion{Na}{1}, \ion{C}{1}, and H$_{\rm tot}$ (Welty \& Hobbs 2001; Jenkins 2009; Jenkins \& Tripp 2011; Welty et al. 2016); for \ion{Ca}{1} and \ion{Fe}{1} (Welty et al. 2003); for \ion{Ti}{2} (Welty \& Crowther 2010); for \ion{Fe}{2}, \ion{P}{2}, and \ion{Zn}{2} (Jenkins 2009); and references therein; an updated compilation is maintained at http://astro.uchicago.edu/$\sim$dwelty/coldens.html).
Some sight lines in the Sco-Oph and Orion Trapezium regions exhibit lower abundances of some trace neutral species (Welty \& Hobbs 2001) and/or enhanced depletions of some refractory elements (e.g., Welty \& Crowther 2010), and are noted with blue squares and blue asterisks, respectively. 
In several cases, the sight line toward HD~62542 (blue star) stands apart from the general trends -- due largely to the unusual characteristics of the main (14 km~s$^{-1}$) component (red star); the abundances in the other components (green star) are more similar to those in the general Galactic ISM.
Differences in the absorption-line profiles of the various atomic species toward HD~62542 (Fig.~\ref{fig:atom}) reflect those component-to-component differences in abundances -- which are due to differences in ionization, depletion, and/or other local physical conditions -- and/or (in some cases) the saturation of specific lines.
Some column density ratios illustrating the effects of those differences for the individual components are given in Table~\ref{tab:arat} and are discussed in the following sections.

\subsubsection{Dominant Species}
\label{sec-dom}

For dominant ions in predominantly neutral atomic or molecular gas, the slopes of the relationships between $N$(X) and $N$(H$_{\rm tot}$) seen in Figure~\ref{fig:dom_htot} reflect (to some degree) the depletion behavior of those species.
For the typically mildly depleted oxygen, for example, the slope $\sim$ 1.0 suggests that the depletion generally does not become much more severe at higher $N$(H$_{\rm tot}$) (where there might be expected to be larger fractions of colder, denser gas).
For the more severely depleted Ti and Fe, however, the slopes $<$ 1.0 suggest that the depletions of those species do become increasingly severe at higher overall column densities.\footnotemark
\footnotetext{While the slope for zinc also appears to be $<$ 1.0, at least part of that trend may be due to underestimation of $N$(\ion{Zn}{2}) at the higher $N$(H$_{\rm tot}$), where the only two detectable \ion{Zn}{2} lines (at 2026.1 and 2062.7 \AA) can be significantly saturated.}
The observed differences in absorption-line profiles (right-hand panel of Fig.~\ref{fig:atom}) then primarily reflect differences in those elemental depletions in the various individual components.
For example, the profiles for the typically mildly depleted \ion{P}{2} and \ion{Zn}{2}, the typically moderately to severely depleted \ion{Mg}{2} and \ion{Si}{2}, and the typically more severely depleted \ion{Fe}{2} and \ion{Ni}{2} exhibit similarities within each pair (most obviously for lines of similar strength), but differences relative to the other pairs.
Those qualitative indications are confirmed by the detailed fits to the line profiles (Table~\ref{tab:acmp}) and by the resulting component-by-component ratios listed in Table~\ref{tab:arat}.
For the mildly depleted \ion{P}{2} and \ion{Zn}{2}, the strongest absorption is in the main component near 14 km~s$^{-1}$; for \ion{Mg}{2} and \ion{Si}{2}, the strongest absorption is in the components near 18 and 27 km~s$^{-1}$; for the severely depleted \ion{Fe}{2} and \ion{Ni}{2}, the strongest absorption is spread over several components, near 12, 14, 18, and 27 km~s$^{-1}$.
While the column densities of the typically similarly depleted elements within each of those three pairs track each other fairly well, the ratios for species with different depletion properties exhibit significant variations.
In the main (14 km~s$^{-1}$) component, for example, the $N_{\rm m}$(\ion{Si}{2})/$N_{\rm m}$(\ion{Zn}{2}) and $N_{\rm m}$(\ion{Fe}{2})/$N_{\rm m}$(\ion{Zn}{2}) ratios are much lower, compared to the corresponding (and more similar) values in all the other components.
As the \ion{Mg}{2} $\lambda\lambda$1239,1240 lines are significantly weaker than the \ion{Si}{2} $\lambda$1808 line, the roughly constant ratio of the column densities of \ion{Mg}{2} and \ion{Si}{2} -- both toward HD~62542 and in other Galactic sight lines (e.g., Fitzpatrick 1997; Cartledge et al. 2006) -- also provides some corroboration for the apparent relative insensitivity of the main component $N_{\rm m}$(\ion{Si}{2}) to the $b$ value adopted in the fits.
Toward HD~62542, the column densities of some of the dominant ions of depleted elements (e.g., \ion{Ti}{2}, \ion{Fe}{2}) fall well below the general trends versus $N$(H$_{\rm tot}$) seen in the local Galactic ISM -- both for the sight line as a whole and (especially) for the main component (lower two panels of Fig.~\ref{fig:dom_htot}).

In principle, the weak semiforbidden $\lambda$2325 \ion{C}{2}] and $\lambda$2335 \ion{Si}{2}] lines can provide accurate gas-phase column densities for those two species -- thus providing important constraints on the composition of interstellar dust grains (e.g., Sofia et al. 2004; Miller et al. 2007).
Both lines are covered in our STIS E230H spectra, but neither is convincingly detected toward HD~62542 at the S/N ratios achieved in that wavelength range.
While there are weak possible absorption features near the expected positions of both of the lines in the summed spectra, the equivalent widths of those weak features are less than 3$\sigma$, they do not appear in all of the individual exposures, and their velocities do not agree with the velocities of the strongest component(s) seen for similarly distributed species.
For \ion{Si}{2}, the column densities listed in Table~\ref{tab:cdatom}, based on the detailed fits to the $\lambda$1808 line, are consistent with the limits derived from the nondetection of the $\lambda$2335 line.
For \ion{C}{2}, the interstellar absorption from the strong $\lambda$1334 line is buried in the core of the strong stellar \ion{C}{2} line, and it is thus very difficult to obtain column densities for \ion{C}{2} directly.
A rough estimate of $N_{\rm m}$(\ion{C}{2}) in the main component may be obtained, however, from the adopted $N_{\rm m}$(H$_{\rm tot}$) and the carbon depletion predicted for the depletion index F$_*$ $\sim$ 1.5 found for that component.
For the depletion coefficients for carbon derived by Jenkins (2009), $N_{\rm m}$(\ion{C}{2}) would be of order 2.5 $\times$ 10$^{17}$ cm$^{-2}$ -- well below the upper limit obtained from the nondetection of the $\lambda$2325 line.

\begin{figure}
\epsscale{0.6}
\plotone{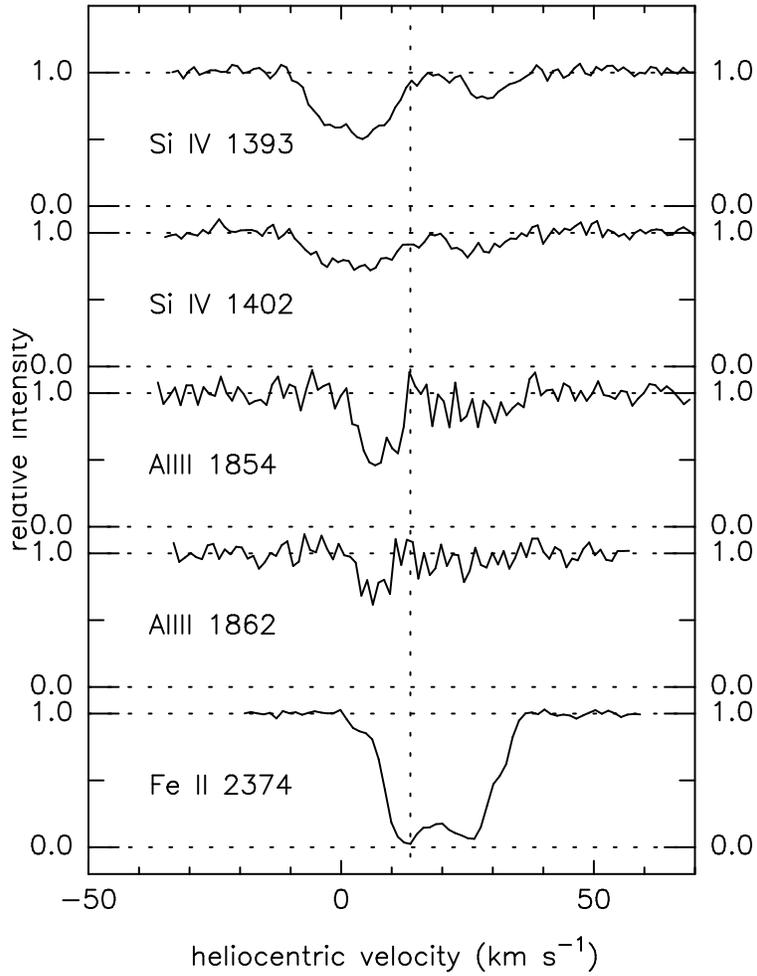}
\caption{Normalized spectra of the $\lambda\lambda$1393, 1402 \ion{Si}{4} and $\lambda\lambda$1854, 1862 \ion{Al}{3} lines toward HD~62542.
The profile of the $\lambda$2374 line of \ion{Fe}{2} (the dominant form of Fe in primarily neutral gas) is shown for comparison.
The absorption from the higher ions is not detected for the main neutral gas component near 14 km~s$^{-1}$, but does overlap in velocity with the absorption from the other, weaker neutral gas components.}
\label{fig:high}
\end{figure}

\subsubsection{Trace Neutral Species}
\label{sec-trace}

The moderately high $N$(H$_{\rm tot}$) toward HD~62542, the steep far-UV extinction, and the quality and coverage of the optical and UV spectra have also yielded detections or fairly stringent limits for a number of trace neutral species -- primarily in the main component at 14 km~s$^{-1}$.
The column densities of \ion{C}{1}, \ion{Na}{1}, \ion{Mg}{1}, and \ion{S}{1}, and the ratio of the column densities of \ion{Mg}{1} and \ion{Mg}{2}, for example, are more than an order of magnitude higher in the main component than in the other components (Tables~\ref{tab:acmp}, \ref{tab:cdatom}, and \ref{tab:arat}).
In the local Galactic ISM, the column densities of trace neutral ions of mildly depleted elements (e.g., \ion{C}{1}, \ion{Na}{1}, \ion{K}{1}) exhibit nearly quadratic relationships with $N$(H$_{\rm tot}$), while the corresponding relationships for the neutral ions of more severely depleted elements (e.g., \ion{Ca}{1}, \ion{Fe}{1}) are somewhat shallower (Welty \& Hobbs 2001; Welty et al. 2003; Fig.~\ref{fig:tr_htot}).
Toward HD~62542, the column densities of \ion{Na}{1}, \ion{Mg}{1}, \ion{Si}{1}, \ion{K}{1}, and \ion{Cl}{1} appear fairly ''normal'', relative to those local Galactic trends; the column densities of \ion{C}{1}, \ion{P}{1}, \ion{S}{1}, and \ion{Zn}{1} are somewhat enhanced; and the column densities of \ion{Ca}{1} and \ion{Fe}{1} are somewhat deficient (upper and lower panels of Fig.~\ref{fig:tr_htot}) -- though the currently available samples for \ion{Mg}{1}, \ion{Si}{1}, \ion{P}{1}, and \ion{Zn}{1} are fairly small.\footnotemark
\footnotetext{To our knowledge, column densities for \ion{P}{1} and \ion{Zn}{1} have not previously been reported.  The values referenced here are from measurements of absorption from those trace neutral species seen in archival {\it HST} spectra (D. E. Welty, in preparation).}
As discussed below (Sec.~\ref{sec-depl}), the observed abundances of the trace neutral species toward HD~62542 -- and the deviations of some of them from the general Galactic trends -- reflect both the ionization behavior and the depletion behavior of those species in the main component.

\subsubsection{Higher Ions}
\label{sec-hion}

Interstellar absorption from several of the commonly observed more highly ionized species is also detected in our STIS spectra of HD~62542 (Fig.~\ref{fig:high}). 
Fits to the line profiles suggest that \ion{Al}{3} has components at 7 and 26 km~s$^{-1}$, with $b$ $\sim$ 3.3 and 7.0 km~s$^{-1}$, respectively, and \ion{Si}{4} has components at $-$4, 5, and 27 km~s$^{-1}$, with $b$ $\sim$ 3.3, 6.9, 5.3 km~s$^{-1}$, respectively; the total column densities for those higher ions are listed in Table~\ref{tab:cdatom}.
Any absorption from interstellar \ion{Si}{3} is blended with a strong stellar line, however, and neither the \ion{C}{4} $\lambda\lambda$1548, 1550 doublet nor the $\lambda$1190 \ion{S}{3} line is covered by our STIS spectra.


\begin{figure}
\epsscale{0.9}
\plotone{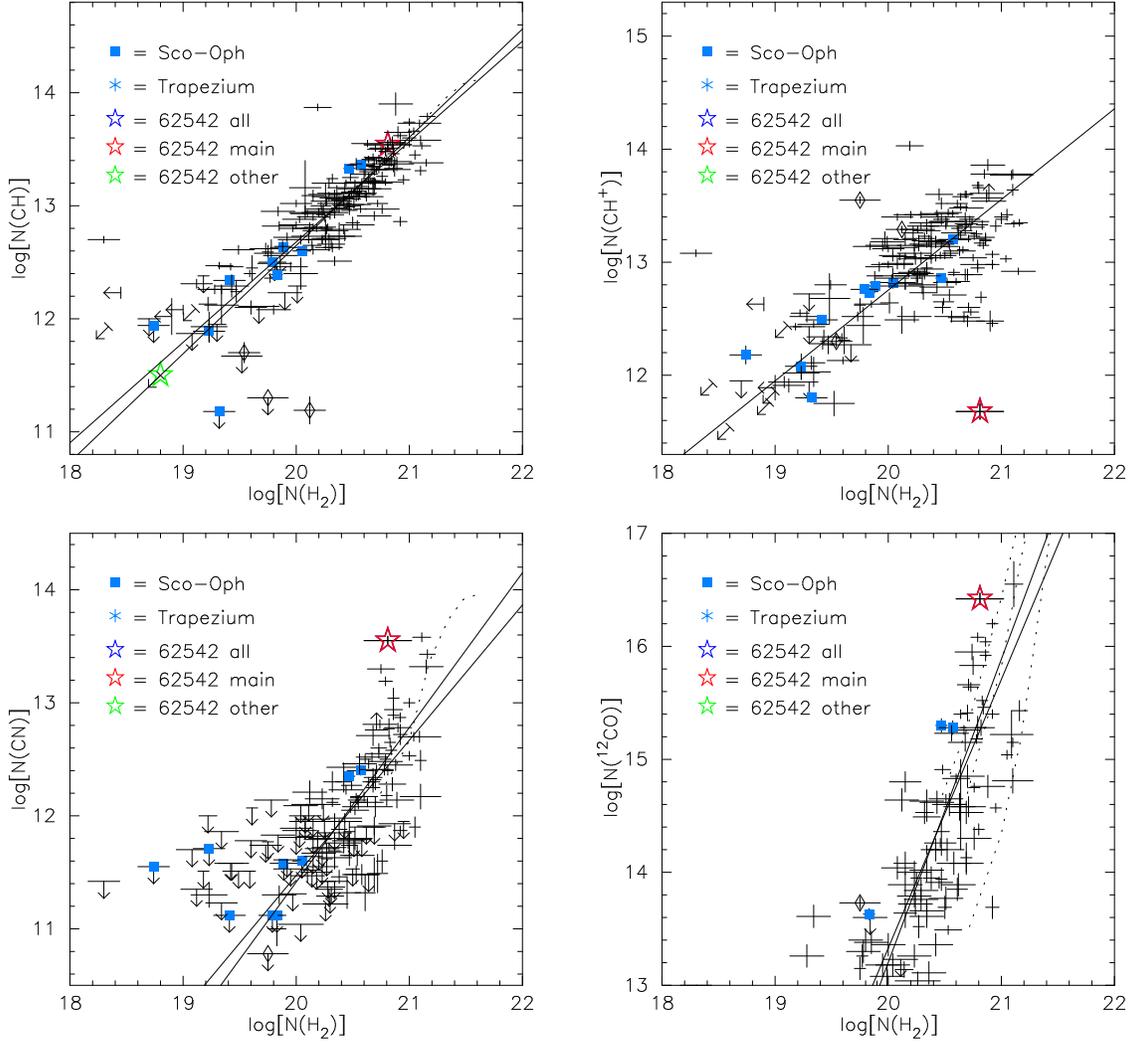}
\caption{Column densities of CH, CH$^+$, $^{12}$CO, and CN vs. $N$(H$_2$), for Galactic sight lines.
The large red star denotes the sight line toward HD~62542 (main component); the open diamonds denote Pleiades sight lines.
The solid lines show weighted and unweighted fits to the data; the dotted lines represent the models of van Dishoeck \& Black (1988, 1989); the three models for $^{12}$CO are (left to right) for $I_{\rm UV}$ = 0.5, 1.0, and 10 times the average local Galactic radiation field.
Relative to the mean trends with $N$(H$_2$), $N_{\rm m}$(CH) is ''normal'' toward HD~62542, $N_{\rm m}$(CH$^+$) is very low, and $N_{\rm m}$($^{12}$CO) and $N_{\rm m}$(CN) are high.}
\label{fig:molvsh2}
\end{figure}

\clearpage

\begin{figure}
\epsscale{0.9}
\plotone{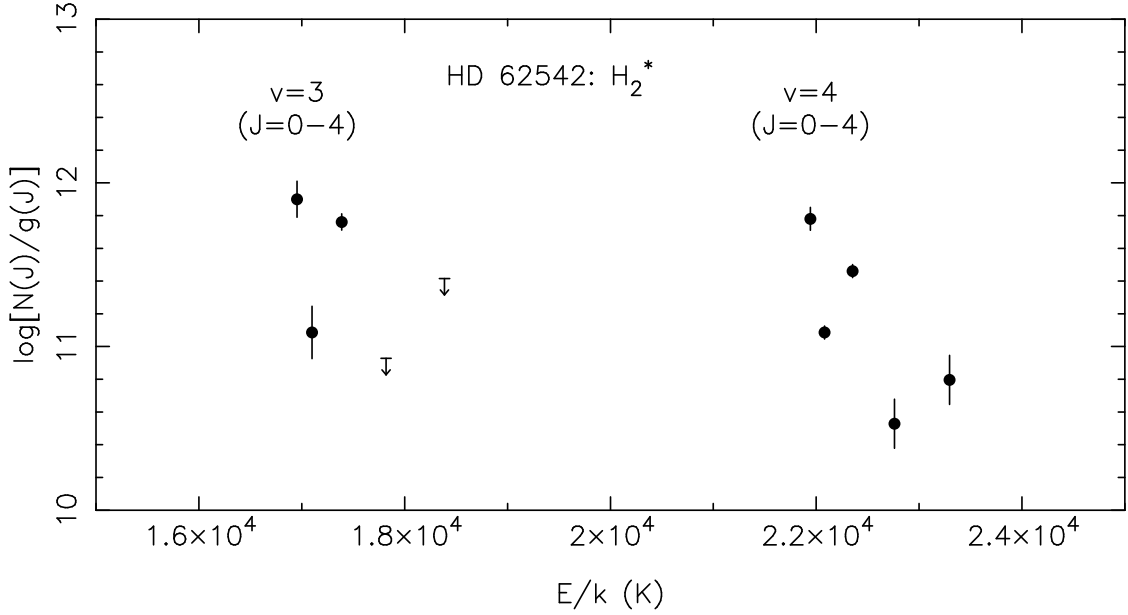}
\caption{Normalized rotational populations for vibrational levels v = 3 and 4 of H$_2$* toward HD~62542.
Only a few weak lines, from rotational levels $J$ = 0--4, are detected (in the main component).
The excitation is qualitatively similar to that seen toward HD~37903 and HD~147888 -- for each v, the normalized populations of the even-$J$ levels lie above those for the odd-$J$ levels, but with similar slopes (and thus similar excitation temperatures) for both.}
\label{fig:h2ex}
\end{figure}

\subsection{Molecular Species}
\label{sec-mol}

The available optical and UV spectra of HD~62542 have also yielded detections of a number of molecular species (H$_2$*, CH, CH$^+$, C$_2$, C$_3$, $^{12}$CN, $^{13}$CN, $^{12}$CO, $^{13}$CO, C$^{18}$O, CS, NH, OH), as well as limits for some others (CH$_2$, C$_2^+$, C$_2^-$, CN$^+$, CO$^+$, NO, NO$^+$, OH$^+$, H$_2$O, SiO, SH, SH$^+$, HCl, AlH).
All of the molecular species are found only in the main (14 km~s$^{-1}$) component (Fig.~\ref{fig:mol}).  
As noted above, fits to the high-S/N ratio Keck spectrum of the strong CH $\lambda$4300 line yield a 3-$\sigma$ upper limit of order 3 $\times$ 10$^{11}$ cm$^{-2}$ for the other components [i.e., less than 1\% of $N_{\rm m}$(CH)], and so we take $N$ = $N_{\rm m}$ for the various molecular species.
Table~\ref{tab:cdmol} lists the main-component column densities for those molecular species, derived from the AOD integrations over the line profiles and/or from detailed fits to the profiles, together with (in several cases) the $b$ values obtained in the fits and the excitation temperatures derived from the relative rotational populations.
Comparisons of the molecular column densities seen toward HD~62542 with general trends found in the local Galactic ISM reveal some distinct differences -- e.g., the very low CH$^+$/CH ratio and the very high CN/CH and C$_3$/C$_2$ ratios known from previous studies (Cardelli et al. 1990; Gredel et al. 1991, 1993; \'{A}d\'{a}mkovics et al. 2003).
For example, the weak absorption now detected for the CH$^+$ $\lambda$4232 line (Fig.~\ref{fig:mol}) indicates that $N_{\rm m}$(CH$^+$) is lower by a factor of $\sim$40 than for other sight lines with comparable $N$(H$_2$) (Fig.~\ref{fig:molvsh2}).
While the abundances of CH, OH, and CS toward HD~62542 are fairly typical, relative to H$_2$, the column densities of C$_2$, C$_3$, NH, $^{12}$CO, $^{13}$CO, and (especially) CN -- all of which trace somewhat denser gas -- are higher than those found for other sight lines with similar $N$(H$_2$) (Fig.~\ref{fig:molvsh2}).

{\bf H$_2$*:}  
A few weak lines from the lower rotational levels of the B-X (0,3) and (0,4) bands of vibrationally excited H$_2$ are detected in the UV spectra (Table~\ref{tab:ewmol}). 
Fits to the line profiles indicate that the lines are fairly narrow ($b$ $\sim$ 1.4 km~s$^{-1}$) and are found at the velocity (13.7$\pm$0.6 km~s$^{-1}$) of the main cloud (Fig.~\ref{fig:mol}) -- as seems to be the case for the few other sight lines in which H$_2$* absorption has been detected in high-resolution UV spectra [HD~37021 (Abel et al. 2016); HD~37061 (Gnaci\'{n}ski 2009); HD~37903 (Meyer et al. 2001); $\rho$ Oph D (Gnaci\'{n}ski 2013)].
While H$_2$* can be produced either by shocks or by radiative pumping (e.g., Shull \& Beckwith 1982), the relatively small $b$ value and the lack of velocity offset (relative to the other molecular species) suggest that the H$_2$* is excited radiatively in relatively cool gas, rather than by shocks.
The H$_2$* excitation in the main component toward HD~62542 (Figure~\ref{fig:h2ex}) is qualitatively similar to that seen toward HD~37903 (Meyer et al. 2001; Gnaci\'{n}ski 2011) and $\rho$~Oph~D (Gnaci\'{n}ski 2013) -- for each vibrational level (v), the normalized populations in the lowest even-$J$ (para) levels are higher than those in the lowest odd-$J$ (ortho) levels, but with similar slopes (and thus similar excitation temperatures) for both even and odd $J$.
As the column densities of the detected individual H$_2$* rotational levels are typically 5--25 $\times$ 10$^{11}$ cm$^{-2}$, the vibrationally excited H$_2$ comprises only a small fraction of the total H$_2$.

 
\begin{figure}
\epsscale{1.0}
\plotone{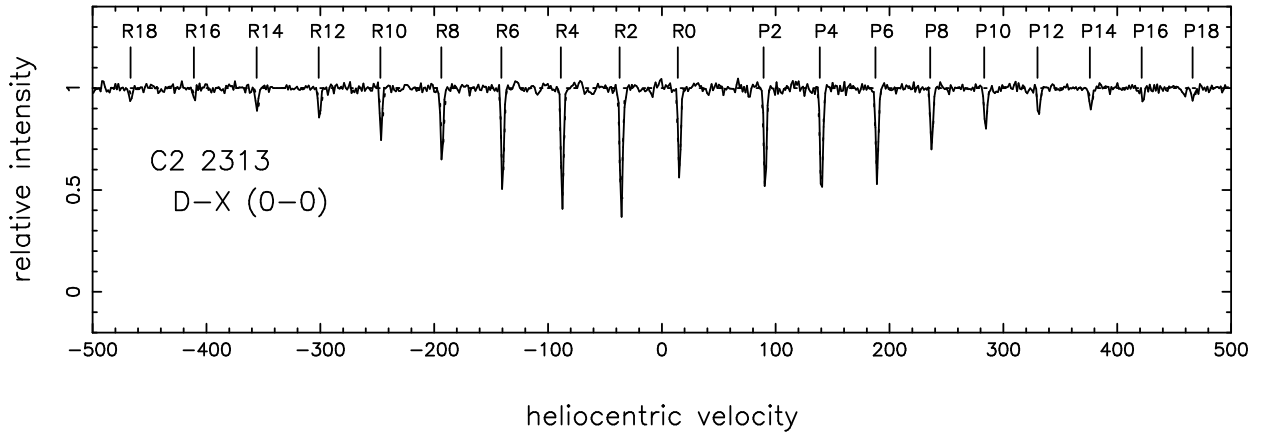}
\caption{Normalized spectrum of the C$_2$ D-X (0-0) band near 2313 \AA\ toward HD~62542.
The solid line is the observed spectrum; the dotted line is the fit for $b$ = 0.7 km~s$^{-1}$.
Absorption from rotational levels $J$ = 0--18 (even $J$ only) may be discerned.}
\label{fig:c2dx}
\end{figure}

\begin{figure}
\epsscale{0.9}
\plotone{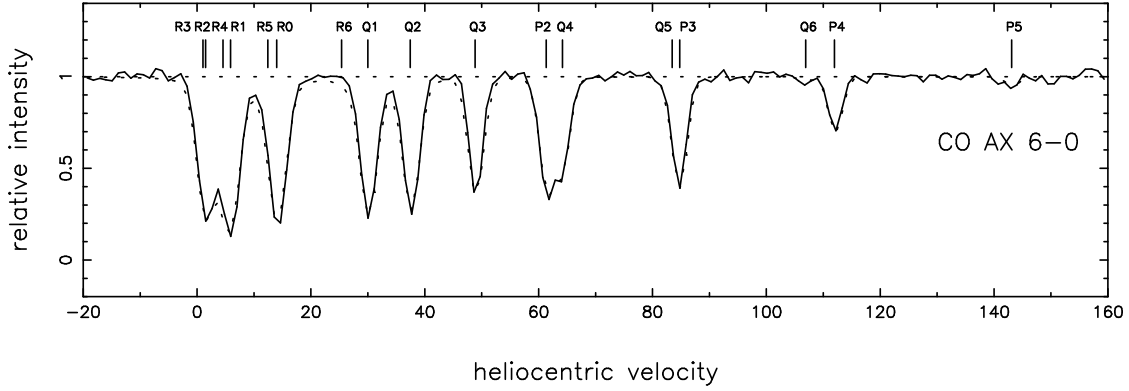}
\caption{Normalized spectrum of the $^{12}$CO A-X (6-0) band near 1367 \AA\ toward HD~62542.
The solid line is the observed spectrum; the dotted line is the fit for $b$ = 0.5 km~s$^{-1}$.
Absorption from rotational levels $J$ = 0--6 may be discerned.}
\label{fig:coax6}
\end{figure}

\begin{figure}
\epsscale{0.9}
\plottwo{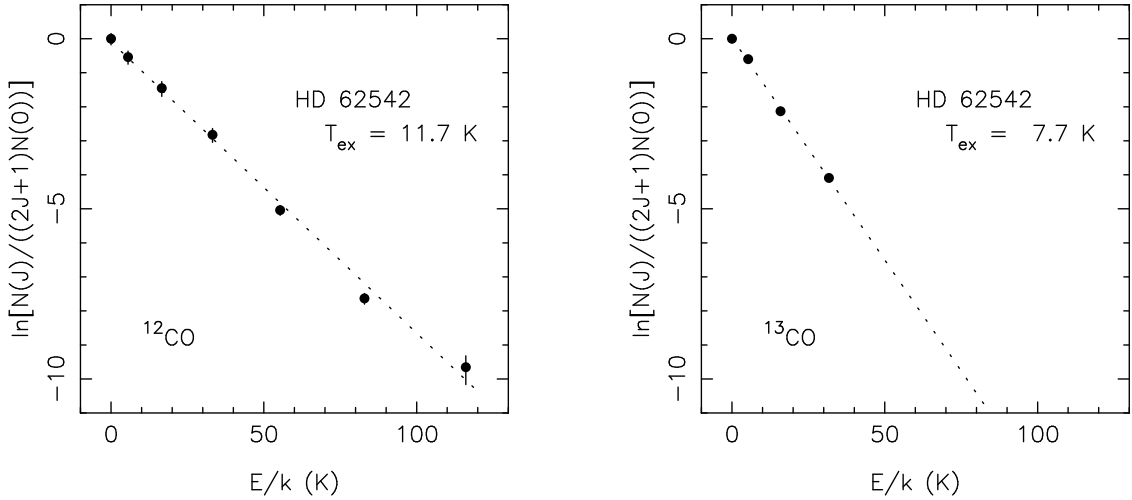}{co13nj.eps}
\caption{Rotational excitation of $^{12}$CO ({\it left}) and $^{13}$CO ({\it right}) toward HD~62542.
The lines represent fits to all detected rotational levels.
For both species, the corresponding excitation temperatures are both independent of rotational level and higher than the 3--5 K values typically seen in the Galactic ISM.}
\label{fig:corot}
\end{figure}

{\bf C$_2$:}  
Six C$_2$ bands are detected: the A-X (3,0), (2,0), and (1,0) bands near 7719, 8757, and 10143 \AA, respectively; the D-X (0,0) band near 2313 \AA; and the F-X (1,0) and (0,0) bands near 1314 and 1341 \AA, respectively.
Fits to the relatively strong, well-separated lines of the D-X (0,0) band (Fig.~\ref{fig:c2dx}) yield the best characterization of the rotational populations, with well-determined column densities for rotational levels $J$ = 0--18 (even $J$ only; Appendix Table~\ref{tab:ewmol}).
Fits to the weaker A-X (2,0) and (3,0) bands yield slightly lower column densities (for $J$=0--8), but the relative $N_{\rm m}$($J$) are similar to those found for the D-X (0,0) band.
Fits to the F-X (1,0) band yield column densities lower by about 30\% (for $J$=0--14), but (again) the relative $N_{\rm m}$($J$) are similar to those found for the D-X (0,0) band.
Perturbations affect the positions and/or relative strengths of some of the lines in the F-X (0,0) band (Hupe et al. 2012), and there may be blending with stellar absorption around the R-branch ''pile-up'' near 1341.4 \AA.
The total $N_{\rm m}$(C$_2$) = 9.5 $\times$ 10$^{13}$ cm$^{-2}$ includes contributions from undetected higher rotational levels, estimated using the temperature $T_{\rm k}$ = 40 K and density of collision partners $n_{\rm c}$ = $n_{\rm H}$ + $n_{\rm H_2}$ = 250 cm$^{-3}$ derived from the observed $J$ = 0--18 levels (see Sec.~\ref{sec-dens} below).
Previously available measurements of the C$_2$ A-X (2,0) band (Gredel et al. 1993) included only $J$=0--8, so that $T_{\rm k}$ and $n_{\rm c}$ were not as well constrained.

{\bf CN:}  
Three CN bands are detected in the optical spectra:  the B-X (1,0) and (0,0) bands, near 3580 and 3875 \AA, respectively; and the A-X (2,0) band near 7906 \AA\ (Appendix Table~\ref{tab:ewmol}). 
Lines from $J$=0 and 1 are detected in all three bands, and weak lines from $J$=2 are detected in the strong B-X (0,0) band.
In the main cloud toward HD~62542, the CN excitation temperature $T_{01}$ = 2.89$\pm$0.15 K (Table~\ref{tab:cdmol}) is slightly higher than both the 2.725$\pm$0.002 K expected from excitation by the cosmic microwave background radiation (CMBR; Mather et al. 1999) and the average value 2.754$\pm$0.002 K for 11 Galactic sight lines in the sample of Ritchey et al. (2011).
The even-higher $T_{01}$ = 3.7$\pm$0.4 K found by Cardelli et al. (1990) was likely due to the higher $b$ value adopted for CN in that study (see also Gredel et al. 1991; Palazzi et al. 1992).
For $^{13}$CN, only absorption from $J$ = 0 is reliably detected; the total column density is estimated assuming the same excitation temperature found for the more abundant $^{12}$CN.
With that assumption, the ratio $N_{\rm m}$($^{12}$CN)/$N_{\rm m}$($^{13}$CN) $\sim$ 79$\pm$13 is consistent with (but perhaps slightly higher than) the average local Galactic ratio $^{12}$C/$^{13}$C $\sim$ 70 (e.g., Stahl \& Wilson 1992; Sheffer et al. 2002b).

{\bf CO:}  
A number of CO bands, from several isotopologs, are detected in the UV spectra (Appendix Table~\ref{tab:cobands} and Figs.~\ref{fig:co12} and \ref{fig:co13}):  for $^{12}$CO, nine of the permitted A-X bands ((4,0) through (12,0)) and some of the much weaker intersystem bands (Morton \& Noreau 1994; Sheffer et al. 2002a; Eidelsberg \& Rostas 2003); for $^{13}$CO, five of the permitted A-X bands ((4,0) through (8,0)) and the d12 intersystem band; and for C$^{18}$O, the two permitted A-X (4,0) and (5,0) bands.
The stronger $^{12}$CO A-X bands reveal absorption from rotational levels up to $J$=6 (Fig.~\ref{fig:coax6}), with the relative populations of all detected levels reasonably described by an excitation temperature $T_{\rm ex}$ = 11.7 K (Fig.~\ref{fig:corot}). 
Rotational levels up to $J$=3 are detected for $^{13}$CO, with a corresponding common $T_{\rm ex}$ = 7.7 K.
Such high excitation temperatures are unusual, as most sight lines with comparable $N$(H$_{\rm tot}$) and $A_{\rm V}$ in which CO has been detected have $T_{\rm ex}$($^{12}$CO) $\sim$ $T_{\rm ex}$($^{13}$CO) $\sim$ 3--5 K; only HD~147888 exhibits a $T_{\rm ex}$($^{12}$CO) comparable to that seen toward HD~62542 (Sonnentrucker et al. 2007; Burgh et al. 2007; Sheffer et al. 2008).
The $N_{\rm m}$($^{12}$CO)/$N_{\rm m}$($^{13}$CO) ratio found toward HD~62542, $\sim$ 62$\pm$8, is consistent with (but perhaps slightly lower than) the average local $^{12}$C/$^{13}$C ratio.

{\bf CS:}  
For the relatively small sample of sight lines in which the $\lambda$1400 CS feature has been detected (Destree et al. 2009), only HD~154368 has a higher $W$(1400) than the 9.3 m\AA\ measured toward HD~62542.
As for several other sight lines noted by Destree et al., the available rotational structure for that feature does not yield a good fit to the observed broad CS profile.

{\bf OH$^+$:}
While a weak possible absorption feature is present in the UVES spectrum near the expected location of the strongest OH$^+$ line at 3583.8 \AA, similar weak features are also seen in the spectra of several lightly reddened stars obtained on the same night; they appear to be residual detector pattern features (Porras et al. 2014).
The 3$\sigma$ upper limit on the equivalent width ($W$ $<$ 0.39 m\AA) obtained after removing that residual feature corresponds to a column density limit $N_{\rm m}$(OH$^+$) $<$ 6.5 $\times$ 10$^{12}$ cm$^{-2}$, using the recently revised $f$ values of Hodges et al. (2018).

{\bf UID:}  
The unidentified line at 1300.45 \AA\ (Destree \& Snow 2009) seems to be present particularly in sight lines with fairly strong CN absorption.
As that unidentified line is stronger toward HD~62542 than toward X~Per by a factor $\sim$ 3 -- similar to the differences observed for CN, C$_2$, and CS in the two sight lines -- it may be due to some molecular species that is preferentially present in relatively dense gas.


{\bf DIBs:}  
While most of the commonly observed DIBs are quite weak toward HD~62542, a number of the C$_2$-DIBs have been detected there (Snow et al. 2002b; \'{A}d\'{a}mkovics et al. 2005).
Equivalent widths for some of the DIBs that could be measured in the UCLES and UVES spectra are listed at the end of Appendix Table~\ref{tab:ewmol}.
In most cases, the equivalent widths agree reasonably well with the values obtained by \'{A}d\'{a}mkovics et al. (2005) and Fan et al. (2017).
Our values for the $\lambda$5780.6 and $\lambda$5797.2 DIBs are somewhat smaller, however, due to different choices for the local continua and extent of the absorption (e.g., Fan et al. 2017), and our somewhat tentative detection of the broader $\lambda$6284.1 DIB is well above the limit given by \'{A}d\'{a}mkovics et al.

\clearpage

\section{DISCUSSION}
\label{sec-disc}

\subsection{Physical Conditions in the Main Cloud}
\label{sec-phys}

In the main cloud that dominates the absorption toward HD~62542, the extensive set of measured atomic and molecular column densities yields multiple diagnostics for the temperature, local hydrogen density, and electron density.
Differences in the properties inferred from the various diagnostics may either suggest shortcomings in our understanding of the diagnostics or provide information about nonuniformities in the structure of the cloud.
This relatively simple, isolated cloud -- which may be representative of the cores of other diffuse molecular clouds (\'{A}d\'{a}mkovics et al. 2005) -- should provide a useful test for cloud models.

\begin{figure}
\epsscale{0.6}
\plotone{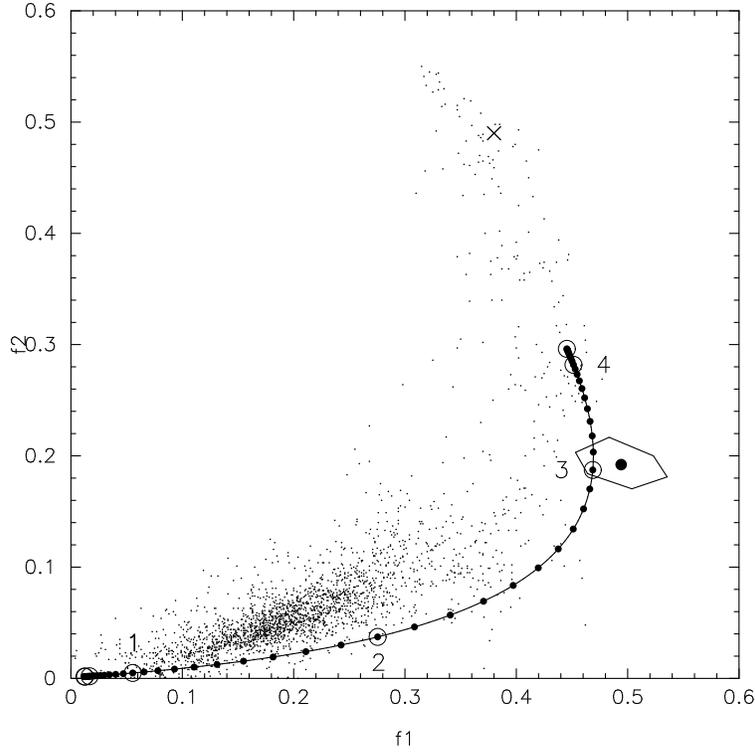}
\caption{Relative \ion{C}{1} fine-structure populations [$f_1$ = $N$(\ion{C}{1}*)/$N$(\ion{C}{1}$_{\rm tot}$); $f_2$ = $N$(\ion{C}{1}**)/$N$(\ion{C}{1}$_{\rm tot}$)] for main component toward HD~62542 (large black dot with 1-$\sigma$ error box), compared to Galactic values from Jenkins \& Tripp (2011) (small black dots).
The curve gives theoretical predictions for $T$ = 43 K (from H$_2$) and the WJ1 radiation field; the small circles along the curve indicate steps in log($n_{\rm H}$) of 0.1 dex; the larger open circles indicate log($n_{\rm H}$) = 4.0, 3.0, 2.0, 1.0, etc.
The slightly stronger radiation field estimated for the main cloud toward HD~62542 would only affect the predictions at the lowest densities.
The excitation in the main component toward HD~62542 is consistent with $n_{\rm H}$ $\sim$ 1500 cm$^{-3}$ (adjusted to the scale of Jenkins \& Tripp).}
\label{fig:c1fs}
\end{figure}

\subsubsection{Temperature}
\label{sec-tk}

Estimates for the local kinetic temperature ($T_{\rm k}$) may be obtained from observations of the rotational excitation of molecules (particularly the homonuclear H$_2$ and C$_2$), of the fine-structure excitation of \ion{O}{1}, and of the widths of absorption (or emission) lines in high-resolution spectra.

{\bf H$_2$ and C$_2$ rotational excitation:} 
Toward HD~62542, the ratio $N_{\rm m}$($J$=1)/$N_{\rm m}$($J$=0) for H$_2$ yields $T_{\rm k}$ = 43$\pm$11 K (Rachford et al. 2002).\footnotemark
\footnotetext{Unfortunately, the S/N ratios in the available {\it FUSE} spectra of HD~62542 are not high enough to permit reliable measurement of the higher H$_2$ rotational levels, which otherwise could yield constraints on the density and ambient UV radiation field (Jensen et al. 2010).}
Analysis of the rotational excitation of C$_2$, for $J$ = 0--18, yields $T_{\rm k}$ $\sim$ 40$\pm$5 K (Sec.~\ref{sec-dens}) -- consistent with a general tendency for the temperature derived from C$_2$ to be less than or equal to the value obtained from H$_2$ (Sonnentrucker et al. 2007).
While these are the best estimates currently available for the average $T_{\rm k}$ in the main component toward HD~62542 (Table~\ref{tab:cdmol}), some variations may be expected within that main cloud.
The somewhat higher excitation temperature derived from the lowest ($J$ = 0--14) rotational levels of C$_3$, $T_{\rm ex}$ $\sim$ 75 K -- which is not uncommon for the relatively small sample of sight lines with data for both C$_2$ and C$_3$ -- may indicate a radiative contribution to the excitation (\'{A}d\'{a}mkovics et al. 2003).
The subthermal excitation temperatures obtained for the polar molecules $^{12}$CO (11.7 K), $^{13}$CO (7.7 K), and CN ($\sim$2.9 K) reflect a minimum excitation due to the CMBR, with additional excitation (from collisions and/or local CO emission) for the two CO isotopologs.

{\bf \ion{O}{1} fine-structure excitation:}
In relatively cool, neutral, primarily atomic clouds, the relative populations in the excited fine-structure states of neutral oxygen are set primarily by a balance between collisional excitation (with \ion{H}{1}) and radiative decay -- and the ratio $N$(\ion{O}{1}*)/$N$(\ion{O}{1}**) can yield an estimate for $T_{\rm k}$ (e.g., Jenkins \& Tripp 2011).
The observed value for that ratio in the main component toward HD~62542, $\sim$ 1.9$\pm$0.3 (for $b$ = 1.0$\pm$0.1 km~s$^{-1}$), corresponds to a temperature of about 110 K (from Fig.~15 in Jenkins \& Tripp 2011); a slightly higher value is obtained from the more recent analysis of Ritchey et al. (2019; see Sec.~\ref{sec-dens}).


{\bf Line widths:}
The widths of interstellar absorption features are usually considered to be due to a combination of thermal and internal turbulent broadening, with the line-width parameter $b$ = ($2kT_k/m + 2v_t^2$)$^{1/2}$, where $m$ is the atomic or molecular weight and $v_t$ is the ''turbulent'' velocity.
While the main-component $b$ values measured for various atomic and molecular species toward HD~62542 are all fairly small -- ranging from about 0.5 km~s$^{-1}$ (for CO) to about 1.5 km~s$^{-1}$ (for \ion{Ca}{2}) -- the strongest constraint on the temperature comes from the weak lines of vibrationally excited H$_2$ (the lightest species for which accurate line widths could be determined).
The width derived from fits to the H$_2$* line profiles, $b$ = 1.4 km~s$^{-1}$, corresponds to a maximum temperature of about 230 K; the actual $T_{\rm k}$ would be smaller if turbulent broadening contributes to the line widths.
For the component near 27 km~s$^{-1}$ the larger $b$ value for \ion{Na}{1} (1.4 km~s$^{-1}$, compared to 0.85 km~s$^{-1}$ for the main component) suggests that the gas there is somewhat warmer, but less than about 2700 K (for purely thermal broadening of the \ion{Na}{1} lines).
While the thermal and turbulent contributions to the line broadening of a particular component may (in principle) be separated by comparing the line widths for species of sufficiently different mass, such comparisons are less straightforward if the various species are not entirely coextensive.

\subsubsection{Hydrogen Density}
\label{sec-dens}

Estimates for the local total hydrogen density $n_{\rm H}$ = $n_{\rm H^0}$ + 2$n_{\rm H_2}$ may be obtained from observations of the fine-structure excitation of \ion{C}{1}, of the rotational excitation of C$_2$ and CO, and (via simple chemical models) of the abundances of CH, C$_2$, CN, and CO.
For clouds with a significant molecular fraction, the densities obtained from \ion{C}{1} fine-structure excitation are often significantly smaller than those estimated from molecular diagnostics such as CN or C$_2$.
Such a difference could arise if \ion{C}{1} traces primarily more diffuse atomic gas, but cloud models (e.g., Warin et al. 1996; Le Petit et al. 2006; Visser et al. 2009) suggest that the \ion{C}{1} abundance rises with the CO abundance, at least to A$_{\rm V}$ $\sim$ 0.8.
In principle, observations of \ion{C}{1}, CO, CN, and C$_2$ in the main component toward HD 62542 -- with minimal contributions from diffuse atomic gas -- could indicate whether the differences seen in other sight lines are due to sampling mixtures of diffuse and denser gas or to inadequate understanding of the various density diagnostics.

{\bf \ion{C}{1} fine-structure excitation:}
In predominantly neutral interstellar clouds, the relative populations in the two upper fine-structure levels in the ground electronic state of \ion{C}{1} (denoted \ion{C}{1}* and \ion{C}{1}**) reflect a balance between collisional excitation, collisional de-excitation, and radiative decays (Jenkins \& Shaya 1979; Jenkins \& Tripp 2001, 2011).
Analyses of the excitation of those \ion{C}{1} levels can yield estimates for the local thermal pressures $p/k$ = $n_{\rm H}T_{\rm k}$ -- and thus $n_{\rm H}$, if $T_{\rm k}$ is known.
The solid curve in Figure~\ref{fig:c1fs} shows the predicted relationship between $f_2$ = $N$(\ion{C}{1}**)/$N$(\ion{C}{1}$_{\rm tot}$) (the fraction of neutral carbon in the second excited state) and $f_1$ = $N$(\ion{C}{1}*)/$N$(\ion{C}{1}$_{\rm tot}$) (the fraction in the first excited state), for $T_{\rm k}$ = 43 K, the average interstellar UV radiation field, and log($n_{\rm H}$) increasing from $-$2 (near the origin) to +5 (Jenkins \& Tripp 2011; Welty et al. 2016).
Corresponding curves for other temperatures follow similar trajectories in the $f_1$--$f_2$ plane; stronger ambient UV radiation fields would move the ''origins'' of the curves (at the lowest densities) to higher $f_1$ (and slightly higher $f_2$), but have little effect at higher densities (see, e.g., Jenkins \& Shaya 1979).
The small filled circles along the curve indicate steps in log($n_{\rm H}$) of 0.1 dex; the larger open circles mark integer values of log($n_{\rm H}$).
The many small black dots show the values found for small velocity intervals along 89 Galactic sight lines in the \ion{C}{1} survey of Jenkins \& Tripp (2011, hereafter JT11).
Their location -- often slightly above and/or to the left of the theoretical curves for ''reasonable'' temperatures for cool, neutral gas -- suggests that some high-pressure gas may generally be associated with the more abundant gas at relatively low pressures along many sight lines (Jenkins \& Shaya 1979; Jenkins \& Tripp 2001, 2011).

\begin{figure}
\epsscale{0.6}
\plotone{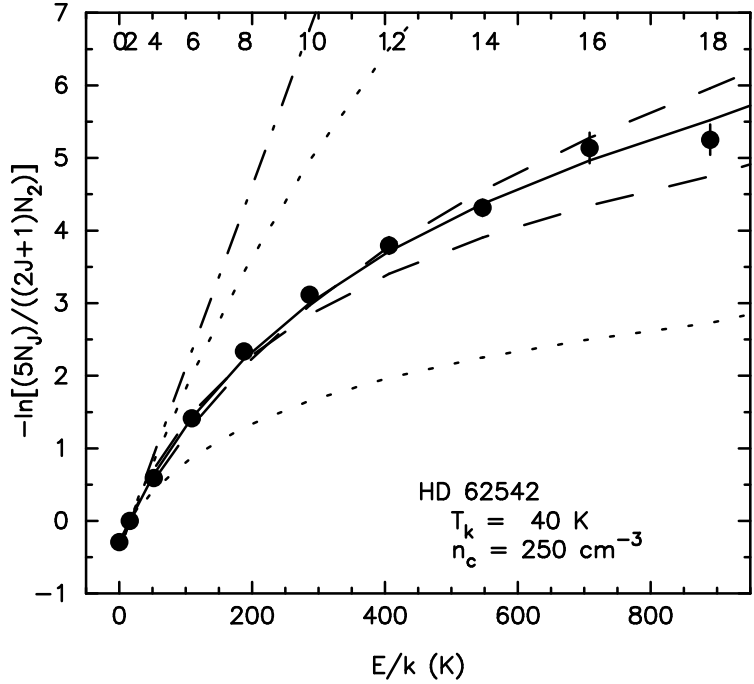}
\caption{Rotational excitation of C$_2$ toward HD~62542, based on the $N_{\rm m}$($J$) derived from the D-X (0-0) band, compared to values predicted by the analysis of van Dishoeck \& Black (1982) (adjusted for our adopted C$_2$ $f$ values; see Sonnentrucker et al. 2007).
The solid curve represents the best-fit theoretical model (to the nearest 5 K in the kinetic temperature $T_{\rm k}$ and 50 cm$^{-3}$ in the density of collision partners (H and H$_2$) $n_{\rm c}$).
The straight dotted-dashed line shows the thermal equilibrium distribution for the best-fit $T_{\rm k}$; the two dotted curves show the distributions for that $T_{\rm k}$ and $n_{\rm c}$ = 1000 and 100 cm$^{-3}$.
The two dashed curves show the distributions for ($T_{\rm k}-10$, $n_{\rm c}-50$) and ($T_{\rm k}+10$, $n_{\rm c}+50$).
If $f$(H$_2$) = 0.81 (and assuming an average near-IR radiation field), then the local density of hydrogen nuclei would be $n_{\rm H}$ $\sim$ 430 cm$^{-3}$.}
\label{fig:c2rot}
\end{figure}

The larger black dot at ($f_1$, $f_2$) = (0.49, 0.19) in Figure~\ref{fig:c1fs} indicates the relative populations found for the main component toward HD~62542 (Table~\ref{tab:acmp}); the hexagonal box enclosing that dot gives the 1$\sigma$ uncertainties (due to the uncertainties in the column densities of the three levels).
Within the uncertainties, the \ion{C}{1} excitation in that main component apparently can be characterized by a single density (and thermal pressure):  $n_{\rm H}$ $\sim$ 1500 cm$^{-3}$ (adjusted to the scale of Jenkins \& Tripp 2011; see Welty et al. 2016).
The corresponding pressure, log($p/k$) = log($n_{\rm H}T$) $\sim$ 4.8 (cm$^{-3}$ K), is more than an order of magnitude higher than the value log($p/k$)$_{\rm low}$ = 3.60 (cm$^{-3}$ K) associated with the ''center of mass'' of the distribution of Galactic points in the Jenkins \& Tripp (2011) survey, which is at ($f_1$, $f_2$) $\sim$ (0.21, 0.07).
The relative excited state populations are lower for the component at 27 km~s$^{-1}$, with ($f_1$, $f_2$) = (0.16, 0.12).
While that lies above most of the Galactic points in Figure~\ref{fig:c1fs} -- suggestive of a mixture of gas at high and low pressures -- the projection onto the curve predicted for a representative $T$ = 80 K (for Galactic diffuse, primarily atomic clouds) would correspond to a density of about 20 cm$^{-3}$ for the majority of the gas at that velocity (and still lower values for higher temperatures).
The densities and thermal pressures derived from \ion{C}{1} fine-structure excitation should be the most accurate values for this sight line -- as they are fairly insensitive to the (relatively poorly constrained) ambient radiation field.


{\bf \ion{O}{1} fine-structure excitation:}
Ritchey et al. (2019) have developed a method for estimating both the temperature and the density from the observed fine-structure excitation of \ion{O}{1}, by comparing the ratio of the excited state populations $N$(\ion{O}{1}*)/$N$(\ion{O}{1}**) with the ratio of excited to total \ion{O}{1} [$N$(\ion{O}{1}*)+$N$(\ion{O}{1}**)]/[$N$(\ion{O}{1})+$N$(\ion{O}{1}*)+$N$(\ion{O}{1}**)].
For the main cloud toward HD~62542, comparison of the two ratios yields $T$ $\sim$ 140 K (dependent primarily on the excited state ratio) and $n_{\rm H}$ $\sim$ 350 cm$^{-3}$ (dependent primarily on the ratio of excited to total \ion{O}{1}), with relatively little sensitivity to $f$(H$_2$).

{\bf C$_2$ rotational excitation:}
As both radiative and collisional processes contribute to the rotational excitation of C$_2$ in interstellar clouds, comparison of the observed $N$($J$) with models for the excitation can yield estimates for both the cloud kinetic temperature $T_{\rm k}$ and the local density of collision partners $n_{\rm c}$ = $n_{\rm H^0}$ + $n_{\rm H_2}$ (van Dishoeck \& Black 1982; van Dishoeck \& de Zeeuw 1984; Sonnentrucker et al. 2007).
In general, the lowest rotational levels provide the strongest constraints on $T_{\rm k}$, while the higher rotational levels provide the strongest constraints on the density.
Figure~\ref{fig:c2rot} compares the observed relative C$_2$ rotational populations, for $J$ = 0--18, with several theoretical curves.
If the ambient near-IR radiation field (which is primarily responsible for the initial excitation of the C$_2$) incident on the main cloud toward HD~62542 is of ''average'' strength, then the observed $N_{\rm m}$($J$) are most consistent with $T_{\rm k}$ $\sim$ 40$\pm$5 K and $n_{\rm c}$ $\sim$ 250$^{+50}_{-25}$ cm$^{-3}$ (solid curve); $n_{\rm c}$ would be proportionally higher (lower) if the ambient near-IR field is stronger (weaker) than average.
The straight dotted-dashed line shows the thermal equilibrium distribution for the best-fit $T_{\rm k}$; the two dotted curves show the distributions for that $T_{\rm k}$ and $n_{\rm c}$ = 1000 and 100 cm$^{-3}$ (steeper and shallower, respectively); the two dashed curves show the distributions for ($T_{\rm k}-10$, $n_{\rm c}-50$) and ($T_{\rm k}+10$, $n_{\rm c}+50$).\footnotemark
\footnotetext{The relationships between the C$_2$ rotational populations and $T_{\rm k}$ and $n_{\rm c}$ are based on the models of van Dishoeck \& Black (1982), as calculated by B. J. McCall (http://dib.uiuc.edu/c2/), on a grid with $\Delta T$ = 5 K and $\Delta n_{\rm c}$ = 60 cm$^{-3}$ -- but with the densities reduced by a factor of 1.2 to reflect the smaller C$_2$ $f$ values adopted in this study (see van Dishoeck \& de Zeeuw 1984 and Sonnentrucker et al. 2007) and using all of the observed rotational populations.
More recent determinations of the C$_2$ collisional rates (Najar et al. 2008, 2009) suggest that the densities obtained assuming a constant collisional de-excitation cross-section $\sigma$ = 2 $\times$ 10$^{-16}$ cm$^2$ may be slightly overestimated (e.g., Casu \& Cecchi-Pestellini 2012; Welty et al. 2013).}
The inferred $T_{\rm k}$ = 40 K and $n_{\rm c}$ = 250 cm$^{-3}$ are fairly typical of the values found for Galactic sight lines where C$_2$ is detected (e.g., Sonnentrucker et al. 2007).
As the hydrogen in the main cloud is mostly molecular [$f$(H$_2$) = 0.81], the local density of hydrogen nuclei $n_{\rm H}$ is then $\sim$ 430 cm$^{-3}$ (a factor of 3--4 lower than the density inferred from \ion{C}{1} fine-structure excitation).

{\bf CO rotational excitation:}
In diffuse molecular gas in the Galactic ISM, the excitation temperatures for both $^{12}$CO and $^{13}$CO are typically 3--5 K (Sonnentrucker et al. 2007; Sheffer et al. 2008), with the minimum $T_{\rm ex}$ set by the CMBR.
Goldsmith (2013) derived estimates for the density in sight lines where the CO excitation (obtained from UV absorption) exceeds that minimum. 
For sight lines in common with the \ion{C}{1} survey of JT11, those density estimates can be compared with the values obtained from \ion{C}{1} fine-structure excitation (assuming $T_{\rm k}$ from H$_2$), for the velocity intervals over which CO absorption is seen (and where the available \ion{C}{1} lines are not severely saturated).
The velocity information comes from component fits tabulated by Sheffer et al. (2008), from our own fits to available high-resolution optical spectra of CH and/or CN, or from simple comparisons of the CO and \ion{C}{1} profiles in STIS echelle spectra.
In general, the density estimates derived from CO and from \ion{C}{1} are consistent within a factor of $\sim$2, though the values from CO appear to be systematically lower where the UV radiation field estimated from the \ion{C}{1} excitation is stronger than average (e.g., toward HD~37903) and higher where the UV field is weaker (e.g., toward HD~27778, HD~206267).
In the main cloud toward HD~62542, however, $T_{\rm ex}$($^{12}$CO) = 11.7 K and $T_{\rm ex}$($^{13}$CO) = 7.7 K.
Those higher than typical CO excitation temperatures -- beyond the range shown by Goldsmith (2013) -- would seem to imply a very high density in the main cloud, but the constancy of $T_{\rm ex}$ with $J$, for both $^{12}$CO and $^{13}$CO, is inconsistent with the predictions of those models (see also Liszt 2007).

Alternatively, the CO in the main cloud may be excited radiatively (Wannier et al. 1997), as there is fairly strong CO emission in the vicinity of the cloud, with $T_{\rm A}$($^{12}$CO) $\sim$ 8 K along the sight line to HD~62542 and $\sim$ 15 K in several nearby peaks in the emission (Gredel et al. 1994).
Figure~2 in Wannier et al. (1997), indicates, for example, that $T_{\rm ex}$ $\sim$ 11.5 K could be obtained for $n$(H$_2$) = 750 cm$^{-3}$, $T_{\rm k}$ = 30 K, a CO(1-0) brightness temperature of 10 K, and a filling factor of order 0.2 -- which seems not unreasonable, given the distribution of CO emission near HD~62542 (Gredel et al. 1994).
The uniform excitation temperatures found for both CO isotopologs and the somewhat lower value found for $^{13}$CO would also be consistent with radiative excitation of the CO (Wannier et al. 1997; Sonnentrucker et al. 2007).

{\bf Chemical models:}
Simple chemical models, assuming steady-state gas-phase chemistry and incorporating only the most significant reactions, can yield estimates for the local hydrogen density from the observed abundances of CH, C$_2$, and CN (e.g., Federman et al. 1994, 1997; see also Welty et al. 2006). 
For example, a simplified equation for CH, balancing creation (initiated by C$^+$ + H$_2$) against destruction (via photodissociation and reactions with C$^+$ and O$^0$), can be written as
\begin{equation}
N({\rm CH}) \sim \frac{0.67~k({\rm C}^+,{\rm H}_2)~x({\rm C}^+)~N({\rm H}_2)~n_{\rm H}}
{I_{\rm UV}~G_{0}({\rm CH})~+~[k({\rm C}^+,{\rm CH})~x({\rm C}^+)~+~k({\rm O}^0,{\rm CH})~x({\rm O}^0)]~n_{\rm H}}.
\end{equation}
The $k$ values are reaction rate coefficients (with updates given by Pan et al. 2001), the $x$ values are fractional abundances (1.6 $\times$ 10$^{-4}$ for C$^+$, 3.0 $\times$ 10$^{-4}$ for O), and $I_{\rm UV}G_0$(CH) is the photodissociation rate for CH (approximately adjusted for dust extinction using $\tau_{\rm UV}$ = 3 $\times$ $A_{\rm V}$, given the steep far-UV extinction seen toward HD~62542).
For relatively low densities, photodissociation dominates the destruction, but the reactions with C$^+$ and O$^0$ can be comparable for $n_{\rm H}$/$I_{\rm UV}$ greater than a few hundred per cm$^3$.
If $I_{\rm UV}$ = 1, and for $A_{\rm V}$ = 0.9--1.0 in the main component toward HD~62542, then the simplified equations for CH, C$_2$, and CN yield estimated $n_{\rm H}$ $\sim$ 750$\pm$100, 450$\pm$60, and 315$\pm$45 cm$^{-3}$, respectively, for that component.
A stronger UV field would imply proportionally higher densities.
The lower density inferred from CN is somewhat surprising -- as the chemical reaction networks (e.g., Federman et al. 1994) generally suggest that CN should trace denser gas than CH.
More detailed modeling of the cloud (e.g., using the specific UV extinction curve for this sight line and incorporating a more extensive reaction network) may be needed to obtain more concordant density estimates from the various molecular species detected there.

{\bf Summary:}
In the main component toward HD~62542, the density obtained from \ion{C}{1} fine-structure excitation ($\sim$ 1500 cm$^{-3}$) is greater, by factors of 2--4, than those obtained from C$_2$ rotational excitation and from the simple chemical models.
Because the latter depend (linearly) on the ambient NIR and UV radiation fields (respectively) but the former does not, the apparent differences in inferred density could be reconciled if the radiation field were enhanced by a factor of 2--4, relative to the average interstellar field.
As discussed below (Secs.~\ref{sec-rad} and \ref{sec-ne}), such an enhancement in the radiation field is quite plausible, given the relative proximity of $\zeta$~Pup and $\gamma^2$~Vel.
The lower density (and higher temperature) inferred from \ion{O}{1} fine-structure excitation may be due to the \ion{O}{1} tracing somewhat warmer, less dense regions of the cloud than the \ion{C}{1} and the molecular species -- an issue which will be explored more quantitatively in constructing detailed models of the cloud.
For a density $n_{\rm H}$ $\sim$ 1500 cm$^{-3}$, the thickness of the main cloud would be $N$(H$_{\rm tot}$)/$n_{\rm H}$ $\sim$ 0.34 pc.

A total density of hydrogen nuclei of order 1500 cm$^{-3}$ would be consistent with three other, independent estimates for the density and/or pressure characterizing the main cloud toward HD~62542.
Velusamy et al. (2017) used {\it Herschel} HIFI observations of the 158 $\mu$m \ion{C}{2}] emission toward HD~62542 to infer a pressure $p/k$ = 3.11$\pm$0.21 $\times$ 10$^4$ K~cm$^{-3}$ and a density of molecular hydrogen $n$(H$_2$) = 720$\pm$50 cm$^{-3}$ (corresponding to $n_{\rm H}$ = 1440$\pm$100 cm$^{-3}$).
van Dishoeck et al. (1991) obtained a ''preferred'' density of 1500 cm$^{-3}$ (with a range of 500--3000 cm$^{-3}$) from the $^{12}$CO (1-0)/(3-2) ratio.
Cardelli \& Savage (1988) obtained a very similar estimate for the pressure for the main cloud, of order 3 $\times$ 10$^4$ K~cm$^{-3}$, based on considerations of pressure balance with the stellar wind from $\gamma^2$~Vel (at an assumed star-cloud distance of 50 pc).
Both the moderately high density and the minimal atomic envelope associated with the main cloud may thus be due to the effects of the winds (and radiation pressure?) from $\zeta$~Pup and $\gamma^2$~Vel (Cardelli et al. 1990).


\begin{deluxetable}{lcrrcrrcrrcc}
\tablecolumns{12}
\tabletypesize{\scriptsize}
\tablecaption{Photoionization and Recombination Coefficients \label{tab:coeffs}}
\tablewidth{0pt}

\tablehead{
\multicolumn{1}{c}{Element}&
\multicolumn{1}{c}{$A_{\odot}$}&
\multicolumn{1}{c}{$\chi_{\rm ion}$}&
\multicolumn{1}{c}{$\Gamma_0$\tablenotemark{a}}&
\multicolumn{1}{c}{$\Gamma_{\rm K50}$/$\Gamma_0$\tablenotemark{b}}&
\multicolumn{1}{c}{$\Gamma$/$\Gamma_0$\tablenotemark{c}}&
\multicolumn{1}{c}{$\alpha_{\rm r}$\tablenotemark{d}}&
\multicolumn{1}{c}{$\eta$\tablenotemark{d}}&
\multicolumn{1}{c}{$\alpha_{\rm d}$\tablenotemark{e}}&
\multicolumn{1}{c}{$\alpha_{\rm t}$\tablenotemark{e}}&
\multicolumn{1}{c}{$\alpha_{\rm g}$\tablenotemark{f}}&
\multicolumn{1}{c}{S\tablenotemark{f}}\\
\multicolumn{2}{c}{ }&
\multicolumn{1}{c}{(eV)}&
\multicolumn{1}{c}{($\times$10$^{-11}$)}&
\multicolumn{2}{c}{ }&
\multicolumn{1}{c}{($\times$10$^{-12}$)}&
\multicolumn{1}{c}{ }&
\multicolumn{1}{c}{($\times$10$^{-12}$)}&
\multicolumn{1}{c}{($\times$10$^{-12}$)}&
\multicolumn{1}{c}{($\times$10$^{-14}$)}&
\multicolumn{1}{c}{ }}
\startdata
C  & 8.46 & 11.26 & 35.0& 4.9 & 0.25 & 8.29 & 0.621 &10.30 & 18.87 & 3.63 & 0.2 \\
Na & 6.37 &  5.14 &  1.4& 2.2 & 0.41 & 5.82 & 0.682 & 0.00 &  6.23 & 1.91 & 0.2 \\
Mg & 7.62 &  7.65 &  6.6& 1.6 & 0.49 & 5.87 & 0.681 & 0.00 &  7.59 & 2.13 & 0.95\\
Si & 7.61 &  8.15 &450.0& 1.9 & 0.44 & 8.42 & 0.617 &81.60 & 89.73 & 1.94 & 0.95\\
P  & 5.54 & 10.49 &190.0& 4.1 & 0.31 & 6.98 & 0.645 &53.30 & 64.17 &[2.16]& 0.2 \\
S  & 7.26 & 10.36 &110.0& 4.1 & 0.27 &10.50 & 0.593 & 0.00 & 10.50 & 2.12 & 0.2 \\
K  & 5.18 &  4.34 &  3.9& 1.9 & 0.45 & 5.54 & 0.683 & 0.00 &  4.31 & 1.43 & 0.2 \\
Ca & 6.41 &  6.11 & 35.0& 1.3 & 0.46 & 5.58 & 0.683 & 0.00 &  5.58 & 1.50 & 0.95\\
Ti & 5.00 &  6.82 & 24.0&[2.0]& 0.39 & 5.50 & 0.684 &      &       &[1.58]& 0.95\\
Cr & 5.72 &  6.77 &160.0&[2.0]& 0.47 & 6.00 &[0.685]& 0.00 &  6.00 &[1.32]& 0.95\\
Mn & 5.58 &  7.43 &  3.3&[2.0]& 0.51 & 5.45 & 0.686 & 0.00 &  5.45 & 1.48 & 0.95\\
Fe & 7.54 &  7.90 & 47.0& 2.2 & 0.47 & 5.45 & 0.686 &      &       & 1.42 & 0.95\\
Ni & 6.29 &  7.63 &  9.8&[2.0]& 0.50 & 5.56 & 0.681 &      &       &[1.37]& 0.95\\
Zn & 4.70 &  9.39 & 41.0& 4.0 & 0.28 & 4.52 & 0.700 & 0.00 &  4.52 &[1.48]& 0.2 \\
\enddata
\tablenotetext{a}{Photoionization rate $\Gamma_0$ at cloud edge, for Draine (1978) radiation field (D78; Heays et al. 2017).}
\tablenotetext{b}{Multiplier for the photoionization rate for the radiation field from a Kurucz (1993) model with $T$ = 50,000 K (K50), scaled to have equal flux to the D78 field at 2000 \AA.}
\tablenotetext{c}{Attenuation factor ($\Gamma$/$\Gamma_0$) due to extinction within the cloud, obtained from the coefficients tabulated by van Dishoeck (1988), for $A_{\rm V}$ = 0.5 (cloud center).}
\tablenotetext{d}{Radiative recombination rate coefficient is given by $\alpha_{\rm r}$($T$/100)$^{-\eta}$ [P\'{e}quignot \& Aldrovandi 1986; S. Nahar 2008, private communication (Cr); Mazzitelli \& Mattioli 2002 (Zn)].}
\tablenotetext{e}{Dielectronic and total recombination rate coefficients, at $T$ = 100 K [Badnell et al. 2003; Badnell 2006 (C, Na, Mg, K); Abdel-Naby et al. 2012 (Si); Kaur et al. 2018 (P)]; P\'{e}quignot \& Aldrovandi 1986 (other elements).}
\tablenotetext{f}{Grain-assisted recombination rate coefficient ($\alpha_{\rm g}$) is calculated from the fitting coefficients tabulated by Weingartner \& Draine (2001), for a grain charging parameter $\psi$ $\sim$ 50 K$^{1/2}$ cm$^3$; values in square braces are estimated (see text).  The sticking parameter S reflects the depletion behavior (0 = no depletion, 1 = complete depletion).}
\end{deluxetable}

\begin{deluxetable}{lcccrrccrrcccc}
\rotate
\tablecolumns{14}
\tabletypesize{\scriptsize}
\tablecaption{Electron Densities (Main Component) \label{tab:ne}}
\tablewidth{0pt}

\tablehead{
\multicolumn{1}{c}{Element}&
\multicolumn{2}{c}{$\Gamma$ ($\times$ 10$^{-11}$)\tablenotemark{a}}&
\multicolumn{1}{c}{$\alpha_{\rm r}$\tablenotemark{b}}&
\multicolumn{1}{c}{$\alpha_{\rm d}$\tablenotemark{b}}&
\multicolumn{1}{c}{$\alpha_{\rm t}$\tablenotemark{b}}&
\multicolumn{1}{c}{$\alpha_{\rm g}$\tablenotemark{c}}&
\multicolumn{1}{c}{S\tablenotemark{c}}&
\multicolumn{1}{c}{$N$(X I)}&
\multicolumn{1}{c}{$N$(X II)}&
\multicolumn{1}{c}{X I/X II}&
\multicolumn{1}{c}{$n_{\rm e}$(rad)\tablenotemark{d}}&
\multicolumn{1}{c}{$n_{\rm e}$(tot)\tablenotemark{d}}&
\multicolumn{1}{c}{$n_{\rm e}$(ga)\tablenotemark{d}}\\
\multicolumn{1}{c}{ }&
\multicolumn{1}{c}{D78}&
\multicolumn{1}{c}{D+2K50}&
\multicolumn{1}{c}{($\times$10$^{-12}$)}&
\multicolumn{1}{c}{($\times$10$^{-12}$)}&
\multicolumn{1}{c}{($\times$10$^{-12}$)}&
\multicolumn{1}{c}{($\times$10$^{-14}$)}&
\multicolumn{1}{c}{ }&
\multicolumn{1}{c}{(cm$^{-2}$)}&
\multicolumn{1}{c}{(cm$^{-2}$)}&
\multicolumn{1}{c}{ }&
\multicolumn{1}{c}{D78}&
\multicolumn{1}{c}{D+2K}&
\multicolumn{1}{c}{(cm$^{-3}$)}}
\startdata
C  &   8.8 &  94.5 & 14.36 &  23.2 &  37.6 & 3.63 & 0.2 &   15.41 &[17.40]&[$-$1.99]&   [0.06]&  [0.26] & \nodata \\
Na &   0.6 &   3.1 & 10.81 &   0.0 &  10.8 & 1.91 & 0.2 &   14.04 &\nodata& \nodata & \nodata & \nodata & \nodata \\
Mg &   3.2 &  13.6 & 12.88 &   0.0 &  12.9 & 2.13 & 0.95&   13.81 & 14.41 & $-$0.60 &    0.62 &   2.65  &    2.06 \\
Si & 198.0 & 950.4 & 13.78 & 102.2 & 116.0 & 1.94 & 0.95&   11.60 & 14.48 & $-$2.88 &    0.19 &   0.11  &    0.10 \\
P  &  58.9 & 541.9 & 16.13 & 131.3 & 147.5 &[2.16]& 0.2 &   12.01 & 13.86 & $-$1.85 &    0.52 &   0.52  &   [0.34]\\
S  &  29.7 & 273.2 & 16.82 &   0.0 &  16.8 & 2.12 & 0.2 &   14.64 &[16.10]&[$-$1.46]&   [0.61]&  [5.63] &   [4.11]\\
K  &   1.8 &   8.4 &  7.92 &   0.0 &   7.9 & 1.43 & 0.2 &   12.08 &\nodata& \nodata & \nodata & \nodata & \nodata \\
Ca &  16.1 &  58.0 &  9.93 &   0.0 &   9.9 & 1.50 & 0.95&    9.10 & 11.41 & $-$2.31 &    0.08 &   0.29  &    0.16 \\
Ti &   9.4 &  46.8 &  9.80 &\nodata& [25.0]&[1.58]& 0.95&$<$10.03 & 10.69 &$<-$0.66 &  $<$2.1 & $<$4.1  & \nodata \\
Cr &  75.2 & 376.0 & 10.70 &   0.0 &  10.7 &[1.32]& 0.95& $<$9.68 &\nodata& \nodata & \nodata & \nodata & \nodata \\
Mn &   1.7 &   8.4 &  9.72 &   0.0 &   9.7 & 1.48 & 0.95&$<$10.67 &$<$12.72&\nodata & \nodata & \nodata & \nodata \\
Fe &  22.1 & 119.3 &  9.72 &\nodata& [25.0]& 1.42 & 0.95&   11.28 & 13.81 & $-$2.53 &    0.07 &  [0.14] &   [0.10]\\
Ni &   4.9 &  24.5 &  9.88 &\nodata& [25.0]&[1.37]& 0.95&$<$10.68 & 12.59 &$<-$1.91 & $<$0.06 &$<$0.12  & \nodata \\
Zn &  11.5 & 103.3 &  8.16 &   0.0 &   8.2 &[1.48]& 0.2 &   12.13 & 12.87 & $-$0.74 &    2.56 &  23.0   &  [20.8] \\
\enddata
\tablenotetext{a}{Photoionization rate $\Gamma$ at cloud center ($A_{\rm V}$ = 0.5), for the Draine (1978) radiation field and for that D78 field plus 2 times the scaled K50 field -- both attenuated by the factor $\Gamma$/$\Gamma_0$ from Table~\ref{tab:coeffs}.}
\tablenotetext{b}{Radiative, dielectronic, and total recombination rate coefficients, for $T$ = 43 K.  C, Na, Mg, Si, P, and K are from Badnell (2006), Abdel-Naby et al. (2012), and Kaur et al. (2018); Cr is from S. Nahar 2008, private communication; Zn is from Mazzitelli \& Mattioli (2002); all others are from P\'{e}quignot \& Aldrovandi (1986).}
\tablenotetext{c}{Grain-assisted recombination rate coefficient ($\alpha_{\rm g}$) and sticking coefficient (S) (Weingartner \& Draine 2001).}
\tablenotetext{d}{Electron densities calculated assuming only radiative recombination (for the D78 field) and radiative plus dielectronic recombination (for the D78 field + 2 $\times$ the K50 field] or including grain-assisted recombination (for the D78 + 2 $\times$ the K50 field).}
\end{deluxetable}

\subsubsection{Radiation Field}
\label{sec-rad}

Indications that the material in the IRAS Vela shell near the main cloud toward HD~62542 has been shaped by the winds from $\zeta$~Pup and $\gamma^2$~Vel (Cardelli \& Savage 1988; Sahu 1992; Churchwell et al. 1996; Pereyra \& Magalhaes 2002) and the differences in the densities derived from \ion{C}{1} fine-structure excitation, C$_2$ rotational excitation, and simple chemical models both suggest that those two bright, early-type stars may also significantly affect the radiation field surrounding that cloud.
The O5 Iaf star $\zeta$~Pup, at a distance of 330 pc, is about 4.5 degrees away from the sight line to HD~62542, and so is at least 26 pc away from the cloud; the WC8+O9 I binary $\gamma^2$~Vel, at a distance of 340 pc, is about 7.1 degrees away from the sight line, and so is at least 42 pc away from the cloud.
While the distance to the cloud itself is not known, there are hints that it is at least slightly closer than those two stars (Churchwell et al. 1996).
UV spectra obtained with {\it IUE} indicate that the two stars have observed fluxes of order 1.0 and 1.4 $\times$ 10$^{-8}$ erg cm$^{-2}$ s$^{-1}$ \AA$^{-1}$ at 2000 \AA, respectively.
Neglecting the modest reddening toward the two stars and scaling those observed fluxes to the respective minimum distances to the cloud yields values of order 1.6 and 0.9 $\times$ 10$^{-6}$ erg cm$^{-2}$ s$^{-1}$ \AA$^{-1}$ at the cloud -- which are quite comparable to the flux characterizing the Draine (1978, hereafter D78) field at 2000 \AA\ (but higher at shorter wavelengths).
Roughly equal contributions from the average (D78) field and from each of the two stars would thus appear to be reasonably consistent with the factor of 2--4 enhancement over the average field suggested by the various estimates for the local hydrogen density.
While an enhanced UV field would tend to depress the abundances of various trace neutral and molecular species, the observed abundances of those species suggest that the relatively high density and the unusually steep far-UV extinction characterizing the cloud have (to some degree) counteracted those effects of the radiation field.

\subsubsection{Electron Density}
\label{sec-ne}

It has often been assumed that the free electrons present in diffuse clouds are due primarily to the photoionization of carbon, so that the fractional electron abundance $n_{\rm e}$/$n_{\rm H}$ can be approximated by the gas-phase carbon abundance (typically $\sim$ 1.4--1.6 $\times$ 10$^{-4}$; e.g., Sofia et al. 2004).
There are indications, however, that photoionization and radiative recombination are not the only processes significantly affecting the ionization balance in such clouds, and that the electron fractions can often be larger than the value that would be expected from carbon alone (e.g., Welty et al. 1999, 2003; Welty \& Hobbs 2001; Weingartner \& Draine 2001; Liszt 2003).
In particular, grain-assisted recombination -- where an atomic ion is neutralized via charge exchange with a neutral or negatively charged small dust grain (or large molecule) -- can be more significant than radiative recombination for some elements (Weingartner \& Draine 2001), and cosmic-ray ionization of H can be a significant source of electrons (e.g., Liszt 2003; Indriolo et al. 2007, 2009).
For the main cloud toward HD~62542, a likely lower limit for the electron density may be obtained from the inferred values of $n_{\rm H}$ $\sim$ 1500 cm$^{-3}$ and $N_{\rm m}$(C$^+$)/$N_{\rm m}$(H$_{\rm tot}$) $\sim$ 1.6 $\times$ 10$^{-4}$ -- yielding $n_{\rm e}$ $\ga$ 0.24 cm$^{-3}$ -- but the available absorption-line data provide several other avenues for estimating the electron density.

{\bf CN excitation:} 
The rotational excitation of CN is typically dominated by the CMBR (e.g., Roth \& Meyer 1995; Ritchey et al. 2011), but collisions (primarily with electrons) can increase the excitation at higher densities (Thaddeus 1972; Black \& van Dishoeck 1991).
The analysis of Ritchey et al. suggests that the excess in excitation temperature is proportional to the local electron density $n_{\rm e}$ times the rate coefficient for electron excitation of CN (e.g., Allison \& Dalgarno 1971).
Using the rate of electron excitation of CN at 40 K (Allison \& Dalgarno 1971), the observed excess excitation ($T_{01}$ $-$ $T_{\rm CMBR}$ = 0.16$\pm$0.15 K) implies $n_{\rm e}$ $\sim$ 0.43$\pm$0.40 cm$^{-3}$ -- somewhat lower than the 1--2 cm$^{-3}$ estimated by Cardelli et al. (1990) from their adopted $T_{01}$ = 3.7$\pm$0.4 K.
Given the large relative uncertainty, we adopt the 3$\sigma$ upper limit $n_{\rm e}$ $<$ 1.2 cm$^{-3}$ from the CN excitation.


{\bf Fine-structure excitation of ionized species:}
Observations of the excited fine-structure states of ions such as \ion{C}{2} and \ion{Si}{2} can yield estimates for $n_{\rm e}$, as collisions with electrons are primarily responsible for the excitation; IR radiation and UV fluorescence may also contribute (Bahcall \& Wolf 1968; York \& Kinahan 1978; Fitzpatrick \& Spitzer 1997; see also Vreeswijk et al. 2007 and Prochaska et al. 2006 for applications to GRB absorbers).
Unfortunately, it is difficult to obtain accurate values for $N_{\rm m}$(\ion{C}{2}), $N_{\rm m}$(\ion{C}{2}*), and $N_{\rm m}$(\ion{Si}{2}*) toward HD~62542, as the interstellar lines are blended with strong, relatively narrow stellar lines, and no excited fine-structure lines are detected for Fe$^+$ or Ni$^+$.


{\bf Ionization equilibrium:} 
Ratios of the column densities of trace neutral species and the corresponding dominant first ions $N$(\ion{X}{1})/$N$(\ion{X}{2}) are often used to estimate $n_{\rm e}$, under the usual assumption of photoionization equilibrium.
Detailed studies of several individual sight lines, however, have found that the $n_{\rm e}$ inferred in that way from different elements can be quite different (e.g., Fitzpatrick \& Spitzer 1997; Welty et al. 1999); moreover, the pattern of those differences can vary from one sight line to another (Welty et al. 2003).
The incorporation of grain-assisted recombination into the ionization balance may reduce some (but not all) of those differences (Weingartner \& Draine 2001; Liszt 2003; Welty et al. 2003).
As many trace neutral species are detected in the main cloud toward HD~62542, a number of different \ion{X}{1}/\ion{X}{2} ratios can be used to further investigate those element-to-element differences in estimated $n_{\rm e}$ in what appears to be a single cloud.
Table~\ref{tab:coeffs} compiles the various parameters used to compute the photoionization rates and recombination rate coefficients for 14 elements detected in the main cloud toward HD~62542.
Table~\ref{tab:ne} lists the adopted photoionization rates (at cloud center) and recombination rate coefficients (for $T$ = 43 K), the measured or estimated column densities for the neutral and singly ionized forms of those elements (Table~\ref{tab:cdatom}), and several estimates for the electron density in the main cloud.

The photoionization rate depends on the shape and amplitude of the radiation field incident on the cloud and on the attenuation of that field within the cloud.
In order to gauge the effects of different radiation fields on the relative photoionization rates for different elements (and thus on the corresponding inferred relative $n_{\rm e}$), we calculated the rates at the outer edge of the cloud ($\Gamma_0$) for the D78 and Mathis et al. (1983, hereafter MMP) fields, and also for the fields produced by Kurucz (1993) model atmospheres with $T_{\rm eff}$ = 12,000, 26,000, and 50,000 K (K12, K26, and K50; corresponding roughly to late-B, early-B, and mid-O stars, respectively), matched to the D78 flux at 2000 \AA.
For most of the elements considered, we used the photoionization cross sections given by Verner et al. (1996) and by Bautista et al. (1998), which are based on fits to experimental or theoretical values, averaged over resonances.
In most cases, the $\Gamma_0$ values calculated for the D78, MMP, and K26 fields are fairly similar [e.g., the $\Gamma_0$(MMP)/$\Gamma_0$(D78) ratio is typically $\sim$ 0.7], and the new values for $\Gamma_0$ generally agree with those calculated for the WJ1 field (e.g., P\'{e}quignot \& Aldrovandi 1986) to within factors of 2--3.
The higher FUV flux produced by the hotter K50 model -- which would be more similar to the spectra of $\zeta$~Pup and $\gamma^2$~Vel -- yields somewhat higher $\Gamma_0$ for neutrals with high $\chi_{\rm ion}$ (C, P, S, Zn) (column 5 of Table~\ref{tab:coeffs}).
The table lists the values for $\Gamma_0$ for the D78 field, as tabulated by Heays et al. (2017), along with a multiplicative factor ($\Gamma_{\rm K50}$/$\Gamma_0$) for the rates that would be due to the K50 field, scaled to have equal flux to the D78 field at 2000 \AA.
The reductions in the photoionization rates at cloud center ($\Gamma$/$\Gamma_0$, for $A_{\rm V}$ $\sim$ 0.5), obtained from the coefficients tabulated by van Dishoeck (1988), range from 0.25 to 0.31 for elements with ionization potentials $\chi_{\rm ion}$ $\sim$ 9--13 eV (i.e., those most sensitive to the steep far-UV extinction) and from 0.39 to 0.51 for elements with lower $\chi_{\rm ion}$.
We note that the increased photoionization in the harder K50 field for those elements with high $\chi_{\rm ion}$ would thus be offset somewhat by the effects of selective extinction within the cloud.

Several different recombination processes can contribute to the total recombination rates for the different elements.
For temperatures below about 1000 K, the radiative recombination rate coefficients can be given by $\alpha_{\rm r}$($T$/100)$^{-\eta}$, with $\eta$ typically about 0.7 [P\'{e}quignot \& Aldrovandi 1986 (most elements); S. Nahar 2008, private communication (Cr); Mazzitelli \& Mattioli 2002 (Zn)].
While the contributions of dielectronic recombination have long been recognized at temperatures above 1000 K (e.g., Burgess 1964; Shull \& van Steenberg 1982; Nussbaumer \& Storey 1983; Arnaud \& Rothenflug 1985), more recent calculations have indicated a significant contribution at much lower temperatures for ions with fine-structure splitting in the ground state (Badnell et al. 2003; Bryans et al. 2009).
Of the elements listed in Tables~\ref{tab:coeffs} and \ref{tab:ne}, dielectronic recombination would thus be expected to be significant in cold interstellar clouds for C, Si, P, Ti, Fe, and Ni, but not for Na, Mg, S, K, Ca, Cr, Mn, or Zn. 
Those expectations are borne out for C, Na, Mg, Si, P, and K by the calculations of Badnell et al. (2003), Badnell (2006), Abdel-Naby et al. (2012), and Kaur et al. (2018), which include the contributions from the excited fine-structure levels.\footnotemark
\footnotetext{Parameters for calculating the radiative and dielectronic rates are tabulated at http://amdpp.phys.strath.ac.uk/tamoc/DATA/}
At $T$ = 100 K, for example, the dielectronic rate is roughly 8--10 times higher than the radiative rate for Si and P, about equal to the radiative rate for C, and negligible for Na, Mg, and K.
As no such calculations have been made as yet (to our knowledge) for S, Ca, Ti, Cr, Mn, Fe, Ni, or Zn, we have assumed that the dielectronic rate is slightly higher than the radiative rate for Ti, Fe, and Ni, but can be neglected for S, Ca, Cr, Mn, and Zn.
As the radiative recombination rates recently calculated by Badnell and collaborators generally agree well with those obtained from the parameters listed by P\'{e}quignot \& Aldrovandi (1986), we have adopted the latter values for elements not considered by Badnell et al.
The grain-assisted recombination rate coefficients ($\alpha_{\rm g}$) are calculated from the fitting coefficients tabulated by Weingartner \& Draine (2001), for a charging parameter $\psi$ = $G$ $T^{1/2}$ / $n_{\rm e}$ $\sim$ 50 K$^{1/2}$ cm$^3$ (assuming a radiation field parameter $G$ $\sim$ 4, $T$ = 43 K, and $n_{\rm e}$ $\sim$ 0.5 cm$^{-3}$).
For P, Ti, Cr, Ni, and Zn (which were not considered by Weingartner \& Draine), we have used the fitting coefficients for the element nearest in ionization potential, scaling $C_0$ (only) by the square root of the ratio of atomic weights (B. Draine 2017, private communication).
The assumed sticking factors ($S$), used in the subsequent calculation of the electron densities, reflect the depletion behavior of each element (0 = no depletion, 1 = complete depletion).

Where information for both \ion{X}{1} and \ion{X}{2} is available, the $n_{\rm e}$(rad) -- obtained by assuming only radiative recombination (at $T$ = 43 K) and using the attenuated $\Gamma$ at the center of the cloud -- range from less than 0.2 cm$^{-3}$ (for C, Si, Ca, Fe, and Ni) to 2.6 cm$^{-3}$ (for Zn); intermediate values are obtained for Mg, P, and S.
If the radiation field is stronger than average (as suggested by the discussions above), then the $n_{\rm e}$(rad) would all be correspondingly higher (with some relative differences, if the shape of the field differs from the standard D78 field).
For both the D78 and MMP fields, the individual values for $n_{\rm e}$(rad) range over a factor of more than 40, but there are no obvious trends with either $\chi_{\rm ion}$ or depletion.
The $n_{\rm e}$ estimates in the next-to-last column of Table~\ref{tab:ne} correspond to an ambient radiation field characterized by roughly equal contributions at 2000 \AA\ from the D78 field, from $\zeta$~Pup, and from $\gamma^2$~Vel (both approximated by the K50 field; see Sec.~\ref{sec-rad} above), but also include the effects of low-temperature dielectronic recombination.
For that stronger composite field, the increases in $n_{\rm e}$ are offset to varying degrees by the dielectronic recombination, and the range in relative $n_{\rm e}$ is larger than for the D78 field (and radiative recombination) alone.
The $n_{\rm e}$ estimates in the last column of the table, which include the effects of grain-assisted recombination according to eqn. 16 in Weingartner \& Draine (2001) for $\psi$ $\sim$ 50 K$^{1/2}$ cm$^3$, are somewhat lower than the values in the previous column, with larger differences for the less-depleted elements.
Iterative solutions for $n_{\rm e}$ for the individual elements yield a wide range in $\psi$ and $n_{\rm e}$, and do not qualitatively change the relative $n_{\rm e}$.
The pattern of relative $n_{\rm e}$ toward HD~62542 seen in the last three columns of Table~\ref{tab:ne} does not resemble those seen for the few other sight lines where such comparisons have been made (Welty et al. 2003); in particular, the $n_{\rm e}$ obtained from the \ion{Ca}{1}/\ion{Ca}{2} ratio is not higher than the values determined from the corresponding ratios for other elements.
The values from Mg, S, and (especially) Zn exceed the upper limit estimated from CN excitation, while the values from Si, Fe, and Ni are below the lower limit estimated from the carbon abundance.


Estimates for the cosmic-ray ionization rate ($\zeta_{\rm p}$) in relatively low-density, predominantly atomic gas may be obtained from observations of OH$^+$ (e.g., Hollenbach et al. 2012; Porras et al. 2014; Bacalla et al. 2019).
The simple chemical model described by Porras et al., for example, which balances the most significant production and destruction processes, leads to $\zeta_{\rm p}$ $\sim$ 1.3 $\times$ 10$^{-10}$ $n_{\rm H}$ $N_{\rm m}$(OH$^+$)/$N_{\rm m}$(H).
If our observed limit for the OH$^+$ abundance toward HD~62542 applies to the outer, primarily atomic layer of the main cloud, with $N_{\rm m}$(H) $\sim$ 3 $\times$ 10$^{20}$ cm$^{-2}$ (Table~\ref{tab:cdatom}) and $n_{\rm H}$ $\sim$ 20 cm$^{-3}$ (as estimated for the diffuse component at 27 km~s$^{-1}$; Sec.~\ref{sec-dens}), and assuming $T$ = 80 K there (as in Porras et al.), then that relationship between $\zeta_{\rm p}$ and the OH$^+$ abundance yields $\zeta_{\rm p}$ $<$ 5.6 $\times$ 10$^{-17}$ s$^{-1}$.
Inclusion of the effects of polycyclic aromatic hydrocarbons (PAHs) on the ionization (Hollenbach et al. 2012; Bacalla et al. 2019) then would imply $\zeta_{\rm p}$ $<$ 1.9--2.8 $\times$ 10$^{-16}$ s$^{-1}$.
These limits fall within the range of ionization rates (from several references) tabulated by Bacalla et al. (2019) (all adjusted for the $f$ value for the OH$^+$ $\lambda$3583.8 line from Hodges et al. (2018)).

\subsubsection{Summary of Physical Conditions}
\label{sec-sumphys}

Snow \& McCall (2006) proposed characterizing cloud regimes (diffuse, diffuse molecular, translucent, dense) on the basis of local chemical properties --- principally $f$(H$_2$) and the relative abundances of \ion{C}{2}, \ion{C}{1}, and CO --- reflecting the importance of those species for chemistry and cooling.
In that picture, the transition from H to H$_2$ takes place in the diffuse molecular gas, and the transition from \ion{C}{2} to \ion{C}{1} and CO takes place in the translucent regime (where $f$(H$_2$) is $\sim$ 1).
While cloud models appear able to reproduce the observed dependence of $N$(CO) on $N$(H$_2$) (Sonnentrucker et al. 2007; Sheffer et al. 2008; Visser et al. 2009), those same models are less successful in describing the relative behavior of \ion{C}{1} and CO, which can be very sensitive to both density and radiation field (Burgh et al. 2010).
In the main cloud which dominates the absorption toward HD~62542, most of the carbon is singly ionized [with $N_{\rm m}$(\ion{C}{2}) $>$ $N_{\rm m}$(CO) $>$ $N_{\rm m}$(\ion{C}{1})]\footnotemark, the depletions of the more refractory elements are severe (remarkably so for Mg and Si; see below), the molecular abundances are generally consistent with expectations for moderately dense gas (e.g, strong C$_2$, C$_3$, CO, and CN, but weak CH$^+$ and OH$^+$), and the extensive set of measured atomic and molecular column densities has yielded multiple diagnostics for the temperature, local hydrogen density, and electron density.
\footnotetext{Based on the estimate for log[$N_{\rm m}$(\ion{C}{2})] $\sim$ 17.4 given in Sec.~\ref{sec-dom} and the measured values for log[$N_{\rm m}$(CO)] = 16.42 and log[$N_{\rm m}$(\ion{C}{1})] = 15.41}
Those diagnostics indicate that $f$(H$_2$) = 0.8--1.0, $T_{\rm k}$ = 40--43 K, and $n_{\rm H}$ $\sim$ 1500 cm$^{-3}$ -- i.e., the cloud is cool, relatively dense, and primarily molecular -- placing it within the translucent regime.
The ambient radiation field appears to be two to four times stronger than the average local Galactic field, with contributions from $\zeta$~Pup and $\gamma^2$~Vel each comparable to that of the average field.
The electron density is rather uncertain, but may be of order 0.5 cm$^{-3}$.

Burgh et al. (2010) have suggested that translucent material can be characterized by the ratios $N$(CO)/$N$(H$_2$) $\ga$ 10$^{-6}$ and $N$(CO)/$N$(\ion{C}{1}) $\ga$ 1.
Both ratios increase with $f$(H$_2$), and they exhibit a fairly tight, nearly linear relationship with each other over nearly three orders of magnitude.
For the main cloud toward HD~62542, the derived log[$N_{\rm m}$(CO)/$N_{\rm m}$(H$_2$)] = $-$4.39 (dex) and $N_{\rm m}$(CO)/$N_{\rm m}$(\ion{C}{1}) = 10.2 are among the highest known values for both ratios -- again suggesting that the cloud may be classified as translucent.
Dirks \& Meyer (2019) obtained $N$(CO) and $N$(\ion{C}{1}) from fits to UV absorption profiles and estimated $N$(H$_2$) for 25 sight lines (including HD~62542) near {\it Planck} Galactic Cold Clump sources (PGCC; Planck Collaboration 2016), then used the two ratios suggested by Burgh et al. (2010) in an attempt to characterize the properties and spatial structure of those sources.
They found systematic declines in $N$(CO), $N$(CO)/$N$(\ion{C}{1}), and $N$(CO)/$N$(H$_2$) with the (relative) separation between the source and the stellar sight line for the eight sight lines with $N$(CO) $>$ 10$^{15}$ cm$^{-2}$ in the sample, but saw no such trend for the 17 sight lines with smaller $N$(CO). 
Examination of the thermal pressures available for three of the high $N$(CO) and nine of the lower $N$(CO) sight lines (Jenkins \& Tripp 2011; this paper) indicates that all but two have log($n_{\rm H}T$) $\sim$ 3.3--3.7 cm$^{-3}$ K at the velocities of the main components seen in CO and \ion{C}{1} -- consistent with typical values in the Galactic ISM (Jenkins \& Tripp 2011) -- with no obvious dependence on separation.
While HD~62542 has both the smallest separation from the nominal clump center and the largest $N$(CO), $N$(\ion{C}{1}), $N$(CO)/$N$(\ion{C}{1}), $N$(CO)/$N$(H$_2$), and thermal pressure among the eight high $N$(CO) sight lines, those values likely are largely driven by the enhanced density (due to the special environmental circumstances) and steep far-UV extinction rather than by cloud structure (as might be characterized by separation/proximity).

Much more limited information is available for assessing the physical conditions in the other, lower $N$(H$_{\rm tot}$) components toward HD~62542, as no molecular species and few of the trace neutral species are detected there.
If the strongest of those lower $N$(H$_{\rm tot}$), primarily atomic components (at $\sim$ 27 km~s$^{-1}$) is taken as representative, then the measured line widths and the \ion{C}{1} fine-structure excitation suggest that they are likely to be warmer and much less dense than the main component.

\begin{figure}
\epsscale{0.8}
\plotone{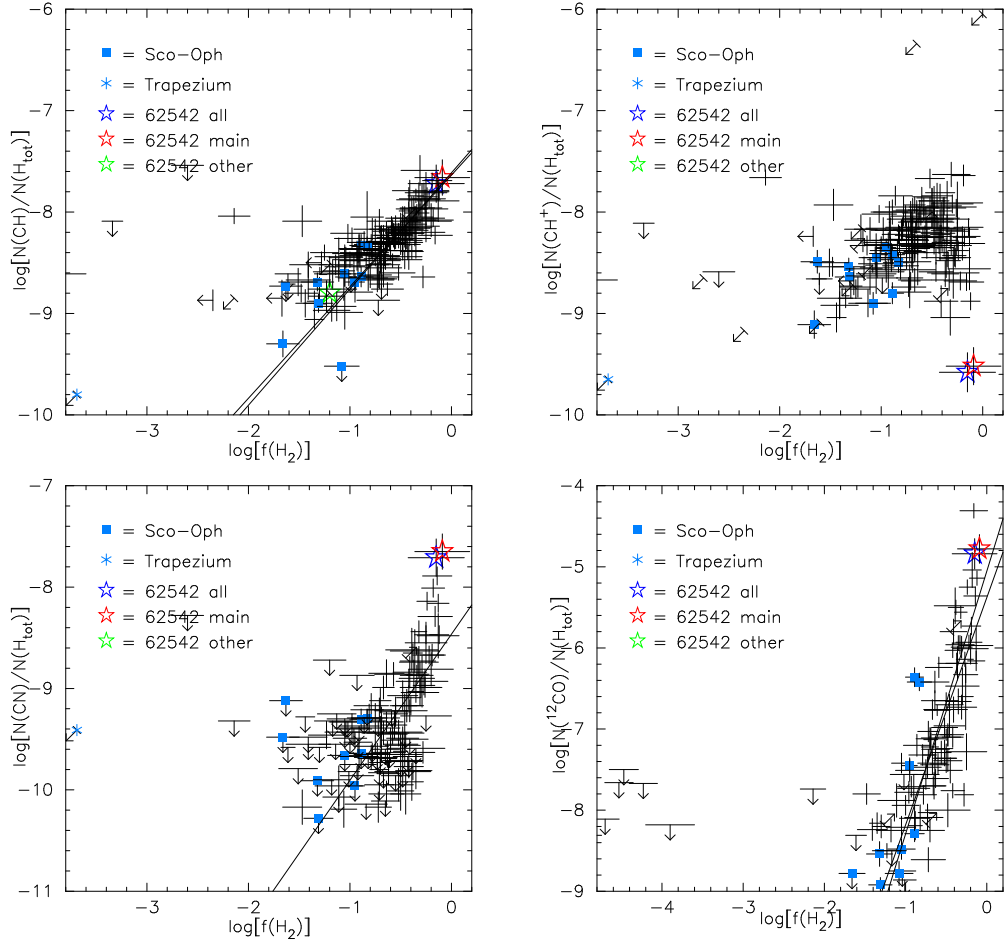}
\caption{Normalized abundances $N$(X)/$N$(H$_{\rm tot}$) for molecular species CH, CH$^+$, CN, and $^{12}$CO versus molecular fraction $f$(H$_2$).
The stars denote the sight line toward HD~62542: red = main (14 km~s$^{-1}$) component; green = all other components; blue = all components.
The other symbols are for sight lines in the Galactic ISM -- mostly from Sonnentrucker et al. (2007) and Sheffer et al. (2008).
The solid lines represent weighted and unweighted fits to the data, for $f$(H$_2$) $>$ 0.1.
The normalized abundances of CH, CN, and CO increase with molecular fraction for $f$(H$_2$) $>$ 0.1; the normalized abundance of CH$^+$ exhibits an initial increase, but then decreases for $f$(H$_2$) $\ga$ 0.3.}
\label{fig:mol_fh2}
\end{figure}

\begin{figure}
\epsscale{0.9}
\plotone{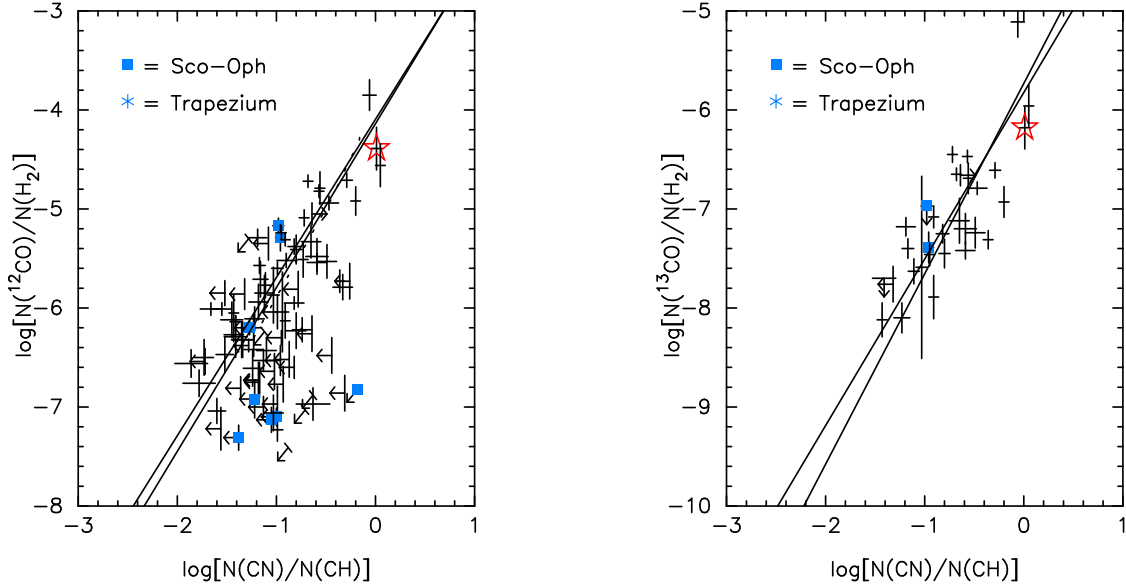}
\caption{$N$($^{12}$CO)/$N$(H$_2$) ({\it left}) and $N$($^{13}$CO)/$N$(H$_2$) ({\it right}) vs. $N$(CN)/$N$(CH).
The sight line to HD~62542 is denoted by the open star.
The $N$(CN)/$N$(CH) ratio is thought to be a density indicator; the abundances of both CO isotopologs increase nearly quadratically with that ratio.}
\label{fig:coh2}
\end{figure}

\begin{figure}
\epsscale{0.9}
\plotone{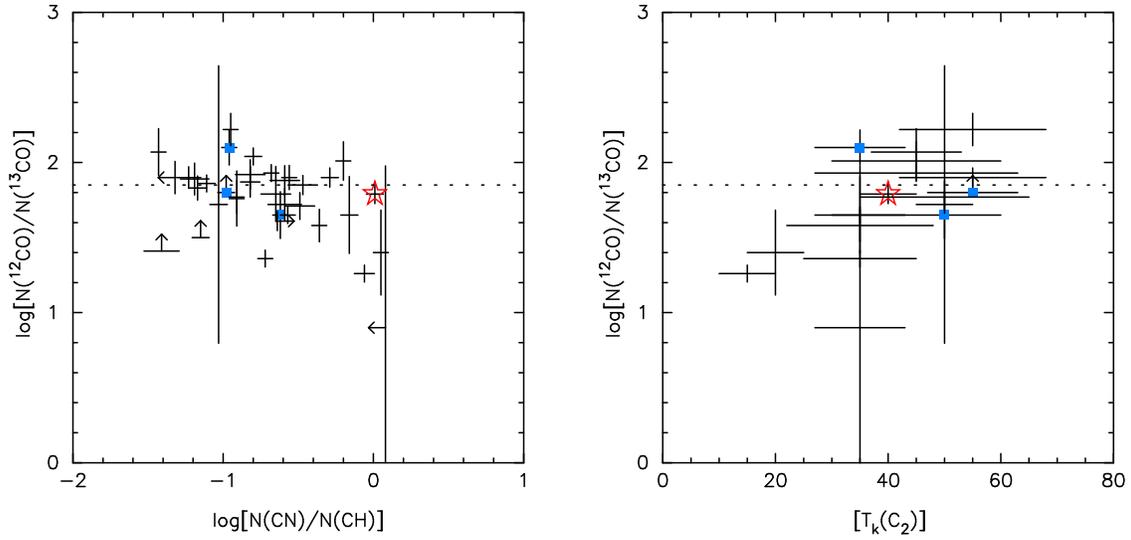}
\caption{$N$($^{12}$CO)/$N$($^{13}$CO) vs. $N$(CN)/$N$(CH) ({\it left}) and $T_{\rm k}$(C$_2$) ({\it right}).
The sight line to HD~62542 is denoted by the open star.
The ratio of the two CO isotopologs decreases with the density indicator $N$(CN)/$N$(CH) and increases with $T_{\rm k}$(C$_2$).
Ratios less than the local Galactic $^{12}$C/$^{13}$C ratio $\sim$ 70 (dotted line) are likely due to isotopic charge exchange, which favors $^{13}$CO at temperatures below about 35 K; higher ratios are likely due to selective photodissociation of $^{13}$CO, as the more abundant $^{12}$CO can self shield.}
\label{fig:co1213}
\end{figure}

\subsection{Molecular Abundances / Chemistry / Cloud Models}
\label{sec-molchem}

As discussed above, the optical/UV absorption lines detected for a number of molecular species in the main cloud toward HD~62542 have indicated high abundances of CH, CN, C$_2$, C$_3$, CS, NH, OH, and CO, but a very low abundance of CH$^+$; those molecular detections have provided both qualitative and quantitative diagnostics for the conditions in and around that cloud.
The fairly high (average) density suggested by both the high $f$(H$_2$) and the relatively high observed C$_3$/C$_2$ and CN/CH ratios in the main cloud has been confirmed both by simple chemical models for the abundances of CH, C$_2$, and CN and by analyses of the excitation of \ion{C}{1}, C$_2$, and CO (Sec.~\ref{sec-dens} above).
The differences in $b$ for CH, CN, C$_2$, and the CO isotopologs may indicate, however, that those molecular species are somewhat stratified within that primarily molecular cloud -- with the species with smaller $b$ progressively more centrally concentrated in colder, less turbulent inner regions of the cloud.

Figure~\ref{fig:mol_fh2} gives a slightly different view of the relationships between CH, CH$^+$, CN, and CO and H$_2$ depicted above in Fig.~\ref{fig:molvsh2}, with both abscissa and ordinate in each case divided by $N$(H$_{\rm tot}$) to yield some discrimination between sight lines containing gas with a significant molecular fraction and sight lines dominated by more diffuse, primarily atomic gas.
The relationships for those normalized abundances of CH, CN, and CO versus $f$(H$_2$) are similar to those seen for the unnormalized abundances versus $N$(H$_2$), and the normalized abundances of those molecules in the main cloud toward HD~62542 are quite consistent with the trends seen for other sight lines with relatively high $f$(H$_2$) (thought to be indicative of higher local densities).
For CH$^+$, however, the normalization reveals the ''lambda-shaped'' behavior noted by Fan et al. (2017) for many of the DIBs and for some atomic and molecular species.
An initial rise in the CH$^+$ abundance with increasing $f$(H$_2$) is followed by a steep decline for $f$(H$_2$) $\ga$ 0.3 -- likely reflecting the rapid destruction of CH$^+$ in denser gas.

The presence of a single narrow, dominant component (at the same velocity) for both the absorption and the emission toward HD~62542 and the agreement between the CO column densities derived from the absorption and emission spectra suggest that both the absorption and emission are due to the same main cloud, foreground to the star.
The availability of abundances for a number of atomic and molecular species (yielding multiple diagnostics), the constraints on the radiation field and the cosmic-ray ionization rate, and the observations of molecular emission from both the line of sight and the surrounding region (Gredel et al. 1994) suggest that detailed models of that main cloud could be both reasonably well constrained and quite informative.
While we defer the exploration and discussion of such detailed physical/chemical models to a subsequent paper, we will briefly note some implications of the unusually low observed CH$^+$ abundance and of the isotopic ratios observed for CO.


\subsubsection{CH$^+$ Abundance}
\label{sec-chp}

Because of its slow formation in steady-state chemical models and its rapid destruction (via collisions with H, H$_2$, and electrons), various nonthermal processes have been proposed to account for the large observed interstellar abundances of CH$^+$ (e.g., Elitzur \& Watson 1978; Draine \& Katz 1986; Falgarone et al. 1995; Federman et al. 1996; Godard et al. 2009, 2014; Agundez et al. 2010; Myers et al. 2015; Valdivia et al. 2017).
While the CH$^+$ abundance observed toward HD~62542 is a factor $\sim$30 below the ''typical'' $N$(CH$^+$)/$N$(H$_{\rm tot}$) $\sim$ 10$^{-8}$ (e.g., Godard et al. 2014), even that small amount of CH$^+$ may be difficult to explain.
Despite the apparent effects of stellar winds on the gas and dust in the vicinity of HD~62542, the lack of velocity differences among the various atomic and molecular species and the observed severe depletions suggest that the CH$^+$ is not produced by shocks.
Both CH$^+$ and (undetected) SH$^+$ can be enhanced in the presence of H$_2$* (e.g., Agundez et al. 2010), but those models suggest that more H$_2$* than is seen toward HD~62542 would be needed to produce the observed CH$^+$.
Turbulent dissipation has been suggested to produce CH$^+$ (e.g., Godard et al. 2014), but that process is inefficient at densities above about 300 cm$^{-3}$. 
While the main cloud toward HD~62542 is much denser ($n_{\rm H}$ $\sim$ 1500 cm$^{-3}$), the weak observed CH$^+$ could be concentrated in its thin, lower density outer layers.

\subsubsection{Isotopic Ratios}
\label{sec-iso}

While CO and its isotopologs are often used to estimate the amount of molecular gas when H$_2$ cannot be measured directly via far-UV absorption, interpretation of the mm-wave CO emission-line data can be uncertain, due to differences in distribution, saturation, and fractionation among the isotopologs.
Interpretation of the CO absorption lines observed in UV spectra is fairly straightforward, however.
In the absence of fractionation, the $^{12}$CO/$^{13}$CO ratio would equal the typical Galactic $^{12}$C/$^{13}$C ratio of about 70.
Because CO is photodissociated via line absorption in various UV and far-UV bands, $N$(CO) will increase rapidly if those lines either become saturated or are shielded by other strong lines (e.g., of \ion{H}{1} or H$_2$).
This shielding is quite effective for the more abundant $^{12}$CO, but not for the other isotopologs.
In denser, colder gas ($T_{\rm k}$ $\la$ 35 K), however, isotope exchange reactions favor the production of $^{13}$CO (Watson et al. 1976; Federman et al. 2003).
Translucent cloud models (Warin et al. 1996; Visser et al. 2009) predict a somewhat complex dependence of the $^{12}$CO/$^{13}$CO ratio on $n_{\rm H}$, $T_{\rm k}$, and $A_{\rm V}$, with $^{12}$CO/$^{13}$CO reaching a minimum value of about one-fifth the typical interstellar value for $n_{\rm H}$ of order 1000 cm$^{-3}$ and $A_{\rm V}$ $\sim$ 1.5--2.0, but then returning gradually to the typical interstellar value for thicker clouds.

Absorption from $^{12}$CO, $^{13}$CO, and C$^{18}$O has been detected in the main component toward HD~62542, with $^{12}$CO/$^{13}$CO ($\sim$ 62$\pm$8) and $^{12}$CO/C$^{18}$O ($\sim$ 3550$\pm$780); both ratios are similar to those found toward X~Per (Sheffer et al. 2002b).
Figures~\ref{fig:coh2} and \ref{fig:co1213} show the observed abundances of $^{12}$CO and $^{13}$CO and the ratio $^{12}$CO/$^{13}$CO versus the CN/CH ratio (thought to be $\propto$ $n_{\rm H}^2$).
The abundances of both CO isotopologs appear to increase fairly strongly with density, so that $^{12}$CO may contain $\sim$ 10\% of the total gas-phase carbon toward HD~62542 and HD~73882 -- both with log(CN/CH) $\sim$ 0 and $f$(H$_2$) $>$ 0.65.
For the three points in Figure~\ref{fig:co1213} with the largest $^{12}$CO/$^{13}$CO ratios (and small CN/CH ratios), Federman et al. (2003) proposed that the $^{12}$CO/$^{13}$CO ratio is controlled largely by selective photodissociation --- dependent on the ratio of $N$(CO) to the strength of the ambient radiation field.
Inclusion of additional data, however, suggests some dependence on the density (from the CN/CH ratio) and $T_{\rm k}$ (from C$_2$ excitation; Sonnentrucker et al. 2007) -- and the importance of isotope exchange reactions -- with a minimum value similar to that predicted by the models.
In both Figures~\ref{fig:coh2} and \ref{fig:co1213}, however, many of the points at highest CN/CH (and lowest $^{12}$CO/$^{13}$CO) are based on CO column densities derived from {\it IUE} data, and thus are somewhat uncertain.
For $T_{\rm k}$ $\sim$ 40 K, isotope exchange should not be significant in the main cloud toward HD~62542. 
The $N_{\rm m}$($^{12}$CO)/$N_{\rm m}$($^{13}$CO) $\sim$ 62$\pm$8 and $N_{\rm m}$($^{12}$CN)/$N_{\rm m}$($^{13}$CN) $\sim$ 79$\pm$13 ratios toward HD~62542 (versus the average $^{12}$C/$^{13}$C $\sim$ 70) are consistent with the general anticorrelation between the two found by Ritchey et al. (2011) and ascribed to a ''competition'' for the available $^{13}$C between CN and the more abundant CO.


\begin{figure}
\epsscale{0.8}
\plotone{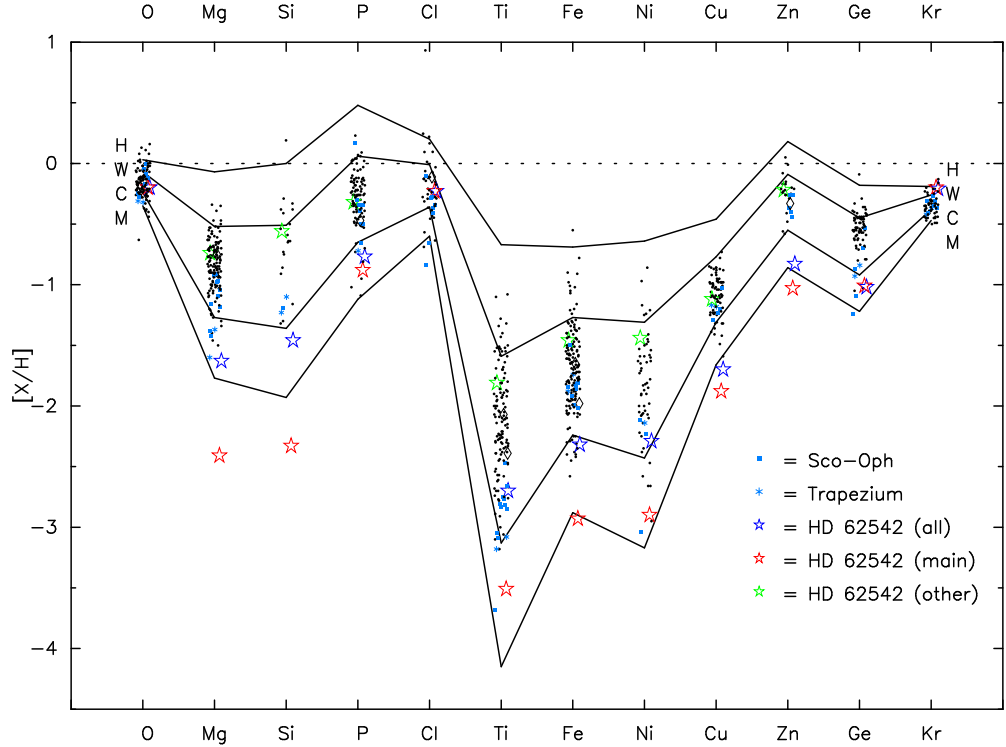}
\caption{Depletions of selected elements ([X/H] = log(X/H)$_{\rm obs}$ $-$ log(X/H)$_{\odot}$) toward HD~62542.
The red stars are for the main component near 14 km~s$^{-1}$; the green stars are for the sum of the other components; the blue stars are for the whole sight line.
The other symbols are for other Galactic sight lines, mostly from Welty \& Crowther (2010; Ti) and Jenkins (2009; all other elements).
The solid lines give representative depletion patterns for depletion parameters F* = $-$0.2, 0.25, 1.0, and 1.5 (Jenkins 2009) -- loosely characterized as ''halo cloud'' (H), ''warm cloud'' (W), ''cold cloud'' (C), and ''molecular cloud'' (M), respectively; the first three of those are similar to previous empirically defined patterns (e.g., Savage \& Sembach 1996; Welty et al. 1999).
The main component is assumed to contain all the observed H$_2$ (and some H); the other components contain primarily atomic hydrogen.
While many of the main-component depletions lie between the ''cold cloud'' and ''molecular cloud'' values, the depletions of Mg and Si are more severe than the 'molecular cloud' values and the depletions of Cl and Kr are less severe than the ''cold cloud'' values.
The average depletions for the other components are generally slightly more severe than the ''warm cloud'' values, with F$_*$ $\sim$ 0.28.}
\label{fig:depl}
\end{figure}

\subsection{Depletions in Dense Gas}
\label{sec-depl}

In the Galactic ISM, the apparent association of more severe depletions with both high $f$(H$_2$) and high average sight-line densities $<n_{\rm H}>$ (Jenkins 1987, 2009; Cardelli 1994; Cartledge et al. 2006) is usually taken to indicate a relationship between the depletions and the {\it local} $n_{\rm H}$.
Moreover, consideration of the time scales for grain formation (e.g., in stellar outflows) and grain destruction (e.g., in interstellar (IS) shocks) seems to require significant (re)construction of grains in the denser IS clouds (e.g., Draine 1990).
It was therefore initially somewhat surprising that the Fe depletions found for sight lines in the {\it FUSE} translucent cloud sample were not more severe than the highest values determined from {\it Copernicus} observations of less-reddened sight lines (Snow et al. 2002a).
It has become apparent, however, that most ''translucent'' sight lines actually contain mixtures of dense and diffuse gas -- and that both the high $f$(H$_2$) and the very severe depletions expected for any dense, translucent gas can be masked by blending in integrated sight-line averages (Rachford et al. 2002, 2009; Sonnentrucker et al. 2002, 2003).
For example, if a sight line contains 20\% diffuse \ion{H}{1} gas with a ''warm, diffuse cloud'' iron depletion D(Fe) = $-$1.4 dex and 80\% dense molecular gas with {\it no} Fe in the gas phase, the average sight-line D(Fe) = $-$2.1 dex --- slightly {\it less} severe than that observed for the ``cold clouds'' toward $\zeta$ Oph.
The sight line toward HD~62542 -- dominated by a single relatively dense, primarily molecular core -- may provide one of the best current opportunities for finding the more extreme depletions expected if grains grow in dense clouds.

In a survey of the abundances of 17 elements in 243 sight lines primarily probing the relatively local Galactic ISM, Jenkins (2009) proposed a single parameter ($F_*$), based on the depletions found for the various elements in a given sight line (and an assumption of similar relative depletion patterns), to describe the overall level of depletion in that sight line.
Ritchey et al. (2018) have extended the sample to include the abundances of five additional heavier neutron-capture elements (Ga, As, Cd, Sn, Pb), plus the light element boron, in 118 sight lines.
In the formulation of Jenkins (2009), the logarithmic depletions of individual elements appear to be roughly linearly related to $F_*$, with slopes $A_{\rm X}$ (obtained from the data for the entire sample) ranging from about 0.0 for little-depleted nitrogen and cadmium to about $-$2.0 for severely depleted titanium (with some scatter).
For well-defined $F_*$, the values range from about $-$0.4 [for the very mild depletions in the low-$N$(H$_{\rm tot}$) sight line toward $\beta$~CMa] to about 1.15 (for the severe depletions found toward HD~37903 and HD~110432).
Previously used representative Galactic ''halo cloud'', ''warm cloud'', and ''cold cloud'' depletion patterns (e.g., Savage \& Sembach 1996; Welty et al. 1999) correspond (roughly) to depletion levels $F_*$ $\sim$ $-$0.20, 0.25, and 1.0, respectively.
For comparison with the more severe depletions seen toward HD~62542, we adopt a fourth pattern, defined by $F_*$ = 1.5, for ''molecular clouds''.

\begin{figure}
\epsscale{0.65}
\plotone{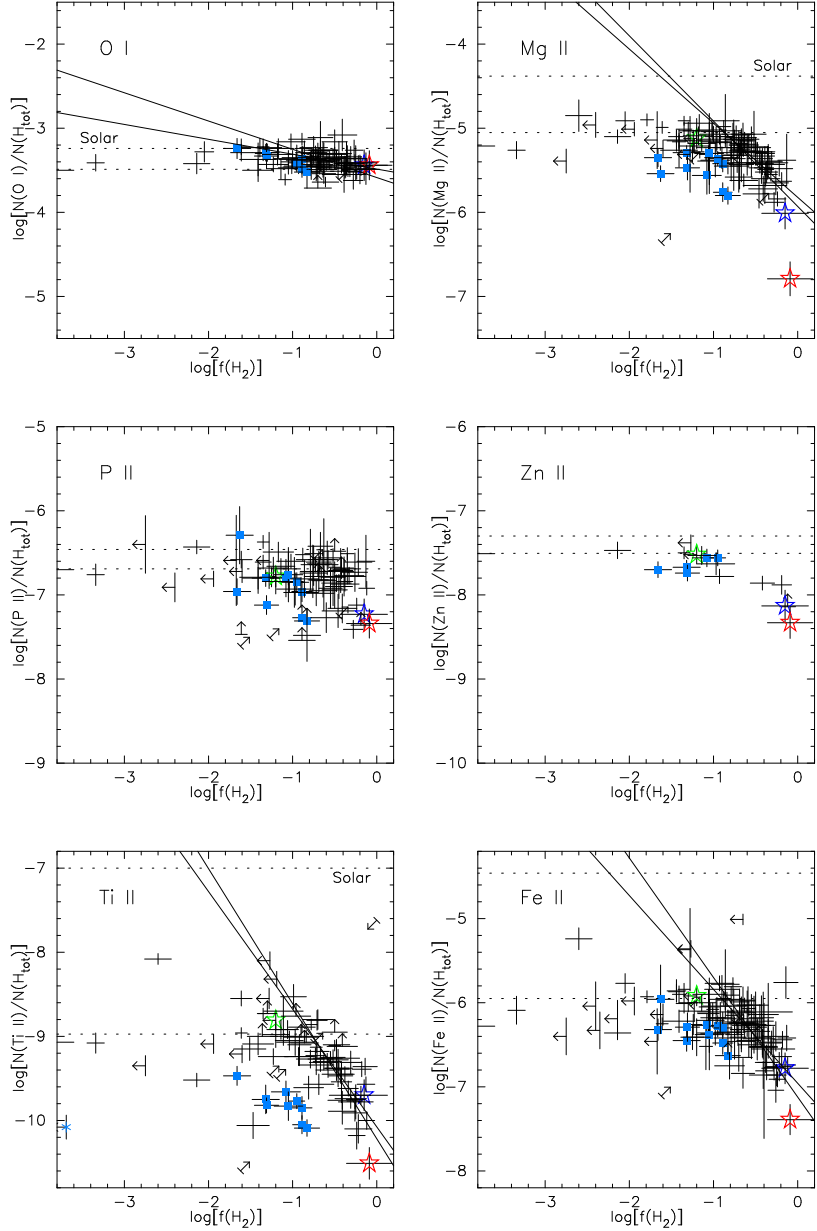}
\caption{Normalized abundances of six dominant atomic species $N$(X)/$N$(H$_{\rm tot}$) versus molecular fraction $f$(H$_2$) (a proxy for the local hydrogen density).
The stars denote the sight line toward HD~62542: red = main (14 km~s$^{-1}$) component; green = all other components; blue = all components.
The other symbols are for sight lines in the Galactic ISM -- mostly from Welty \& Crowther (2010) for \ion{Ti}{2} and from Jenkins (2009) for the other species.
The solid lines represent weighted and unweighted fits to the trends for $f$(H$_2$) $>$ 0.1 (not including the Sco-Oph or Trapezium sight lines); the upper horizontal dotted lines mark the protosolar photospheric abundances (Lodders 2003); the lower horizontal dotted lines give the average values for $f$(H$_2$) $<$ 0.01.
The declines in the normalized abundances for $f$(H$_2$) $>$ 0.1 reflect increasingly severe depletions in denser gas -- particularly for Mg, Ti, and Fe.
The depletions of several of these elements are more severe than usual for some Sco-Oph sight lines (blue squares) -- and are even more severe in the main component toward HD~62542 (red stars).}
\label{fig:dom_fh2}
\end{figure}

For this study, we have also adopted a somewhat different depletion slope for chlorine than that derived by Jenkins (2009).
While Cl$^+$ is the dominant form of chlorine in primarily atomic gas (and was the only form of chlorine considered in Jenkins 2009), Cl$^0$ can become dominant in gas with an appreciable molecular fraction, due to a reaction between Cl$^+$ and H$_2$ (e.g., Jura 1974).
Using a set of sight lines with data for both Cl$^0$ and Cl$^+$, we have therefore estimated the depletion slope for the total amount of chlorine in the predominantly neutral gas in those sight lines.
The resulting value, $A_{\rm Cl}$ = $-$0.60, fits the general trend of $A_{\rm X}$ with condensation temperature (Fig.~15 in Jenkins 2009) much better than the previous value ($-$1.24).
In view of the high $f$(H$_2$) in the main component toward HD~62542, we assume that Cl$^0$ is the dominant form of chlorine there; any Cl$^+$ in that component would only increase the total chlorine abundance (making the depletion even milder).
In addition, consideration of larger samples of data for \ion{Ti}{2} (Welty \& Crowther 2010) and \ion{Kr}{1} (Ritchey et al. 2018) suggests that the $A_{\rm X}$ derived by Jenkins (2009) for those two elements may also be too small, by $\sim$ 0.2 and 0.1, respectively -- especially if the apparently somewhat anomalous Sco-Oph sight lines (e.g., Figs.~\ref{fig:dom_htot} and \ref{fig:dom_fh2}; Sec.~\ref{sec-deplfh2}) are excluded from the fits.
Slightly higher $A_{\rm X}$ values for Ti and Kr would be in better agreement with the slopes for the depletions of those two elements versus $f$(H$_2$), as seems to be the case for a number of other elements (see Fig.~\ref{fig:slopes} and Sec.~\ref{sec-deplfh2} below).  

\subsubsection{HD~62542 -- Dominant Species}
\label{sec-depldom}

Figure~\ref{fig:depl} shows the depletions toward HD~62542 for most of the elements considered by Jenkins (2009) -- for the main component near 14 km~s$^{-1}$ (red star), for the sum of the other predominantly neutral components (green star), and for the sight line as a whole (blue star; Table~\ref{tab:cdatom}), compared with those found for other Galactic sight lines and with the four representative Galactic depletion patterns.
The small black circles represent the integrated sight-line values for the other Galactic sight lines (Jenkins 2009; Welty \& Crowther 2010); their concentration between the ''warm cloud'' and ''cold cloud'' patterns is likely due (at least in part) to the averaging over all the individual components in each sight line.
For the main component toward HD~62542, in which most of the hydrogen is in molecular form, the depletions range from very mild ($\sim$ $-$0.3 dex for O, Cl, and Kr) to very severe ($\la$ $-$2.3 dex for the more refractory elements Mg, Si, Ti, Fe, and Ni).
Using the coefficients listed in Table~4 of Jenkins (2009; except for Cl) and Table~8 of Ritchey et al. (2018), the depletions toward HD~62542 may be converted to corresponding $F_*$ values for elements exhibiting significant ranges in depletion in those surveys.
For the main component P, Ti, and Ni, $F_*$ ranges from about 1.2 to 1.3, corresponding to ''cold cloud'' depletions.
For the main component Mg, Si, Fe, Cu, Ga, and Ge, however, F$_*$ ranges from about 1.5 to 2.1 -- more consistent with the ''molecular cloud'' depletions.
The ratio $N_{\rm m}$(\ion{Ca}{1})/$N_{\rm m}$(\ion{Ca}{2}) $\sim$ 0.005 for the main component suggests that most of the calcium there is in Ca$^+$ -- and thus that calcium is severely depleted there as well (Table~\ref{tab:cdatom}; see Welty et al. 2003).
Taken together, the depletions of the more refractory elements in the main cloud toward HD~62542 thus are more severe than those found (so far) in typical Galactic ''cold clouds'' -- by nearly a factor of 10 for Mg and Si -- and are more severe than nearly all of the depletions for those elements in the Jenkins (2009) and Ritchey et al. (2018) samples.
For most of the elements detected in the main component -- including undepleted Cd and modestly depleted O, Zn, and Sn -- the depletions are reasonably consistent with the $F_*$ $\sim$ 1.52 estimated above (Sec.~\ref{sec-isnh}; Fig.~\ref{fig:depl_fh}); the upper limits for B, As, and Pb are also consistent (but not very restrictive), given the trends for those elements found by Ritchey et al. (2018).
Curiously, however, the main-component depletions of both Mg and Si are more severe, and those of Cl and Kr are milder, than would be expected for that F$_*$.
Those differences for Mg, Si, Cl, and Kr (and, to a lesser extent, for P, Ti, and Ni) indicate that the detailed pattern of depletions in the main component toward HD~62542 differs somewhat from those previously defined from general trends in the Galactic interstellar abundances.
The depletions in the other components toward HD~62542 are much less severe than those in the main component -- generally slightly more severe than the ''warm cloud'' pattern, consistent with the $F_*$ = 0.28 estimated above (Sec.~\ref{sec-isnh}; Fig.~\ref{fig:depl_fh}).

The significant differences in depletions between the main component and the other components toward HD~62542 suggest that care must be taken in interpreting total, integrated sight-line values for this sight line.
While the depletions for the sight line as a whole are broadly similar to the 'cold cloud'' pattern (with estimated $F_*$ $\sim$ 0.85 -- i.e., between the depletions found for the main component and for the other components), the $N$(H$_{\rm tot}$) predicted from the total-sight line column densities via the procedure described in Jenkins (2009) is nearly a factor of 2 smaller than the sum of the values obtained separately for the main component and for the other components (Sec.~\ref{sec-isnh}).
In this particular case, the total column densities for the least-depleted elements (O, Kr, Cd) are dominated by the contributions from the main component, while the total column densities for the more severely depleted elements (Ti, Fe, Ni) are dominated by the contributions from the other, lower $N$(H$_{\rm tot}$) components -- so that the relationship used to estimate $F_*$ and $N$(H$_{\rm tot}$) (as in Fig.~\ref{fig:depl_fh}) is not consistent with a single value of the slope $F_*$ (see also section 7.2 and figure 9 in Jenkins 2009).
Even though the main component dominates the total H$_{\rm tot}$ toward HD~62542, both the unusually severe depletions and the deviations from the typical Galactic patterns found for that component might not have been suspected if only the integrated sight-line values were available.

\begin{figure}
\epsscale{0.8}
\plotone{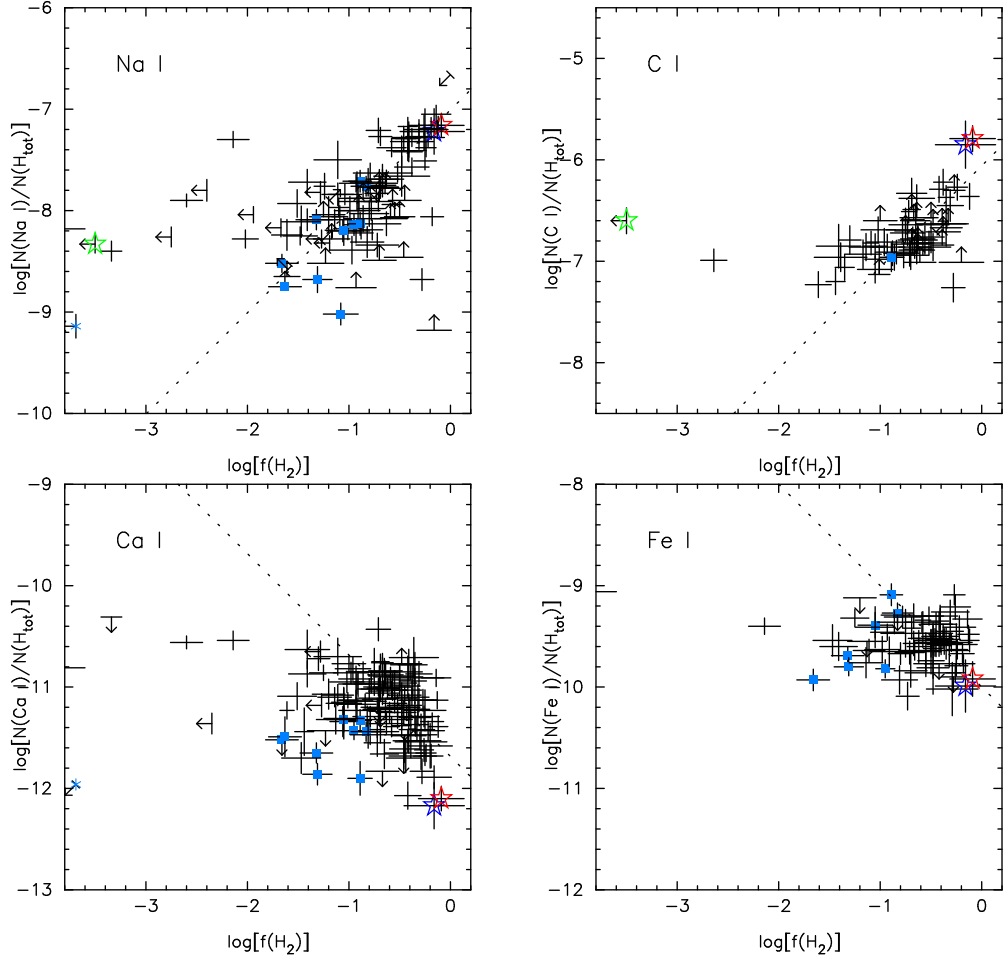}
\caption{Normalized abundances of four trace neutral atomic species $N$(X)/$N$(H$_{\rm tot}$) versus molecular fraction $f$(H$_2$).
The stars denote the sight line toward HD~62542: red = main (14 km~s$^{-1}$) component; green = all other components; blue = all components.
The other symbols are for sight lines in the Galactic ISM -- mostly from Welty \& Hobbs (2001), Jenkins \& Tripp (2011), and Welty et al. (2016) for \ion{Na}{1} and \ion{C}{1} and from Welty et al. (2003) for \ion{Ca}{1} and \ion{Fe}{1}.
The dotted lines have slopes of 1 (\ion{Na}{1}, \ion{C}{1}) or $-$1 (\ion{Ca}{1}, \ion{Fe}{1}).
The differences in the behavior of those trace species for $f$(H$_2$) $>$ 0.1 reflect differences in their ionization and depletion in denser gas.}
\label{fig:tr_fh2}
\end{figure}


\begin{figure}
\epsscale{0.9}
\plottwo{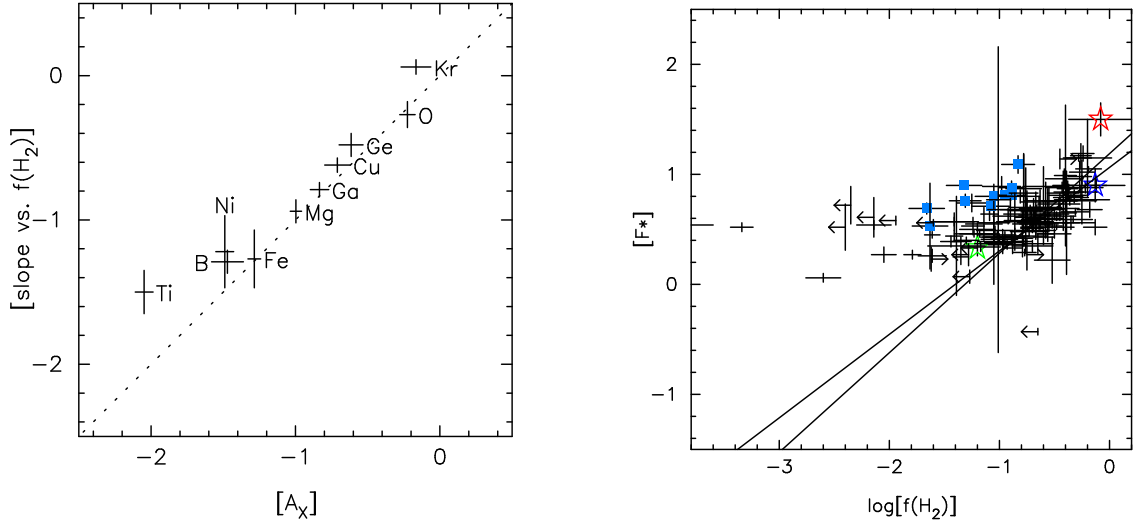}{f_fh2.eps}
\caption{Comparisons of the slopes of the normalized column densities $N$(X)/$N$(H$_{\rm tot}$) with respect to the molecular fraction $f$(H$_2$) vs. the depletion slope parameter $A_{\rm X}$ (Jenkins 2009) ({\it left}) and the depletion index $F_*$ (Jenkins 2009) vs. $f$(H$_2$) ({\it right}).
For the left-hand panel, the slopes for $N$(X)/$N$(H$_{\rm tot}$) were determined for sight lines with $f$(H$_2$) $\ge$ 0.1 (as in Fig.~\ref{fig:dom_fh2}); the somewhat anomalous Sco-Oph sight lines were excluded from the fits.
The values for most elements fall close to the dotted line denoting equality; the $A_{\rm X}$ values for Ti and Kr may be slightly too small (see Sec.~\ref{sec-depl}).
Elements not shown (e.g., C, Si, Mn, Zn) have too few accurate abundance determinations at high $f$(H$_2$) for reliable slopes to be determined.
In the right-hand panel, the open stars denote the values for HD~62542:  main component (red), the sum of the other components (green), all components (blue).
The fits (weighted and unweighted) again are for $f$(H$_2$) $\ge$ 0.1, where the two quantities appear to be fairly well correlated.
The Sco-Oph sight lines (solid blue squares) either have more severe depletions than other sight lines with comparable $f$(H$_2$) or lower molecular fractions than other sight lines with comparable depletions.}
\label{fig:slopes}
\end{figure}

\begin{deluxetable}{lrrrrrcrrrrc}      
\tablecolumns{12}
\tabletypesize{\scriptsize}
\tablecaption{Ratios of Trace Neutral Species X I/K I (Main Component) \label{tab:x1k1}}
\tablewidth{0pt}

\tablehead{
\multicolumn{1}{c}{Element}&
\multicolumn{1}{c}{$A_{\odot}$}&
\multicolumn{1}{c}{$\chi_{\rm ion}$}&
\multicolumn{2}{c}{-~-~-~$\Gamma$/$\alpha_{\rm t}$\tablenotemark{a}~-~-~-}&
\multicolumn{1}{c}{$N_{\rm m}$(X I)}&
\multicolumn{1}{c}{$N_{\rm m}$(X II)}&
\multicolumn{1}{c}{X I/K I}&
\multicolumn{2}{c}{X I/K I (predicted)\tablenotemark{b}}&
\multicolumn{1}{c}{Diff\tablenotemark{c}}&
\multicolumn{1}{c}{D(main)\tablenotemark{d}}\\
\multicolumn{2}{c}{ }&
\multicolumn{1}{c}{(eV)}&
\multicolumn{1}{c}{D78}&
\multicolumn{1}{c}{D78+2K50}&
\multicolumn{1}{c}{(cm$^{-2}$)}&
\multicolumn{1}{c}{(cm$^{-2}$)}&
\multicolumn{1}{c}{Obs}&
\multicolumn{1}{c}{(D78)}&
\multicolumn{1}{c}{(D78+2K50)}&
\multicolumn{2}{c}{ }}
\startdata
C  & 8.46 & 11.26 &  2.3& 25.1&   15.41 &[17.40] &    3.33 &   3.26 &    2.91 &   0.42 &[$-$0.25]\\ 
Na & 6.37 &  5.14 &  0.5&  2.9&   14.04 &\nodata &    1.96 &   1.81 &    1.76 &   0.20 & \nodata \\
Mg & 7.62 &  7.65 &  2.5& 10.5&   13.81 & 14.41  &    1.73 &   2.39 &    2.44 &$-$0.71 & $-$2.40 \\ 
Si & 7.61 &  8.15 & 17.1& 81.9&   11.60 & 14.48  & $-$0.48 &   1.54 &    1.54 &$-$2.02 & $-$2.32 \\ 
P  & 5.54 & 10.49 &  4.0& 36.7&   12.01 & 13.86  & $-$0.07 &   0.10 & $-$0.18 &   0.11 & $-$0.87 \\ 
S  & 7.26 & 10.36 & 17.7&162.4&   14.64 &[16.10] &    2.56 &   1.18 &    0.90 &   1.66 &[$-$0.35]\\ 
K  & 5.18 &  4.34 &  2.2& 10.6&   12.08 &\nodata & \nodata & \nodata& \nodata & \nodata& \nodata \\
Ca & 6.41 &  6.11 & 16.2& 58.4&    9.10 & 11.41  & $-$2.98 &   0.37 &    0.49 &$-$3.47 & \nodata \\
Ti & 5.00 &  6.82 &  3.7& 18.7&$<$10.03 & 10.95  &$<-$2.05 &$-$0.41 & $-$0.43 &$<-1.62$& $-$3.50 \\
Cr & 5.72 &  6.77 & 70.3&351.4& $<$9.68 &\nodata &$<-$2.40 &$-$0.96 & $-$0.98 &$<-1.42$& \nodata \\
Mn & 5.58 &  7.43 &  1.7&  8.7&$<$10.67 &$<$12.72&$<-$1.41 &   0.51 &    0.49 &$<-1.90$&$<-$2.05 \\
Fe & 7.54 &  7.90 &  8.8& 47.7&   11.28 & 13.81  & $-$0.80 &   1.76 &    1.71 &$-$2.51 & $-$2.92 \\ 
Ni & 6.29 &  7.63 &  2.0&  9.8&$<$10.68 & 12.59  &$<-$1.40 &   1.16 &    1.15 &$<-2.55$& $-$2.89 \\
Zn & 4.70 &  9.39 & 14.1&126.6&   12.13 & 12.87  &    0.05 &$-$1.28 & $-$1.56 &   1.61 & $-$1.02 \\ 
\enddata
\tablenotetext{a}{Ratio of photoionization rate to radiative plus dielectronic recombination rate, at cloud center and for $T$ = 43 K, for the Draine (1978) and the D78 plus 2 $\times$ K50 radiation fields.}
\tablenotetext{b}{Observed ratios X~I/K~I and predicted ratios (assuming photoionization equilibrium and no depletion), for the D78 and D78+2K50 fields.}
\tablenotetext{c}{Difference between observed X~I/K~I and predicted value for D78+2K50 field.}
\tablenotetext{d}{Observed depletion in the main component (Table~\ref{tab:cdatom}).}
\end{deluxetable}

\subsubsection{Depletions versus $f$(H$_2$)}
\label{sec-deplfh2}

Other aspects of the behavior of the depletions of various elements may be seen via plots of the normalized abundances $N$(X)/$N$(H$_{\rm tot}$) versus the molecular fraction $f$(H$_2$), which can be taken as a rough proxy for the local density $n_{\rm H}$ (e.g., Cardelli 1994; Sofia et al. 1999; Cartledge et al. 2006; Fan et al. 2017).
Figures~\ref{fig:dom_fh2} and \ref{fig:tr_fh2} show examples of such comparisons for selected dominant and trace species, respectively.
At low molecular fractions [$f$(H$_2$) $\la$ 0.1], the depletions exhibited by the dominant species are relatively constant (as marked by the lower horizontal dotted lines; though with considerable scatter), at levels broadly similar to the ''warm cloud'' values -- so that $f$(H$_2$) is not a good discriminant in that regime.
For higher molecular fractions [$f$(H$_2$) $\ga$ 0.1], however, the depletions generally become more severe with increasing $f$(H$_2$) -- so that the molecular fraction does appear to provide a good discriminant where H$_2$ is significant -- whether or not there is a direct relationship between the two quantities.
Moreover, steeper slopes [fitting only sight lines with $f$(H$_2$) $\ge$ 0.1] are found for the more refractory elements (compare \ion{Mg}{2}, \ion{Ti}{2}, \ion{Fe}{2} versus \ion{O}{1}, \ion{P}{2}).
Intriguingly, those slope values for $f$(H$_2$) $\ge$ 0.1 are generally very similar to the corresponding $A_{\rm X}$ coefficients (the slopes of depletion versus $F_*$) in the Jenkins (2009) formalism (left-hand panel of Fig.~\ref{fig:slopes}) -- though the samples are still small for some elements.
That similarity likely reflects the fairly good correlation between F$_*$ and $f$(H$_2$) over that range in $f$(H$_2$) (right-hand panel of Fig.~\ref{fig:slopes}) -- though that relationship flattens out for smaller $f$(H$_2$) (as in Fig.~16 of Jenkins 2009\footnotemark).
\footnotetext{Note that the values of $f$(H$_2$) in that figure should all be multiplied by 2 (E. B. Jenkins 2020, private communication).}
Figure~\ref{fig:tr_fh2} shows that the normalized abundances of little-depleted trace species (\ion{C}{1}, \ion{Na}{1}) tend to increase monotonically with increasing molecular fraction for $f$(H$_2$) $\ga$ 0.1, while the normalized abundances of severely depleted trace species (\ion{Ca}{1}, \ion{Fe}{1}) tend to decline with molecular fraction (though not as severely as the corresponding dominant species; e.g., \ion{Fe}{1} in Fig.~\ref{fig:tr_fh2} versus \ion{Fe}{2} in Fig.~\ref{fig:dom_fh2}).
The figures also reveal the distinct behavior of the depletions and/or the molecular fractions in the Sco-Oph region (solid blue squares), where either the depletions of the more refractory species are more severe than for other sight lines with similar $f$(H$_2$) or the molecular fractions are smaller than for other sight lines with similar levels of depletion.
Some of the scatter seen in these comparisons is thus likely to reflect local/regional differences in environmental conditions.
The locations of the HD~62542 main, other, and total sight-line values provide vivid illustrations of both the increasingly severe depletions at higher molecular fractions (higher densities?) and the tendency for the variations in the depletions among the individual components within a given sight line to be obscured in the integrated sight-line values.



\subsubsection{HD~62542 -- Trace Species}
\label{sec-depltr}

Additional information on the relative depletions of some elements may be obtained from ratios of the respective trace neutral species, which should preferentially trace the denser parts of interstellar clouds.
In general, the relative abundances $N$(\ion{X}{1})/$N$(\ion{Y}{1}) should be relatively insensitive to $n_{\rm e}$ (which largely cancels in taking the ratios), $T_{\rm k}$ (due to most elements having similar temperature dependences for radiative recombination), and the overall strength (but perhaps not the shape) of the ambient UV radiation field (e.g., Snow 1984; Welty \& Hobbs 2001; Welty et al. 2003).
The column densities of the various trace neutral species are generally fairly well correlated with each other in the Galactic ISM (e.g., Jenkins \& Shaya 1979), with slopes in log[$N$(\ion{X}{1})] versus log[$N$(\ion{K}{1})] (for example) of order 1.0$\pm$0.5 (Welty \& Hobbs 2001; Welty et al. 2003).
The shallower slopes seen for \ion{Fe}{1} and \ion{Ca}{1} likely reflect the increasingly severe depletions of Fe and Ca in sight lines with higher $N$(H$_{\rm tot}$).
In the main component toward HD~62542, the trace neutral species with high ionization potentials $\chi_{\rm ion}$ $>$ 9 eV (\ion{C}{1}, \ion{P}{1}, \ion{S}{1}, \ion{Zn}{1}) are all enhanced, relative to both \ion{K}{1} (Table~\ref{tab:x1k1}) and H$_{\rm tot}$ (Fig.~\ref{fig:tr_htot}) -- likely due to selective suppression of ionizing photons by the steep far-UV extinction.\footnotemark
\footnotetext{While \ion{Cl}{1} also has a fairly high ionization potential, its abundance is controlled by a reaction between H$_2$ and \ion{Cl}{2} (Jura 1974).
Given the high column density of \ion{Cl}{1} (and the corresponding apparent negligible depletion of chlorine), the high molecular fraction in the main component appears to have driven most of the chlorine into \ion{Cl}{1}.}
The trace neutral ions of the more refractory elements (\ion{Mg}{1}, \ion{Si}{1}, \ion{Ca}{1}, \ion{Ti}{1}, \ion{Cr}{1}, \ion{Mn}{1}, \ion{Fe}{1}, \ion{Ni}{1}), however, are all less abundant than ''expected'' or ''usual'' -- consistent with the enhanced depletions of those elements inferred from the corresponding dominant first ions.
The abundances of the trace neutral species thus reflect the effects of both the unusually severe depletions and the unusually steep far-UV extinction in the main component toward HD~62542.


\subsubsection{Depletions and UV Extinction}
\label{sec-deplext}

There have been a number of attempts to tie unusual UV extinction to specific environmental conditions and/or other properties of the sight lines, in order to understand the size distribution and/or composition of interstellar dust grains.
In a homogeneous survey of 417 Galactic sight lines observed with {\it IUE}, Valencic et al. (2004) identified four sight lines whose UV extinction curves could not be well fitted with the $R_V$-dependent parameterization developed by Cardelli et al. (1989):  HD~29647, HD~62542, HD~204827, and HD~210121.
All four of those ''discrepant'' sight lines exhibit both steep far-UV extinction and strong absorption from various molecular lines -- suggestive of relatively high densities in the clouds.
They appear to sample rather different environments, however -- from apparently quiescent clouds in the disk (HD~29647) or low halo (HD~210121) to clouds that may have been subjected to shocks and/or stronger than average radiation fields (HD~62542, HD~204827).
Because such steep UV extinction curves have been difficult to model with ''standard'' mixtures of graphite and silicate grains, they have been used to test more complex models for the interstellar dust -- e.g., incorporating polycyclic aromatic hydrocarbons (PAHs) (e.g., Zonca et al. 2011) or nanodiamonds (Rai \& Rastogi 2012).
For HD~62542, those more complex models appear to prefer rather different mixtures of the silicate and total carbonaceous components.
The observed very severe depletion of silicon in the main cloud toward HD~62542 indicates that silicates are likely to be major constituents of the dust there.
Voshchinnikov \& Henning (2010) noted a possible association between higher dust-phase abundances (i.e., more severe depletions) of Mg and Si, increased $R_V$, and flatter far-UV extinction for a few sight lines in the Sco-Oph region, which might be taken as evidence for growth of Mg-Si grains due to accretion.
Although the depletions of Mg and Si are very severe in the main component toward HD~62542 -- and coagulation and/or mantling of the grains might be expected in that relatively dense cloud -- the overall $R_V$ $\sim$ 2.9 is close to the average local value, and the far-UV extinction is steep.
Such steep far-UV extinction is often assumed to reflect an enhanced population of small grains -- perhaps due to erosion of the grains by shocks or strong UV radiation -- but the severe depletions of many elements (especially Mg and Si) and the absence of velocity offsets among the various atomic and molecular absorption lines suggests that the main cloud toward HD~62542 has not been strongly shocked, and the ambient UV field there appears to be only mildly enhanced.
Voshchinnikov et al. (2012) then reported a weak correlation between the dust-phase abundance of Si and the interstellar polarization, suggesting that the silicate component of the grains would be primarily responsible for the polarization.
Despite the severe depletions of Mg and Si in the main cloud toward HD~62542 (and the relative simplicity of the sight line), however, the measured polarization (P = 1.42$\pm$0.01 \%; Voshchinnikov et al. 2012) due to that main cloud is unremarkable.

\clearpage

\section{SUMMARY / CONCLUSIONS}
\label{sec-sum}

High-resolution optical and UV spectra of the moderately reddened B3-5 V star HD~62542, located behind a ridge of gas and dust in the {\it IRAS} Vela shell, have yielded detections of interstellar absorption from many neutral and singly ionized atomic and molecular species.
The interstellar material in this unusual sight line is dominated by an apparently single narrow ''main'' component near $v_{\odot}$ = 14 km~s$^{-1}$ containing nearly 90\% of the total $N$(H$_{\rm tot}$) (seen most prominently in the molecular and trace neutral atomic species), but with a number of ''other'' components between about 4 and 33 km~s$^{-1}$ (seen primarily in some of the dominant singly ionized atomic species).
The main component contains at least 99\% of the H$_2$ (and other molecules) and roughly half of the atomic hydrogen (as estimated from several indicators) in the sight line, and is characterized by a very high molecular fraction $f$(H$_2$) $\sim$ 0.81; it appears to be dominated by a relatively dense, mostly molecular core, with much of the usual more diffuse outer, primarily atomic gas having been stripped away by stellar winds and/or radiation from $\zeta$~Pup and $\gamma^2$~Vel.
The rest of the atomic hydrogen in the sight line is distributed among the other components, with average $f$(H$_2$) $<$ 0.06.
Striking differences in the absorption-line profiles for lines from the various dominant species suggest large component-to-component differences in the relative abundances -- and thus significant differences in the depletions of the more refractory elements.

The many atomic and molecular species detected in the main component provide multiple, independent diagnostics of the physical conditions in that primarily molecular cloud: 
\begin{enumerate}
\item{Analyses of the rotational excitation of H$_2$ (Rachford et al. 2002) and C$_2$ yield similar values for the kinetic temperature $T_{\rm k}$ = 40--43 K; the higher value estimated from the low-$J$ levels of C$_3$ (\'{A}d\'{a}mkovics et al. 2003) likely reflects a radiative contribution to the excitation.
Consistent upper limits on $T_{\rm k}$ are obtained from the $N$(\ion{O}{1}*)/$N$(\ion{O}{1}**) fine-structure ratio and from the $b$ value (1.4 km~s$^{-1}$) obtained from fits to the profiles of several lines from vibrationally excited H$_2$.
The unusually high (but subthermal) and uniform excitation temperatures $T_{\rm ex}$ found for both $^{12}$CO (11.7 K; for $J$=0--6) and $^{13}$CO (7.7 K; for $J$ = 0--3) may be due to radiative (rather than collisional) excitation.}
\item{Analysis of the \ion{C}{1} fine-structure excitation yields a well-constrained total hydrogen density $n_{\rm H}$ $\sim$ 1500 cm$^{-3}$, consistent with recent observations of \ion{C}{2}] emission at 158 $\mu$m (Velusamy et al. 2017).
The C$_2$ rotational excitation, for $J$ = 0-18, is most consistent with $T_{\rm k}$ $\sim$ 40 K and a density of collision partners (H and H$_2$) $n_{\rm c}$ $\sim$ 250 cm$^{-3}$ (for the average interstellar NIR radiation field); for $f$(H$_2$) = 0.81, that $n_{\rm c}$ implies $n_{\rm H}$/$I_{\rm NIR}$ $\sim$ 430 cm$^{-3}$.
Simple chemical models for the column densities of CH, C2, and CN suggest values for $n_{\rm H}$/$I_{\rm UV}$ ranging from 315--750 cm$^{-3}$.
The lower densities inferred from C$_2$ and the chemical models may be reconciled with the 1500 cm$^{-3}$ obtained from \ion{C}{1} if the ambient radiation field is two to four times stronger than the average interstellar field (both in the near-IR and in the UV).
Such an enhancement would be consistent with estimated contributions from $\zeta$~Pup and $\gamma^2$~Vel (whose stellar winds also appear to have shaped the ISM in the region).}
\item{The various available \ion{X}{1}/\ion{X}{2} ratios yield a wide range in implied $n_{\rm e}$, from $<$0.1 to 2.6 cm$^{-3}$ (assuming photoionization equilibrium in the average interstellar field); somewhat larger values of $n_{\rm e}$ -- and larger ranges in $n_{\rm e}$ -- are found if grain-assisted recombination, low-temperature dielectronic recombination, and an enhanced radiation field are considered. 
The pattern of relative $n_{\rm e}$ is not like those seen in many other sight lines, however; e.g., the value inferred from Ca (which is often anomalously high) is at the low end of the range toward HD~62542.
The excitation of CN may be slightly higher than that expected from the CMBR, and yields an upper limit for $n_{\rm e}$ $<$ 1.2 cm$^{-3}$; the observed $n_{\rm H}$ $\sim$ 1500 cm$^{-3}$ would imply a lower limit of $n_{\rm e}$ $\ga$ 0.24 cm$^{-3}$, assuming that most of the electrons come from ionization of carbon.
The non-detection of OH$^+$ yields an upper limit on the cosmic-ray ionization rate, $\zeta_{\rm p}$ $<$ 3 $\times$ 10$^{-16}$ s$^{-1}$, that falls within the range seen for a number of other Galactic sight lines.}
\end{enumerate}

The strong observed absorption from CH, CN, C$_2$, C$_3$, and CO had hinted that the main cloud is relatively dense; the high CO/H$_2$ and CO/\ion{C}{1} ratios (as well as the high molecular fraction) suggest that it may be categorized as translucent (Burgh et al. 2010).
The carbon in this largely molecular core is still primarily singly ionized, however, with $N$(\ion{C}{2}):$N$(CO):$N$(\ion{C}{1}) roughly 100:10:1.
While the absorption from CH$^+$ is quite weak, relative to H$_2$, even that small amount cannot be explained by reactions involving H$_2$* (there is not enough present) or turbulent diffusion (which is ineffective at the high density characterizing the main cloud); the weak CH$^+$ may be concentrated in the remnant lower density, mostly atomic outer layers of the cloud.
The $^{12}$C/$^{13}$C isotopolog ratios for CO ($\sim$ 62) and CN ($\sim$ 79) appear to bracket the average local Galactic ratio ($^{12}$C/$^{13}$C $\sim$ 70), consistent with the anticorrelation between those two isotopolog ratios noted by Ritchey et al. (2011) and ascribed to a competition between CO and CN for the available $^{13}$C.
The derived $T_{\rm k}$ $\sim$ 40-43 K is not low enough for $^{13}$CO to be significantly enhanced via isotope exchange.

For a number of the more refractory elements, the abundances of both trace and dominant ions indicate that the depletions characterizing the main component toward HD~62542 are more severe than those previously seen in any other Galactic sight line.
We have thus adopted a new ''molecular cloud'' pattern for the depletions, characterized by a depletion index F$_*$ = 1.5 (in the formalism of Jenkins 2009) -- but find that the detailed pattern of depletions in the main component is somewhat unusual (with respect to that pattern), with even more severe depletions for Mg and Si but very mild depletions for Cl and Kr.
Neither those extreme depletions and nor the deviations from the typical Galactic pattern would have been noticed if we had considered only the total sight line abundances -- for which the depletions are reasonably consistent with the well known ''cold cloud'' pattern.
Detailed, component-by-component analyses of high-resolution optical and UV spectra are needed to gauge the full range of relative abundances in the ISM (as well as to determine accurate physical conditions from the various diagnostics based on those abundances).
Declining trends of the normalized abundances (X/H$_{\rm tot}$) with $f$(H$_2$) (taken as a rough proxy for the density), for $f$(H$_2$) $\ga$ 0.1, suggest that the depletions are more severe at higher densities; the slopes of those trends agree remarkably well with the depletion slopes determined by Jenkins (2009) and Ritchey et al. (2018).
There is no obvious connection between the very severe depletions for Mg and Si in the main component and either the extinction characterizing the sight line (steep far-UV rise, but normal $R_{\rm V}$) or the optical/UV polarization (which is unremarkable).
The severe depletions and the lack of velocity offsets among the various atomic and molecular species suggest that strong shocks have not significantly affected the main cloud; the normal $R_{\rm V}$ suggests that coagulation has not significantly enlarged the dust grains (despite the fairly high local density in that cloud).

The relative simplicity of this sight line and the wealth of available information (and diagnostics) make the main cloud toward HD~62542 a good candidate for detailed modeling.
The differences in the $b$ values determined for CO (0.5 km~s$^{-1}$), CN and C$_2$ (0.7 km~s$^{-1}$), and CH (1 km~s$^{-1}$) suggest that those molecular species are somewhat stratified within the main cloud -- with the species with smaller $b$ more concentrated in the denser and/or less turbulent part(s) of the cloud -- so that inhomogeneous models will be needed.
Detailed consideration of the radiation field (UV, near-IR, submm) will also be necessary, given the likely contributions from $\zeta$~Pup and $\gamma^2$~Vel, the unusually steep far-UV extinction, and the possible radiative excitation of CO and C$_3$.
Such models, using the Cloudy and Meudon PDR codes, are being explored; it will be very interesting to see if those models can both account for the molecular abundances and reconcile the various observed \ion{X}{1}/\ion{X}{2} ratios -- to better understand both the various diagnostics and the chemistry and ionization equilibrium in such interstellar clouds.

\acknowledgments

Support for {\it HST} guest observer program 12277 was provided by NASA through grant HST-GO-12277.008-A from the Space Telescope Science Institute, which is operated by the Association of Universities for Research in Astronomy, Inc., under NASA contract NAS 5-26555.
We thank Ed Jenkins for pointing out recent work on low-temperature dielectronic recombination and for providing his code for computing \ion{O}{1} fine-structure excitation, Bruce Draine for advice regarding the estimation of grain-assisted recombination rates, Adam Ritchey for comments on \ion{O}{1}, and Steve Federman for alerting us to an updated $f$ value for CS and for useful comments.

{\it Facilities:} \facility{{\it HST} (STIS)}, \facility{ESO: 3.6m (CES)}, \facility{VLT: Kueyen (UVES)}, \facility{Keck:I (HIRES)}, \facility{AAT (UCLES)}, \facility{Magellan: Clay (MIKE)}

\clearpage

\appendix

\section{Spectra of \ion{C}{1}}
\label{sec-c1spec}

\begin{figure}
\epsscale{0.9}
\plotone{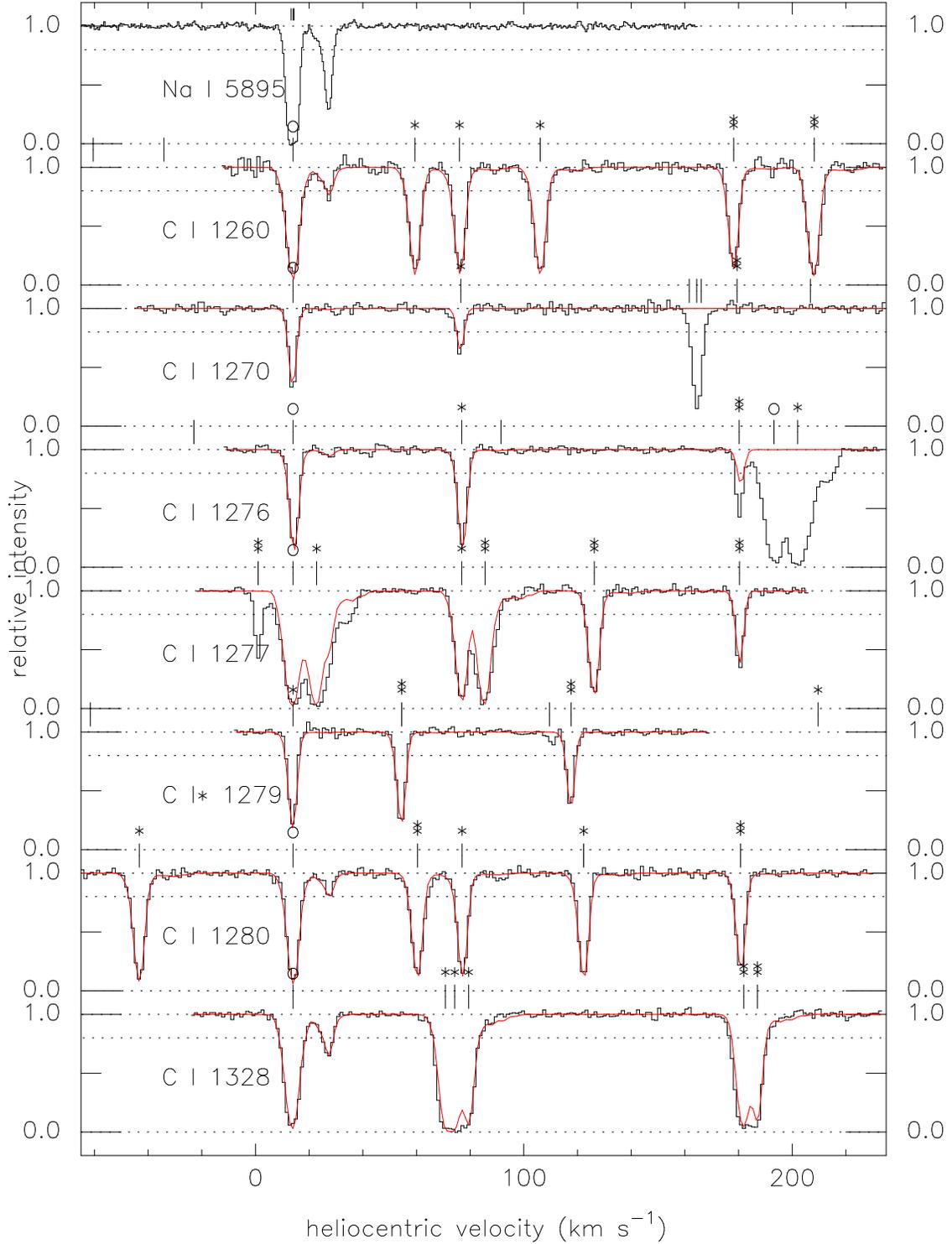}
\caption{Normalized spectra of \ion{C}{1} absorption multiplets toward HD~62542 (black histograms).
The smooth red curves show the theoretical profiles based on the component structures derived from simultaneous fits to the unblended lines of each of the three fine-structure states.
At the top is the corresponding higher resolution spectrum of the \ion{Na}{1} D1 line.
Tick marks above the \ion{Na}{1} D1 line denote the detailed component structure (used also to fit the \ion{C}{1} lines); tick marks above the \ion{C}{1} lines denote the locations of the main component (14 km~s$^{-1}$) for the ground (o) and excited fine-structure states (* and **); unmarked ticks denote lines of other species (\ion{S}{1} or H$_2$*).}
\label{fig:c1}
\end{figure}

Figure~\ref{fig:c1} shows the fits (smooth red lines) to the normalized profiles of the various \ion{C}{1} multiplets observed toward HD~62542 (black histograms).
At the top is the corresponding higher resolution profile of the \ion{Na}{1} $\lambda$5895 line, which is used to define the component structure for the \ion{C}{1} lines.
The tick marks show the location of the main component ($v_{\odot}$ = 14 km~s$^{-1}$) for each line; for \ion{C}{1}, the ''o'' denotes the ground-state transitions, while ''*'' and ''**'' denote the transitions from the excited fine-structure states.
Unattributed tick marks denote lines from other species (either \ion{S}{1} or H$_2$*).
Separate simultaneous fits were performed to the isolated lines from each of the three fine-structure levels, with the relative velocities and $b$ values for most of the components fixed to those determined from \ion{Na}{1}; only the individual component column densities, the $b$ value for the main component, and the overall velocity offset were varied for the \ion{C}{1} lines.
The adopted column densities and main-component $b$ values are given in Table~\ref{tab:acmp}.

We used the $f$ values determined by Jenkins \& Tripp (2001, 2011) in order to obtain column densities consistent with the values in those surveys, but note that some uncertainties remain.\footnotemark
\footnotetext{Three very weak lines were not considered by Jenkins \& Tripp (2001, 2011):  \ion{C}{1}* $\lambda$1287.6, \ion{C}{1}** $\lambda$1274.1, $\lambda$1288.1. 
Revised $f$ values for the first two of those lines are suggested in Sec.~\ref{sec-apew} below.}
To set the absolute scale for the $f$ values, Jenkins \& Tripp (2001) adopted the well-determined values listed by Wiese et al. (1996) for multiplet 2 (the strongest \ion{C}{1} multiplet in $f\lambda$, near 1656 \AA).
They then estimated the $f$ values for successively weaker transitions by requiring consistency in the measured apparent optical depths, considering the absorption profiles for all the stars in their sample.
For the progressively weaker transitions, the derived $f$ values are increasingly higher than the previously reported values -- by nearly a factor of 10 for the weakest transitions (Fig.~3 in Jenkins \& Tripp 2001).
While Jenkins \& Tripp (2001, 2011) conducted various tests to try to determine if those differences could be due to the effects of unresolved saturation in the profiles, no satisfactory explanation for the differences has yet emerged.
Fortunately, the determination of the relative fine-structure populations (used to estimate the thermal pressures and local densities) should be less affected by the uncertainties in the $f$ values than the absolute column densities.

Contributions from $^{13}$C were added to the fitted profiles of the \ion{C}{1} $\lambda$1328 multiplet (for which the difference in velocity (relative to the $^{12}$C lines) is of order $-$2 km~s$^{-1}$; Morton 2003; Berengut et al. 2006), but those contributions had only minor effects on the profiles.
Given the average interstellar $^{12}$C/$^{13}$C ratio $\sim$ 70 (Sec.~\ref{sec-iso}) and the much smaller velocity differences for the other \ion{C}{1} multiplets ($\la$ 0.25 km~s$^{-1}$; Berengut et al. 2006), contributions from $^{13}$C were not included for the other multiplets.

In general, the \ion{C}{1} profiles corresponding to the adopted component structures agree quite well with the observed profiles.
The most noticeable differences are for several blended features in the strongest $\lambda$1277 and $\lambda$1328 multiplets and for the very weak \ion{C}{1}** line in the $\lambda$1276 multiplet (at 1277.2 \AA).
Increasing the $b$ values for the excited state lines (from 0.85 to 1.0-1.1 km~s$^{-1}$) would yield noticeably improved fits to the blended features near 1277.5, 1329.1, and 1329.6 \AA, but not to the blend of \ion{C}{1} $\lambda$1277.2 and \ion{C}{1}* $\lambda$1277.3.
Increasing the $f$ value of the weak \ion{C}{1}** $\lambda$1277.2 line (by a factor of 2.0-2.5) would improve the fit there -- and would be more consistent with the differences (new vs.previous) in $f$ values for the other very weak transitions.

\section{Spectra of CO}
\label{sec-cospec}

\begin{figure}
\epsscale{0.9}
\plotone{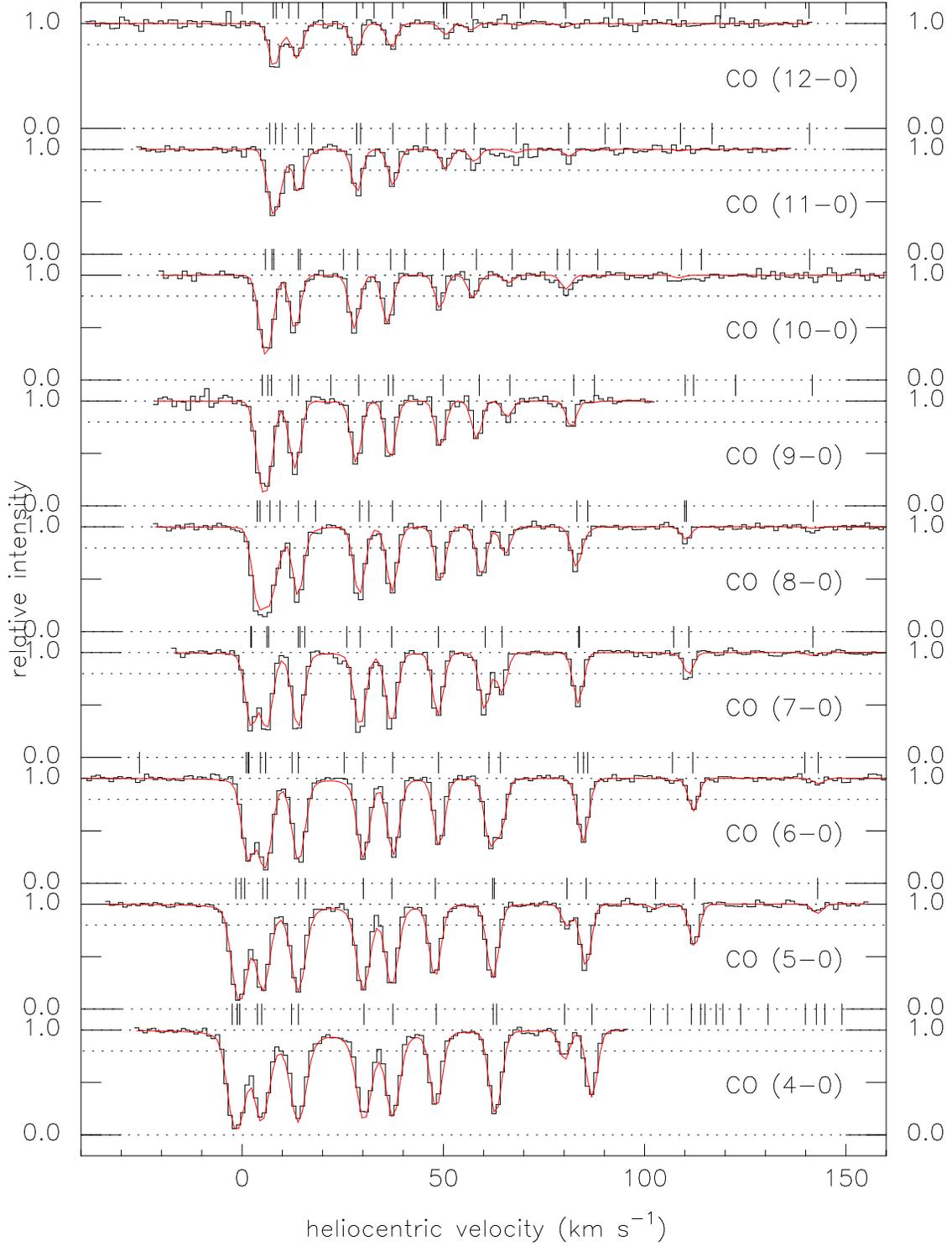}
\caption{Normalized spectra of $^{12}$CO A-X bands toward HD~62542 (black histograms).
The smooth red curves show the theoretical profiles based on the component structure adopted from fits to the ensemble of bands (a single component at 14 km~s$^{-1}$, with $b$ = 0.5 km~s$^{-1}$).
Tick marks above the spectra denote the locations of that component for the individual rotational levels ($J$ = 0--6) from the R, Q, and P branches.}
\label{fig:co12}
\end{figure}

\begin{figure}
\epsscale{0.9}
\plotone{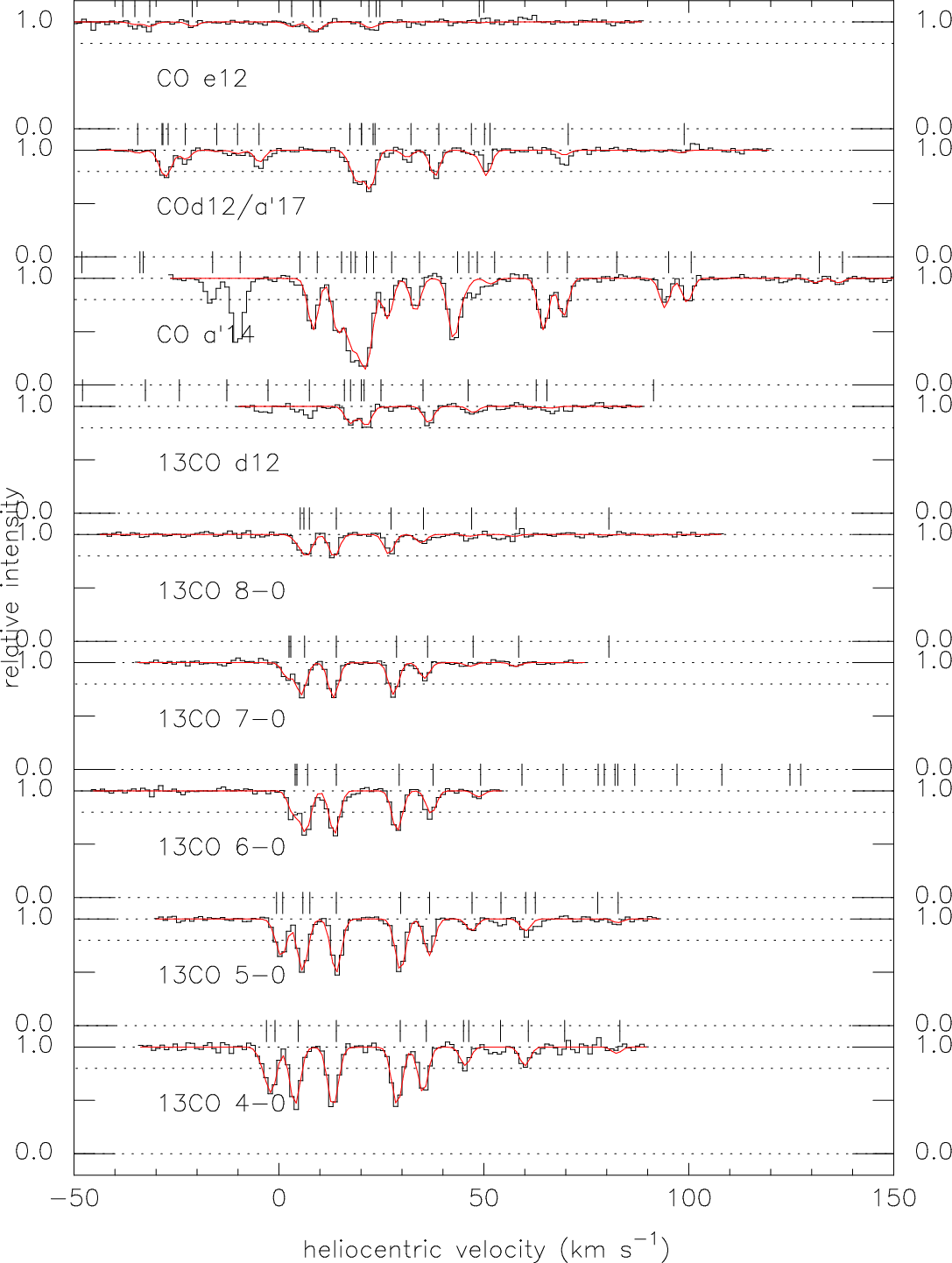}
\caption{Spectra of $^{13}$CO A-X and selected $^{12}$CO intersystem bands toward HD~62542 (black histograms).
The smooth red curves show the theoretical profiles based on the component structure adopted from fits to the ensemble of bands (a single component at 14 km~s$^{-1}$, with $b$ = 0.5 km~s$^{-1}$).
Tick marks above the spectra denote the locations of that component for the individual rotational levels ($J$ = 0--3) from the R, Q, and P branches.
Absorption that is marked, but not fitted, is due to lines of CO, C$^{18}$O, or H$_2$*.}
\label{fig:co13}
\end{figure}

Figures~\ref{fig:co12} and \ref{fig:co13} show the fits (smooth red lines) to the normalized profiles of the various $^{12}$CO and $^{13}$CO bands observed toward HD~62542 (black histograms).
As for the other molecules found in this sight line, absorption from CO is detected only from the main component at $v_{\odot}$ = 14 km~s$^{-1}$.
Both permitted A-X bands and semiforbidden intersystem bands are detected for both isotopologs, from rotational levels $J$=0-6 for $^{12}$CO and $J$=0-3 for $^{13}$CO; for each band, the tick marks give the locations of the various rotational lines (from the R, Q, and P branches) at the velocity of the main component.
For the permitted A-X CO bands, we adopted the rest wavelengths and $f$ values of Morton \& Noreau (1994); for the intersystem bands, we used the values tabulated by Eidelsberg \& Rostas (2003), except for $^{12}$CO a'17 and $^{13}$CO d12 (Morton \& Noreau 1994) and $^{12}$CO e5 (Sheffer et al. 2002a).
The fits to the profiles were performed for a series of choices of the main-component $b$ value, varying only the column densities of the different rotational levels and the overall velocity offset.
A $b$ value of 0.5 km~s$^{-1}$ gave the most consistent fits to both the strong permitted bands and the weaker intersystem bands; the fits for $b$ = 0.4 and 0.6 km~s$^{-1}$ were noticeably less consistent.
The fits to the individual bands and the adopted column densities $N$($J$), for both $^{12}$CO and $^{13}$CO, are summarized in Table~\ref{tab:cobands}i; the adopted values are given in bold type.


\section{Equivalent Widths of Atomic and Molecular Lines}
\label{sec-apew}

Table~\ref{tab:ewatom} lists the total equivalent widths (all components, integrated over the line profile) measured for various atomic species toward HD~62542.
Unless otherwise noted (by a number in the reference column), the wavelengths and $f$ values are from Morton (2003; most species) or Jenkins \& Tripp (2001, 2011; \ion{C}{1} only); the wavelengths are vacuum values for $\lambda$ $<$ 3000 \AA\ and in air for $\lambda$ $>$ 3000 \AA.
The ''Source'' column contains a code specifying which spectrum was used for the measurement.
The last two columns contain the measured equivalent width and the column density estimated by integrating over the apparent optical depth (AOD) in the profile.
In general, the $N$(AOD) for a given line is larger than the column density that would be estimated just from the equivalent width (assuming the line to be optically thin), but is smaller than the true value (especially for the stronger lines affected by saturation).
Where there are multiple lines covering a range in strength for a given species, we have not attempted to obtain a better estimate for the column density by using the method described by Jenkins (1996) -- preferring instead to obtain the column densities from the detailed fits to the line profiles -- which explicitly account for saturation and blending effects.
The uncertainties on the equivalent widths are 1$\sigma$ values, and include contributions from both photon noise (Jenkins et al. 1973) and the fits to the local continuum (Sembach \& Savage 1992), added in quadrature; the upper limits are 3$\sigma$.  
Because the projected rotational velocity of HD~62542 is fairly low, however, the local ''continuum'' set by the stellar absorption lines can in some cases be difficult to define precisely (Fig.~\ref{fig:stel}), so that the strengths of some of the broader interstellar lines may be more uncertain.

For a few very weak lines of \ion{C}{1}, \ion{P}{1}, and \ion{S}{1}, $f$ values either were not available or the values listed in Morton (2003) yielded column densities inconsistent with those derived from the profile fits; these are noted by asterisks is the ''Ref'' column of Table~\ref{tab:ewatom}.
For those lines, we estimate new/revised $f$ values as: \ion{C}{1}* $\lambda$1287.6 (0.00023), \ion{C}{1}** $\lambda$1274.1 (0.00114), \ion{P}{1} $\lambda$1283.9 (0.14), \ion{P}{1} $\lambda$1286.4 (0.14), \ion{P}{1} $\lambda$1377.9 (0.04), \ion{S}{1} $\lambda$1381.6 (0.00018), \ion{S}{1} $\lambda$1388.4 (0.00034).

Table~\ref{tab:ewmol} lists the equivalent widths (main component only) measured for various molecular species toward HD~62542.
The references for the wavelengths and $f$ values are given at the end of the table; the sources of the spectra are specified as in Table~\ref{tab:ewatom}.
The last two columns contain the measured equivalent width and the AOD column density estimate for each line; the uncertainties and upper limits are 1$\sigma$ and 3$\sigma$, respectively.
We note that the $N$(AOD) refer to the lower level involved in each transition; total molecular column densities (as given in Table~\ref{tab:cdmol}) cannot be estimated unless sufficient information on the excitation is available.
Values for the unidentified line near 1300 \AA\ and for several of the DIBs are included at the end of the table.

Table~\ref{tab:ewc2} lists the equivalent widths (main component only) measured for individual rotational lines in four of the C$_2$ bands detected toward HD~62542.
The wavelengths and band $f$ values are those adopted by Sonnentrucker et al. (2007); the sources of the spectra are specified as in Table~\ref{tab:ewatom}.
The last column gives the column densities for each rotational level, as determined via simultaneous fits to all the lines in each band, using the adopted $b$ = 0.7 km~s$^{-1}$.
The AOD column density estimates (next-to-last column) are generally in good agreement with the values obtained from the profile fits for the optical/NIR A-X bands, but not for the lower $J$ levels in the D-X (0,0) band, where the stronger UV lines are affected by saturation.

%
%
%
%
%

\begin{deluxetable}{lcrrcrrr}
\tablecolumns{8}
\tabletypesize{\scriptsize}
\tablecaption{Equivalent Widths -- Atomic Lines \label{tab:ewatom}}
\tablewidth{0pt}

\tablehead{
\multicolumn{1}{c}{Species}&
\multicolumn{1}{c}{$\lambda$}&
\multicolumn{1}{c}{f}&
\multicolumn{1}{c}{Ref\tablenotemark{a}}&
\multicolumn{1}{c}{Source\tablenotemark{b}}&
\multicolumn{1}{c}{$W_{\lambda}$}&
\multicolumn{1}{c}{log[$N$(AOD)]}&
\multicolumn{1}{c}{log[$N$(fit)]}}

\startdata
B II        & 1362.4600 & 0.99600 & &s&$<$1.8        &$<$11.04          & \nodata \\
            & 1362.4730 & 0.99600 & &s& blend        &   \nodata        \\
C I         & 1260.7351	& 0.05880 &1&s&  25.6$\pm$1.2&   13.75$\pm$0.01 & 14.92$\pm$0.04 \\
            & 1270.1430	& 0.00216 &1&s&  10.1$\pm$0.5&   14.67$\pm$0.02 \\
            & 1276.4822	& 0.01180 &1&s&  14.9$\pm$0.4&   14.16$\pm$0.01 \\
            & 1277.2452	& 0.13140 &1&s& blend        &   \nodata        \\
            & 1280.1352	& 0.04806 &1&s&  22.8$\pm$0.7&   13.79$\pm$0.01 \\
            & 1328.8333	& 0.08985 &1&s&  31.2$\pm$0.5&   13.68$\pm$0.01 \\
C I*        & 1260.9262	& 0.02608 &1&s&  20.1$\pm$0.6&   13.98$\pm$0.01 & 15.11$\pm$0.03 \\
            & 1260.9961	& 0.02204 &1&s&  18.4$\pm$0.5&   14.01$\pm$0.01 \\
            & 1261.1224	& 0.02731 &1&s&  22.4$\pm$0.7&   14.01$\pm$0.01 \\
            & 1270.4081	& 0.00041 &1&s&   6.4$\pm$0.6&   15.10$\pm$0.03 \\
            & 1276.7498	& 0.00824 &1&s&  14.3$\pm$0.4&   14.28$\pm$0.01 \\
            & 1277.2827	& 0.08142 &1&s& blend        &   \nodata        \\
            & 1277.5131	& 0.03950 &1&s& blend        &   \nodata        \\
            & 1279.0562	& 0.00610 &1&s&  12.4$\pm$0.5&   14.34$\pm$0.01 \\
            & 1279.8907	& 0.02817 &1&s&  21.0$\pm$0.5&   13.98$\pm$0.10 \\
            & 1280.4043	& 0.01320 &1&s&  17.6$\pm$0.5&   14.19$\pm$0.01 \\
            & 1280.5975	& 0.01777 &1&s&  16.7$\pm$0.4&   14.06$\pm$0.01 \\
            & 1287.6076	&[0.00023]&*&s&   4.4$\pm$0.4&  [15.11]         \\
            & 1329.0849	& 0.03601 &1&s& blend        &   \nodata        \\
            & 1329.1004	& 0.04350 &1&s& blend        &   \nodata        \\
            & 1329.1233	& 0.03086 &1&s& blend        &   \nodata        \\
C I**       & 1261.4255	& 0.02071 &1&s&  16.5$\pm$0.7&   13.98$\pm$0.01 & 14.71$\pm$0.03 \\
            & 1261.5519	& 0.04417 &1&s&  22.7$\pm$0.8&   13.83$\pm$0.01 \\
            & 1274.1090	&[0.00114]&*&s&   8.4$\pm$0.3&  [14.71]         \\
            & 1277.1900	& 0.00075 &1&s&   7.4$\pm$0.2&   14.95$\pm$0.01 \\
            & 1277.5501	& 0.08622 &1&s& blend        &   \nodata        \\
            & 1277.7233	& 0.02764 &1&s&  18.8$\pm$0.5&   13.92$\pm$0.01 \\
            & 1277.9539	& 0.00307 &1&s&  10.4$\pm$0.4&   14.51$\pm$0.01 \\
            & 1279.2290	& 0.00900 &1&s&  12.8$\pm$0.4&   14.19$\pm$0.01 \\
            & 1279.4980	& 0.00362 &1&s&   9.5$\pm$0.3&   14.39$\pm$0.01 \\
            & 1280.3331	& 0.02759 &1&s&  21.2$\pm$0.6&   13.94$\pm$0.01 \\
            & 1280.8471	& 0.01329 &1&s&  15.0$\pm$0.5&   14.09$\pm$0.01 \\
            & 1288.0553	& 0.00002 & &s&$<$1.4        &[$<$15.68]        \\
            & 1329.5775	& 0.05785 &1&s& blend        &   \nodata        \\
            & 1329.6004	& 0.03064 &1&s& blend        &   \nodata        \\
C II        & 1334.5323	& 0.12800 & &s& stellar      &   \nodata        & \nodata \\
            & 2325.4029	& 4.78E-08& &s&$<$2.7        &$<$18.08          \\
C II*       & 1335.6627	& 0.01280 & &s& stellar      &   \nodata        & \nodata \\
            & 1335.7077	& 0.11500 & &s& stellar      &   \nodata        \\
N I         & 1200.2233	& 0.08690 & &s& 141.9$\pm$4.7&$>$14.60          & \nodata \\
            & 1200.7098	& 0.04320 & &s& 142.6$\pm$6.0&$>$15.02          \\
O I         & 1355.5977	& 1.16E-06& &s&   6.8$\pm$0.3&   17.65$\pm$0.02 & [17.82] \\
O I*        & 1304.8576	& 0.04780 & &s&  11.2$\pm$0.6&   13.32$\pm$0.02 & [13.65] \\
O I**       & 1306.0286	& 0.04780 & &s&   7.9$\pm$0.6&   13.16$\pm$0.02 & [13.37] \\
Na I        & 3302.3690	& 0.00897 & &u&  31.3$\pm$1.1&   13.65$\pm$0.01 & 14.05$\pm$0.08 \\
            &           &         & &U&  30.1$\pm$0.3&   13.63$\pm$0.01 \\
            & 3302.9780	& 0.00448 & &u&  22.1$\pm$1.0&   13.78$\pm$0.01 \\
            &           &         & &U&  25.4$\pm$0.3&   13.84$\pm$0.01 \\
            & 5889.9510	& 0.65500 & &c& 225.9$\pm$2.4&$>$12.49          \\ 
            &          	&         & &u& 222.4$\pm$1.3&$>$12.25          \\
            &          	&         & &U& 230.0$\pm$1.5&$>$12.24          \\
            & 5895.9242	& 0.32700 & &c& 184.3$\pm$2.3&$>$12.70          \\ 
            &          	&         & &u& 164.5$\pm$1.3&$>$12.41          \\
            &          	&         & &U& 182.6$\pm$1.4&$>$12.42          \\
Mg I        & 1827.9351	& 0.02420 & &s&  15.8$\pm$0.9&   13.51$\pm$0.02 & 13.84$\pm$0.04 \\
            & 2026.4768	& 0.11300 & &s&  29.4$\pm$0.7&   13.11$\pm$0.01 \\
Mg II       & 1239.9253	& 0.00063 & &s&  13.9$\pm$1.5&   15.25$\pm$0.04 & 15.25$\pm$0.05 \\
            & 1240.3947	& 0.00036 & &s&   6.4$\pm$1.4&   15.15$\pm$0.08 \\
Al I        & 2367.7750 & 0.10600 & &s&$<$1.8        &$<$11.54          & \nodata \\
            & 3944.0060 & 0.11700 & &u&$<$1.2        &$<$10.86          \\
            &           &         & &U&$<$0.7        &$<$10.62          \\
Al III      & 1854.7184	& 0.55900 & &s&  40.1$\pm$6.5&   12.47$\pm$0.06 & 12.45$\pm$0.05 \\
            & 1862.7910	& 0.27800 & &s&  19.1$\pm$6.7&   12.41$\pm$0.13 \\
Si I        & 1845.5202	& 0.27000 & &s&   3.1$\pm$1.0&   11.61$\pm$0.10 & [11.61] \\
            & 2208.6666	& 0.05750 & &s&   1.0$\pm$0.4&   11.61$\pm$0.10 \\
Si II       & 1260.4221	& 1.18000 & &s& stellar      &   \nodata        & 15.39$\pm$0.02 \\
            & 1304.3702	& 0.08630 & &s& 173.0$\pm$1.3&$>$14.75          \\
            & 1808.0129	& 0.00208 & &s&  92.6$\pm$4.0&   15.37$\pm$0.02 \\
            & 2335.1230	& 4.25E-06& &s&$<$3.6        &$<$16.25          \\     
Si II*      & 1264.7377	& 1.05000 & &s& stellar      &   \nodata        & \nodata \\
            & 1265.0020	& 0.11700 & &s& stellar      &   \nodata        \\
Si III      & 1206.5000	& 1.63000 & &s& stellar      &   \nodata        & \nodata \\
Si IV       & 1393.7602	& 0.51300 & &s&  45.1$\pm$2.6&   12.80$\pm$0.03 & 12.83$\pm$0.03 \\
            & 1402.7729	& 0.25400 & &s&  29.1$\pm$3.4&   12.86$\pm$0.05 \\
P I         & 1283.8810	&[0.14]   &*&s&   2.2$\pm$0.5&  [12.02]         & [12.02] \\
            & 1286.4440	&[0.14]   &*&s&   2.2$\pm$0.5&  [12.02]         \\
            & 1377.0730	& 0.11000 & &s&   0.8$\pm$0.3&   11.66$\pm$0.12 \\
            & 1377.9343	&[0.04]   &*&s&   0.7$\pm$0.2&  [12.02]         \\
            & 1379.4282	& 0.21900 & &s&   3.2$\pm$0.4&   11.96$\pm$0.04 \\
            & 1381.4760	& 0.31600 & &s&   4.1$\pm$0.5&   11.93$\pm$0.04 \\
            & 1782.8291	& 0.11300 & &s&   2.2$\pm$0.7&   11.87$\pm$0.09 \\
            & 1787.6481	& 0.06040 & &s&   3.0$\pm$1.0&   12.27$\pm$0.10 \\
P II        & 1301.8743	& 0.01270 & &s&  13.8$\pm$1.2&   13.94$\pm$0.03 & 14.03$\pm$0.03 \\
S I         & 1241.9050	& 0.00164 & &s&   5.3$\pm$0.6&   14.47$\pm$0.03 & 14.65$\pm$0.02 \\
            & 1247.1342	& 0.00064 & &s&   2.1$\pm$0.4&   14.41$\pm$0.05 \\
            & 1247.1602	& 0.00430 & &s&   6.2$\pm$0.1&   14.11$\pm$0.04 \\
            & 1262.8599	& 0.00205 & &s&   8.0$\pm$0.5&   14.56$\pm$0.02 \\
            & 1270.7686	& 0.00011 & &s& blend        &   \nodata        \\
            & 1270.7804	& 0.00837 & &s&  16.9$\pm$0.7&   14.37$\pm$0.01 \\
            & 1270.7874	& 0.00158 & &s& blend        &   \nodata        \\
            & 1295.6531	& 0.08700 & &s&  20.2$\pm$0.5&   13.44$\pm$0.01 \\
            & 1296.1739	& 0.02200 & &s&  14.9$\pm$0.4&   13.88$\pm$0.01 \\
            & 1303.4300	& 0.00436 & &s&   9.5$\pm$0.7&   14.31$\pm$0.02 \\
            & 1316.5425	& 0.02790 & &s&  15.2$\pm$0.4&   13.78$\pm$0.01 \\
            & 1316.6150	& 0.00530 & &s&  14.6$\pm$0.4&   14.44$\pm$0.01 \\
            & 1316.6219	& 0.00036 & &s& blend        &   \nodata        \\
            & 1381.5527	&[0.00018]&*&s&   1.4$\pm$0.3&  [14.65]         \\
            & 1388.4358	&[0.00034]&*&s&   2.6$\pm$0.2&  [14.65]         \\
            & 1401.5142	& 0.01280 & &s&  14.5$\pm$0.4&   14.03$\pm$0.01 \\
            & 1425.0300	& 0.12500 & &s&  24.2$\pm$0.4&   13.33$\pm$0.01 \\
            & 1425.1879	& 0.02230 & &s&  17.4$\pm$0.3&   13.85$\pm$0.01 \\
            & 1425.2191	& 0.00148 & &s&  10.4$\pm$0.4&   14.72$\pm$0.01 \\
            & 1807.3113	& 0.09050 & &s&  32.0$\pm$1.3&   13.39$\pm$0.01 \\
S II        & 1250.5780	& 0.00543 & &s& 105.7$\pm$1.2&$>$15.62          & \nodata \\
            & 1253.8050	& 0.01090 & &s& 125.3$\pm$1.4&$>$15.42          \\
            & 1259.5180	& 0.01660 & &s& 140.6$\pm$1.3&$>$15.37          \\
Cl I        & 1335.7258	& 0.03130 & &s& blend        &   \nodata        & [14.30] \\
            & 1347.2396	& 0.15300 & &s&  15.9$\pm$0.3&$>$13.03          \\
            & 1379.5279	& 0.00275 & &s&   7.6$\pm$0.5&   14.32$\pm$0.02 \\
            & 1389.6920	& 0.00013 & &s&   0.7$\pm$0.3&   14.53$\pm$0.14 \\
K I         & 4044.1422	& 0.00568 & &u&$<$2.0        &$<$12.39          & 12.09$\pm$0.07 \\
            &           &         & &U&   1.0$\pm$0.2&   12.07$\pm$0.09 \\
            &           &         & &h&   0.5$\pm$0.1&   11.77$\pm$0.08 \\
            & 4047.2132 & 0.00264 & &U&   0.3$\pm$0.1&   11.89$\pm$0.13 \\
            &           &         & &h&   0.3$\pm$0.1&   11.93$\pm$0.09 \\
            & 7698.9740 & 0.33930 & &U&  70.8$\pm$0.7&$>$11.69          \\
            &          	&         & &m&  67.7$\pm$0.2&$>$11.63          \\
            &           &         & &a&  88.0$\pm$3.8&$>$11.75          \\
Ca I        & 4226.7280	& 1.77000 & &u&$<$1.7        & $<$9.78          &  [9.12] \\
            &           &         & &U&$<$0.4        & $<$9.19          \\
            &           &         & &h&   0.4$\pm$0.1&    9.10$\pm$0.09 \\
Ca II       & 3933.6614 & 0.62670 & &c&  68.1$\pm$2.0&   11.96$\pm$0.01 & 11.97$\pm$0.02 \\ 
            &          	&         & &u&  68.2$\pm$2.0&   11.96$\pm$0.01 \\
Ti I        & 3635.4620 & 0.25200 & &u&$<$2.6        &$<$10.95          & \nodata \\
            &           &         & &U&$<$0.3        &$<$10.03          \\								
Ti II       & 3241.9829 & 0.23200 & &U&   7.0$\pm$0.5&   11.51$\pm$0.03 & 11.56$\pm$0.03 \\
            & 3383.7588	& 0.35800 & &u&  10.7$\pm$1.7&   11.48$\pm$0.06 \\
            &           &         & &U&  12.8$\pm$0.4&   11.56$\pm$0.02 \\
V II        & 2138.8270 & 0.08200 & &s&$<$3.3        &$<$12.00          & \nodata \\
Cr I        & 3578.6840	& 0.36600 & &u&$<$2.4        &$<$10.76          & \nodata \\
            &           &         & &U&$<$0.3        & $<$9.68          \\
            & 4254.3320 & 0.10990 &2&h&$<$0.4        &$<$10.37          \\
Mn I        & 4030.7530 & 0.05650 & &h&$<$0.4        &$<$10.67          & \nodata \\								
Mn II       & 1201.1180	& 0.12100 & &s&$<$10.7       &$<$12.84          & \nodata \\
Fe I        & 2167.4534	& 0.15000 & &s&   1.2$\pm$0.5&   11.28$\pm$0.15 & [11.30] \\
            & 3440.6057 & 0.02360 & &U&   1.0$\pm$0.2&   11.59$\pm$0.07 \\
            & 3719.9347	& 0.04110 & &u&   1.2$\pm$0.5&   11.36$\pm$0.14 \\
            &           &         & &U&   1.1$\pm$0.1&   11.33$\pm$0.04 \\
            & 3859.9114	& 0.02170 & &u&   1.3$\pm$0.4&   11.68$\pm$0.09 \\
            &           &         & &U&   0.7$\pm$0.1&   11.41$\pm$0.08 \\
Fe II       & 1260.5330	& 0.02400 & &s& blend        &   \nodata        & 14.48$\pm$0.02 \\
            & 1901.7730	& 0.00007 & &s&$<$4.4        &$<$15.29          \\
            & 2234.4472	& 0.00003 &3&s&$<$2.7        &$<$15.38          \\
            & 2249.8768	& 0.00182 & &s&  14.0$\pm$1.5&   14.26$\pm$0.04 \\
            & 2260.7805	& 0.00244 & &s&  26.5$\pm$1.4&   14.42$\pm$0.02 \\
            & 2344.2139	& 0.11400 & &s& 251.6$\pm$1.9&$>$14.25          \\
            & 2367.5905	& 0.00002 &4&s&$<$2.7        &$<$15.41          \\
            & 2374.4612	& 0.03130 & &s& 171.2$\pm$1.3&   14.41$\pm$0.01 \\
Fe II*      & 2338.7248 & 0.08970 & &s&$<$1.4        &$<$11.50          & \nodata \\
            & 2345.0011 & 0.15300 & &s&$<$1.7        &$<$11.35          \\
            & 2349.0223 & 0.08980 & &s&$<$1.7        &$<$11.58          \\
            & 2389.3582 & 0.08250 & &s&$<$1.3        &$<$11.51          \\
Co II       & 2012.1664 & 0.03680 & &s&$<$5.5        &$<$12.60          & \nodata \\
Ni I        & 2320.7468 & 0.68500 & &s&$<$2.1        &$<$10.80          & \nodata \\
Ni II       & 1317.2170	& 0.05710 &5&s&  10.3$\pm$0.9&   13.09$\pm$0.03 & 13.26$\pm$0.04 \\
            & 1370.1320	& 0.05880 &5&s&  16.2$\pm$1.0&   13.26$\pm$0.03 \\
Ni II*      & 2217.1670 & 0.30000 &6&s&$<$1.9        &$<$11.16          & \nodata \\
Cu II       & 1358.7730	& 0.26300 & &s&   2.9$\pm$0.9&   11.84$\pm$0.14 & 11.90$\pm$0.04 \\
Zn I        & 2139.2477	& 1.47000 & &s&  18.6$\pm$0.7&   11.67$\pm$0.01 & [12.14] \\
Zn II       & 2026.1370	& 0.50100 & &s&  89.0$\pm$2.3&$>$12.93          & 13.13$\pm$0.05 \\
Ga II       & 1414.4020	& 1.77000 & &s&   2.5$\pm$1.0&   10.91$\pm$0.15 & 10.92$\pm$0.07 \\
Ge II       & 1237.0591	& 1.23200 &7&s&   7.0$\pm$1.8&   11.72$\pm$0.11 & 11.94$\pm$0.06 \\
As II       & 1263.7700 & 0.25900 &8&s&$<$1.6        &$<$11.64          & \nodata \\
Kr I        & 1235.8380	& 0.20400 &8&s&   5.6$\pm$1.5&   12.36$\pm$0.11 & 12.41$\pm$0.05 \\
Cd II       & 2145.0697 & 0.51000 &8&s&   2.8$\pm$1.8&   11.14$\pm$0.14 & 11.18$\pm$0.06 \\
Sn II       & 1400.4400	& 1.03000 &8&s&   1.4$\pm$0.6&   10.91$\pm$0.15 & 10.89$\pm$0.08 \\
Pb II       & 1433.9050	& 0.32100 &9&s&$<$2.0        &$<$11.54          & \nodata \\
\enddata
\tablecomments{Uncertainties are 1-$\sigma$; limits are 3-$\sigma$.
AOD column densities are lower limits to the true values when derived from stronger lines.
Values in square braces are assumed, in order to estimate new $f$ values.}
\tablenotetext{a}{Transition $f$ values are from Morton (2003), unless specified otherwise: 1 = Jenkins \& Tripp (2001, 2011; C only); 2 = Morton 1991; 3 = Miller et al. 2007; 4 = Welty et al. 1999; 5 = Jenkins \& Tripp 2006; 6 = Gnaci\'{n}ski 2009; 7 = Heidarian et al. 2017; 8 = Morton 2000; 9 = Heidarian et al. 2015; * = estimated in this study (see Sec.~\ref{sec-apew})}
\tablenotetext{b}{Sources for the spectra are:  a = AAT/UCLES, c = ESO/CES, h = Keck/HIRES, m = Magellan/MIKE, s = {\it HST}/STIS, u/U = ESO/UVES.}
\end{deluxetable}

\begin{deluxetable}{llcllrcrr}
\tablecolumns{9}
\tabletypesize{\scriptsize}
\tablecaption{Equivalent Widths -- Molecular Lines \label{tab:ewmol}}
\tablewidth{0pt}

\tablehead{
\multicolumn{1}{c}{Species}&
\multicolumn{1}{c}{Band}&
\multicolumn{1}{c}{Transition}&
\multicolumn{1}{c}{$\lambda$}&
\multicolumn{1}{c}{f}&
\multicolumn{1}{c}{Ref\tablenotemark{a}}&
\multicolumn{1}{c}{Source\tablenotemark{b}}&
\multicolumn{1}{c}{$W_{\lambda}$}&
\multicolumn{1}{c}{log[$N$(AOD)]}}

\startdata
H$_2$     & B-X(0,3) &R0& 1274.537 & 0.09118& 1&s&  0.9$\pm$0.3& 11.84$\pm$0.15 \\
          &          &R1& 1274.922 & 0.06035& 1&s&  1.0$\pm$0.6& 12.06$\pm$0.27 \\
          &          &P1& 1276.812 & 0.03058& 1&s&  0.6$\pm$0.3& 12.15$\pm$0.25 \\
          &          &R2& 1276.325 & 0.05385& 1&s&  1.7$\pm$0.3& 12.34$\pm$0.08 \\
          &          &P2& 1279.464 & 0.03674& 1&s&  1.4$\pm$0.3& 12.41$\pm$0.11 \\
          &          &R3& 1278.734 & 0.05074& 1&s&$<$1.3       & $<$12.25       \\
          &          &P3& 1283.111 & 0.03935& 1&s&$<$1.4       & $<$12.37       \\
          &          &R4& 1282.141 & 0.04868& 1&s& blend       & \nodata        \\
          &          &P4& 1287.732 & 0.04072& 1&s&$<$1.4       & $<$12.37       \\
          & B-X(0,4) &R0& 1333.475 & 0.12500& 1&s&  1.1$\pm$0.3& 11.74$\pm$0.13 \\
          &          &R1& 1333.797 & 0.08352& 1&s&  1.6$\pm$0.4& 12.08$\pm$0.10 \\
          &          &P1& 1335.868 & 0.04150& 1&s& blend       & \nodata        \\
          &          &R2& 1335.131 & 0.07531& 1&s&  1.6$\pm$0.8& 12.12$\pm$0.24 \\
          &          &P2& 1338.568 & 0.04972& 1&s&  0.9$\pm$0.2& 12.06$\pm$0.12 \\
          &          &R3& 1337.467 & 0.07190& 1&s&  0.8$\pm$0.3& 11.85$\pm$0.15 \\
          &          &P3& 1342.256 & 0.05322& 1&s& blend       & \nodata        \\
          &          &R4& 1340.791 & 0.07000& 1&s&  0.6$\pm$0.3& 11.72$\pm$0.20 \\
          &          &P4& 1346.908 & 0.05518& 1&s&  0.6$\pm$0.4& 11.82$\pm$0.30 \\
 & \\
CH        & G-X      &  & 1368.74  & 0.019  & 2&s&  3.3$\pm$0.3& 13.05$\pm$0.03 \\
          &          &  & 1369.13  & 0.030  & 2&s&  3.8$\pm$0.4& 12.90$\pm$0.06 \\
          &          &  & 1370.87  & 0.009  & 2&s&  1.5$\pm$0.3& 13.03$\pm$0.10 \\
          & C-X(0,0) &  & 3137.559 & 0.0021 & 3&U&  2.6$\pm$0.2& 13.15$\pm$0.04 \\
          &          &  & 3143.160 & 0.0064 & 3&U&  8.3$\pm$0.4& 13.19$\pm$0.02 \\
          &          &  & 3145.992 & 0.0043 & 3&U&  5.8$\pm$0.2& 13.20$\pm$0.02 \\
          & B-X(1,0) &Q$_2$(1)+$^{\rm Q}$R$_{12}$(1)  & 
                          3633.289 & 0.00104& 4&u&  1.5$\pm$0.6& 13.09$\pm$0.18 \\
          &          &  &          &        &  &U&  2.0$\pm$0.2& 13.21$\pm$0.04 \\
          &          &$^{\rm P}$Q$_{12}$(1)  & 
                          3636.222 & 0.00069& 4&u&  1.2$\pm$0.5& 13.17$\pm$0.19 \\
          &          &  &          &        &  &U&  1.1$\pm$0.1& 13.15$\pm$0.05 \\
          & B-X(0,0) &R$_2$(1)  & 
                          3878.774 & 0.0011 & 5&u&  2.0$\pm$0.6& 13.14$\pm$0.06 \\
          &          &  &          &        &  &U&  2.8$\pm$0.2& 13.28$\pm$0.03 \\
          &          &Q$_2$(1)+$^{\rm Q}$R$_{12}$(1)  & 
                          3886.409 & 0.0032 & 5&u&  7.2$\pm$0.7& 13.23$\pm$0.03 \\
          &          &  &          &        &  &U&  7.0$\pm$0.3& 13.22$\pm$0.02 \\
          &          &$^{\rm P}$Q$_{12}$(1)  & 
                          3890.217 & 0.0022 & 5&u&  4.9$\pm$0.6& 13.24$\pm$0.03 \\
          &          &  &          &        &  &U&  4.7$\pm$0.3& 13.21$\pm$0.03 \\
          & A-X(0,0) &R$_{\rm 2e}$(1)+R$_{\rm 2f}$(1)  & 
                          4300.3132& 0.00506& 6&u& 23.2$\pm$0.8& 13.50$\pm$0.01 \\
          &          &  &          &        &  &U& 22.1$\pm$0.3& 13.47$\pm$0.01 \\
          &          &  &          &        &  &h& 22.0$\pm$0.1&$>$13.47        \\
CH$^+$    & A-X(0,0) &R0& 4232.548 & 0.00545& 7&u&$<$1.7	    &$<$12.29        \\
          &          &  &          &        &  &h&  0.4$\pm$0.1& 11.68$\pm$0.07 \\
CH$_2$    & D-X      &  & 1396.8   & 0.0050 & 8&s&$<$0.9       &$<$13.01        \\ 
C$_2^+$   & B-X(1,0) &R1& 4711.400 & \nodata&  &U&$<$0.4       & \nodata        \\
          & B-X(0,0) &R1& 5065.200 & \nodata&  &U&$<$0.4       & \nodata        \\
C$_2^-$   & B-X(1,0) &R0& 4894.200 & \nodata&  &U&$<$0.4       & \nodata        \\
CN        & B-X(1,0) &R1& 3579.453 & 0.00196& 9&u&  2.0$\pm$0.8& 12.97$\pm$0.13 \\
          &          &  &          &        &  &U&  2.1$\pm$0.1& 12.99$\pm$0.03 \\
          &          &R0& 3579.963 & 0.00294& 9&u&  6.8$\pm$0.9& 13.33$\pm$0.04 \\
          &          &  &          &        &  &U&  6.4$\pm$0.1& 13.30$\pm$0.01 \\
          &          &P1& 3580.940 & 0.00098& 9&u&$<$2.4       &$<$13.32        \\
          &          &  &          &        &  &U&	 0.7$\pm$0.1& 12.78$\pm$0.07 \\
          & B-X(0,0) &R2& 3873.370 & 0.0203 &10&u&  1.2$\pm$0.6& 11.66$\pm$0.15 \\
          &          &  &          &        &  &U&  1.2$\pm$0.2& 11.63$\pm$0.07 \\
          &          &R$_1$(1)+R$_2$(1)+$^{\rm R}$Q$_{21}$(1)& 
                          3873.999 & 0.0225 &10&u& 19.0$\pm$0.7& 12.85$\pm$0.01 \\ 
          &          &  &          &        &  &U& 18.7$\pm$0.3& 12.84$\pm$0.01 \\
          &          &R$_1$(0)+$^{\rm R}$Q$_{21}$(0)& 
                          3874.607 & 0.0337 &10&u& 27.9$\pm$0.7& 12.87$\pm$0.01 \\ 
          &          &  &          &        &  &U& 27.7$\pm$0.2& 12.86$\pm$0.01 \\
          &          &P$_1$(1)+$^{\rm P}$Q$_{12}$(1)& 
                          3875.764 & 0.0112 &10&u& 11.7$\pm$0.8& 12.92$\pm$0.02 \\ 
          &          &  &          &        &  &U& 11.5$\pm$0.2& 12.91$\pm$0.01 \\
          &          &P2& 3876.312 & 0.0135 &10&U&  0.6$\pm$0.2& 11.55$\pm$0.11 \\
          & A-X(2,0) &$^{\rm R}$Q$_{21}$(0)& 
                          7874.847 & 2.54e-4&11&U&  3.4$\pm$0.3& 13.39$\pm$0.04 \\
          &          &R$_1$(1)& 
                          7903.896 & 2.54e-4&11&U&  1.4$\pm$0.2& 13.00$\pm$0.08 \\
          &          &  &          &        &  &m&  1.5$\pm$0.6& 13.04$\pm$0.17 \\
          &          &R$_1$(0)& 
                          7906.601 & 4.01e-4&11&U&  4.7$\pm$0.3& 13.33$\pm$0.03 \\
          &          &  &          &        &  &m&  6.0$\pm$0.5& 13.43$\pm$0.03 \\ 
          &       & $^{\rm Q}$R$_{12}$(1)+Q$_1$(1)& 
                          7908.955 & 2.33e-4&11&U&  1.7$\pm$0.3& 13.11$\pm$0.08 \\
$^{13}$CN & B-X(0,0) & R$_1$(1)+R$_2$(1)+$^{\rm R}$Q$_{21}$(1)&
                          3874.176 & 0.0225 &10&U&  1.0$\pm$0.2& 11.52$\pm$0.09 \\
          &          &R$_1$(0)+$^{\rm R}$Q$_{21}$(0)&
                          3874.783 & 0.0337 &10&U&  1.7$\pm$0.2& 11.59$\pm$0.04 \\
          &          &  &          &        &  &m&  1.3$\pm$0.2& 11.47$\pm$0.06 \\
CN$^+$    & f-a      &  & 2181.78  &\nodata &  &s&$<$1.7       & \nodata        \\
CO$^+$    & A-X(2,0) &  & 4250.94  & 0.00064&12&u&$<$1.4       &$<$13.15        \\
          &          &  &          &        &  &h&$<$0.4       &$<$12.59        \\
CS        & C-X(0,0) &  & 1400.88  & 0.45   &13&s&  9.3$\pm$0.6& 12.12$\pm$0.03 \\ 
NH        & A-X(0,0) &Q0& 3353.9253& 0.0024 &14&u&  1.7$\pm$0.7& 12.87$\pm$0.13 \\
          &          &  &          &        &  &U&  1.8$\pm$0.2& 12.87$\pm$0.05 \\
          &          &R0& 3358.0525& 0.0041 &14&u&  3.1$\pm$0.8& 12.89$\pm$0.08 \\
          &          &  &          &        &  &U&  3.3$\pm$0.2& 12.91$\pm$0.03 \\
N$_2$     & a-X(6,0) &R0& 1273.30  & \nodata&  &s&$<$1.1       & \nodata        \\ 
NO        & A-X(0,0) &  & 2262.98  & 0.00052&15&s&$<$1.6       &$<$13.86        \\ 
NO$^+$    & A-X(2,0) &R0& 1312.94  & 0.0017 &15&s&$<$0.9       &$<$13.54        \\ 
OH        & D-X(0,0) &Q$_1$(3/2)+$^{\rm Q}$P$_{21}$(3/2)& 
                          1222.07  & 0.0115 &16&s&  5.0$\pm$2.2& 13.58$\pm$0.13 \\ 
          &          &P$_1$(3/2)& 
                          1222.52  & 0.0071 &16&s&  5.2$\pm$2.2& 13.81$\pm$0.12 \\
          & A-X(0,0) &Q$_1$(3/2)& 
                          3078.4398& 6.20e-4&17&U&  4.6$\pm$0.4& 13.73$\pm$0.04 \\
          &          &$^{\rm Q}$P$_{21}$(3/2)& 
                          3078.4720& 4.31e-4&17&U&  blend      &                \\
          &          &P$_1$(3/2)& 
                          3081.6644& 6.48e-4&17&U&  3.0$\pm$0.4& 13.75$\pm$0.06 \\
OH$^+$    & A-X(1,0) &R0& 3346.9556& 3.52e-4&18&U&$<$0.47      &$<$13.13        \\
          & A-X(0,0) &Q0& 3572.6519& 3.12e-4&18&U&$<$0.52      &$<$13.17        \\
          &          &R0& 3583.7557& 5.27e-4&18&u&$<$2.1       &$<$13.54        \\
          &          &  &          &        &  &U&$<$0.39      &$<$12.81        \\ 
H$_2$O    & C-X      &  & 1239.382 & 0.0043 &19&s&$<$2.2       &$<$13.58        \\ 
          &          &  & 1239.728 & 0.0180 &19&s&$<$2.2       &$<$12.95        \\ 
          &          &  & 1240.153 & 0.0057 &19&s&$<$2.2       &$<$13.43        \\ 
SiO       & J-X      &  & 1310.01  &\nodata &  &s&$<$1.6       & \nodata        \\ 
SH        & G-X(0,0) &  & 1257.34  &\nodata &  &s&$<$1.8       & \nodata        \\ 
SH$^+$    & A-X(0,0) &R0& 3363.49  & 0.00062&20&u&$<$2.4       &$<$13.59        \\
          &          &  &          &        &  &U&$<$0.9       &$<$13.14        \\
HCl       & C-X(0,0) &R0& 1290.257 & 0.141  &21&s&$<$1.4       &$<$11.83        \\ 
AlH       & C-X(0,0) &R0& 2242.295 &\nodata &  &s&$<$1.8       & \nodata        \\ 
 & \\
UID       &          &  & 1300.45  &\nodata &  &s&  3.8$\pm$0.4& \nodata        \\
 & \\
DIBs      &          &  & 4963.9   &\nodata &  &o&    6$\pm$1  & \nodata        \\
          &          &  & 5780.6   &\nodata &  &o&   27$\pm$3  & \nodata        \\
          &          &  & 5797.2   &\nodata &  &o&   12$\pm$1  & \nodata        \\
          &          &  & 6196.0   &\nodata &  &o&    4$\pm$1  & \nodata        \\
          &          &  & 6203.7   &\nodata &  &o&   15$\pm$4  & \nodata        \\
          &          &  & 6283.8   &\nodata &  &o&   50$\pm$20 & \nodata        \\
          &          &  & 6613.7   &\nodata &  &o&    7$\pm$1  & \nodata        \\
\enddata
\tablecomments{Uncertainties are 1$\sigma$; limits are 3$\sigma$.}
\tablenotetext{a}{References for $f$ values are: 1 = Abgrall et al. 1994; 2 = Sheffer \& Federman 2007; 3 = S. R. Federman 2012, private communication; 4 = Weselak et al. 2011; 5 = Lien 1984; 6 = Larsson \& Siegbahn 1983a; 7 = Larsson \& Siegbahn 1983b; 8 = Lyu et al. 2001; 9 = Meyer \& Roth 1990; 10 = Brooke et al. 2014; 11 = Gredel et al. 1991; 12 = Herbig 1968; 13 = Xu et al. 2019; 14 = Meyer \& Roth 1991; 15 Jenkins et al. 1973; 16 = Roueff 1996; 17 = Felenbok \& Roueff 1996; 18 = Hodges et al. 2018; 19 = Smith \& Parkinson 1978; 20 = Pineau des For\^{e}ts et al. 1986; 21 = Federman et al. 1995}
\tablenotetext{b}{Sources of the spectra are: h = Keck/HIRES, m = Magellan/MIKE, o = combination of available optical spectra; s = {\it HST}/STIS, u/U = ESO/UVES.}
\end{deluxetable}

\begin{deluxetable}{lcrrcrrr}
\tablecolumns{8}
\tabletypesize{\scriptsize}
\tablecaption{Equivalent Widths -- C$_2$ \label{tab:ewc2}}
\tablewidth{0pt}

\tablehead{
\multicolumn{1}{c}{Band}&
\multicolumn{1}{c}{Transition}&
\multicolumn{1}{c}{$\lambda$}&
\multicolumn{1}{c}{f}&
\multicolumn{1}{c}{Source}&
\multicolumn{1}{c}{$W_{\lambda}$}&
\multicolumn{1}{c}{$N_{12}$(AOD)}&
\multicolumn{1}{c}{$N_{12}$(fit)}}

\startdata
D-X(0,0) & R0 & 2313.163 & 0.0545 & s &  9.7$\pm$0.5 &  4.6$\pm$0.2 &  6.9$\pm$0.5 \\
         & R2 & 2312.769 &        & s & 15.7$\pm$0.6 & 13.6$\pm$0.3 & 25.8$\pm$1.4 \\
         & P2 & 2313.745 &        & s & 11.0$\pm$0.7 & 13.5$\pm$0.5 \\
         & R4 & 2312.371 &        & s & 12.9$\pm$0.5 & 12.0$\pm$0.3 & 25.7$\pm$1.4 \\
         & P4 & 2314.126 &        & s & 12.8$\pm$0.6 & 14.4$\pm$0.4 \\
         & R6 & 2311.967 &        & s & 10.8$\pm$0.5 &  9.9$\pm$0.3 & 16.3$\pm$0.8 \\
         & P6 & 2314.503 &        & s &  9.9$\pm$0.3 & 10.2$\pm$0.2 \\
         & R8 & 2311.560 &        & s &  8.4$\pm$0.5 &  7.2$\pm$0.3 &  8.5$\pm$0.4 \\
         & P8 & 2314.874 &        & s &  6.7$\pm$0.4 &  6.3$\pm$0.2 \\
         & R10& 2311.147 &        & s &  4.5$\pm$0.4 &  3.7$\pm$0.2 &  4.8$\pm$0.3 \\
         & P10& 2315.240 &        & s &  4.9$\pm$0.4 &  4.4$\pm$0.3 \\
         & R12& 2310.730 &        & s &  3.2$\pm$0.4 &  2.5$\pm$0.2 &  2.9$\pm$0.3 \\
         & P12& 2315.601 &        & s &  2.9$\pm$0.6 &  2.5$\pm$0.2 \\
         & R14& 2310.309 &        & s &  2.5$\pm$0.5 &  1.9$\pm$0.3 &  2.0$\pm$0.2 \\
         & P14& 2315.956 &        & s &  2.6$\pm$0.5 &  2.2$\pm$0.3 \\
         & R16& 2309.884 &        & s &  1.1$\pm$0.4 &  0.8$\pm$0.2 &  1.0$\pm$0.2 \\
         & P16& 2316.307 &        & s &  1.4$\pm$0.5 &  1.2$\pm$0.3 \\
         & R18& 2309.455 &        & s &  1.7$\pm$0.4 &  1.3$\pm$0.2 &  1.0$\pm$0.2 \\
         & P18& 2316.652 &        & s &  1.3$\pm$0.4 &  1.1$\pm$0.2 \\
 & \\
A-X(3,0) & R0 & 7719.329 & 0.00065& U &  2.6$\pm$0.2 &  7.6$\pm$0.7 &  7.7$\pm$0.7 \\
         & R2 & 7716.528 &        & U &  telblend    &              & 28.8$\pm$1.6 \\
         & Q2 & 7722.095 &        & U &  4.5$\pm$0.3 & 26.8$\pm$1.8 &              \\
         & P2 & 7725.819 &        & U &  1.2$\pm$0.2 & 36.4$\pm$5.7 &              \\
         & R4 & 7714.944 &        & U &  2.8$\pm$0.3 & 24.4$\pm$2.3 & 25.8$\pm$1.3 \\
         & Q4 & 7724.219 &        & U &  3.8$\pm$0.2 & 22.5$\pm$1.2 &              \\
         & P4 & 7731.663 &        & U &  2.2$\pm$0.3 & 38.6$\pm$5.1 &              \\
         & R6 & 7714.575 &        & U &  1.9$\pm$0.3 & 18.1$\pm$2.6 & 16.3$\pm$1.3 \\
         & Q6 & 7727.557 &        & U &  2.6$\pm$0.3 & 15.4$\pm$1.5 &              \\
         & R8 & 7715.415 &        & U &  telblend    &              &  7.2$\pm$1.4 \\
         & Q8 & 7732.117 &        & U &  1.1$\pm$0.2 &  6.6$\pm$1.3 &              \\
 & \\
A-X(2,0) & R0 & 8757.686 & 0.00140& m &  6.1$\pm$0.6 &  6.5$\pm$0.6 &              \\ 
         &    &          &        & U &  5.9$\pm$0.3 &  6.3$\pm$0.4 &  7.1$\pm$0.3 \\
         & R2 & 8753.949 &        & m &  8.3$\pm$0.5 & 22.0$\pm$1.0 &              \\ 
         &    &          &        & U &  7.8$\pm$0.2 & 20.9$\pm$0.5 & 24.2$\pm$0.5 \\
         & Q2 & 8761.194 &        & m & 12.5$\pm$0.3 & 27.0$\pm$1.0 &              \\ 
         &    &          &        & U &  9.9$\pm$0.3 & 21.4$\pm$0.7 &              \\
         & P2 & 8766.031 &        & m &  2.8$\pm$0.4 & 29.0$\pm$4.0 &              \\ 
         &    &          &        & U &  2.4$\pm$0.3 & 25.8$\pm$2.9 &              \\
         & R4 & 8751.685 &        & U &  6.5$\pm$0.4 & 20.7$\pm$1.1 & 21.7$\pm$0.5 \\
         & Q4 & 8763.751 &        & m &  9.0$\pm$0.4 & 19.0$\pm$1.0 &              \\ 
         &    &          &        & U &  9.1$\pm$0.3 & 19.6$\pm$0.5 &              \\
         & P4 & 8773.430 &        & U & [2.7$\pm$0.3]&[16.8$\pm$1.7]&              \\
         & R6 & 8750.848 &        & U &  3.9$\pm$0.2 & 13.4$\pm$0.7 & 13.0$\pm$0.5 \\
         & Q6 & 8767.759 &        & m &  5.3$\pm$0.5 & 11.0$\pm$1.0 &              \\ 
         &    &          &        & U &  4.9$\pm$0.3 & 10.5$\pm$0.5 &              \\
         & R8 & 8751.486 &        & U & [1.6$\pm$0.4]& [5.8$\pm$1.1]&  6.5$\pm$0.4 \\
         & Q8 & 8773.221 &        & U &  3.3$\pm$0.3 &  7.0$\pm$0.7 &              \\
 & \\
A-X(1,0) & R0 &10143.723 & 0.00231& U &  8.0$\pm$0.7 &  3.9$\pm$0.3 &  4.2$\pm$0.3 \\
         & R2 &10138.540 &        & U & 12.3$\pm$0.8 & 15.0$\pm$1.0 & 15.1$\pm$0.5 \\
         & Q2 &10148.351 &        & U & 12.5$\pm$0.5 & 12.2$\pm$0.5 &              \\
         & P2 &10154.897 &        & U &  3.4$\pm$0.4 & 16.3$\pm$1.9 &              \\
         & R4 &10135.149 &        & U & 12.1$\pm$0.8 & 17.6$\pm$1.2 & 16.4$\pm$0.5 \\
         & Q4 &10151.523 &        & U & 15.7$\pm$0.5 & 15.3$\pm$0.5 &              \\
         & P4 &10164.763 &        & U &  6.7$\pm$0.6 & 19.1$\pm$1.7 &              \\
         & R6 &10133.603 &        & U &  5.8$\pm$0.5 &  9.0$\pm$0.9 &  8.7$\pm$0.5 \\
         & Q6 &10156.515 &        & U &  7.9$\pm$0.5 &  7.6$\pm$0.4 &              \\
         & R8 &10133.854 &        & U &  5.9$\pm$0.8 &  5.9$\pm$0.8 &  6.1$\pm$0.5 \\
         & Q8 &10163.323 &        & U &  6.0$\pm$0.7 &  5.8$\pm$0.6 &              \\
         & Q10&10171.963 &        & U &  3.2$\pm$0.6 &  3.0$\pm$0.6 &  3.5$\pm$0.5 \\
\enddata
\tablecomments{Sources are m = Magellan/MIKE, s = {\it HST}/STIS, U = ESO VLT/UVES.
Column densities are in units of 10$^{12}$ cm$^{-2}$.}
\end{deluxetable}
\begin{deluxetable}{lclrccccccccr}
\rotate
\tablecolumns{13}
\tabletypesize{\scriptsize}
\tablecaption{Fits to CO Bands \label{tab:cobands}}
\tablewidth{0pt}

\tablehead{
\multicolumn{1}{c}{Species}&
\multicolumn{1}{c}{Band}&
\multicolumn{1}{c}{$\lambda$}&
\multicolumn{1}{c}{log($f\lambda$)}&
\multicolumn{1}{c}{Ref\tablenotemark{a}}&
\multicolumn{1}{c}{$N_{15}$(0)}&
\multicolumn{1}{c}{$N_{15}$(1)}&
\multicolumn{1}{c}{$N_{15}$(2)}&
\multicolumn{1}{c}{$N_{15}$(3)}&
\multicolumn{1}{c}{$N_{14}$(4)}&
\multicolumn{1}{c}{$N_{13}$(5)}&
\multicolumn{1}{c}{$N_{12}$(6)}&
\multicolumn{1}{c}{$N_{15}$(tot)}}

\startdata
$^{12}$CO & (12,0) & 1246.059 & $-$0.950 & 1 &  5.2$\pm$0.7 & 10.5$\pm$1.2 & 7.8$\pm$0.9 & 3.5$\pm$0.5 &
                                               [4.0]        & [3.0]        &[4.0]        &27.4$\pm$1.7 \\
          & (11,0) & 1263.433 & $-$0.643 & 1 &  4.5$\pm$0.7 & 11.0$\pm$1.5 & 7.2$\pm$0.8 & 2.9$\pm$0.4 &
                                               (8.5$\pm$2.4 & [3.0]        &[4.0]        &26.5$\pm$1.9 \\
          & (10,0) & 1281.866 & $-$0.279 & 1 &  7.0$\pm$1.7 & 11.9$\pm$2.4 & 6.0$\pm$0.6 & 2.5$\pm$0.3 &
                                                4.8$\pm$1.3 & [3.0]        &[4.0]        &27.9$\pm$3.0 \\
          &  (9,0) & 1301.403 &    0.092 & 1 &  7.1$\pm$2.6 & 10.0$\pm$2.7 & 5.8$\pm$0.8 & 2.1$\pm$0.3 &
                                                4.7$\pm$0.8 & $<$4.6       &[4.0]        &25.5$\pm$3.8 \\
          &  (8,0) & 1322.150 &    0.423 & 1 &(10.2$\pm$1.8)&(18.7$\pm$2.6)& 8.7$\pm$1.0 & 2.4$\pm$0.2 &
                                                3.3$\pm$0.3 &  5.2$\pm$1.6 &[4.0]        &40.4$\pm$3.3 \\
          &  (7,0) & 1344.186 &    0.745 & 1 &  5.6$\pm$1.0 & 14.1$\pm$2.0 & 7.4$\pm$0.9 & 2.3$\pm$0.3 &
                                                3.7$\pm$0.4 &  1.8$\pm$2.5 &(4.0$\pm$8.1)&29.8$\pm$2.4 \\
          &  (6,0) & 1367.623 &    1.036 & 1 &  5.2$\pm$0.8 &  7.3$\pm$0.9 & 5.9$\pm$0.6 & 1.9$\pm$0.5 &
                                                3.0$\pm$0.2 &  2.7$\pm$0.7 & 8.6$\pm$3.4 &20.6$\pm$1.4 \\
          &  (5,0) & 1392.525 &    1.305 & 1 & (2.7$\pm$0.4)& (5.8$\pm$0.6)&(3.7$\pm$0.5)& 2.7$\pm$0.5 &
                                                2.6$\pm$0.2 &  3.1$\pm$0.3 & 7.5$\pm$0.2 &15.2$\pm$1.0 \\
          &  (4,0) & 1419.044 &    1.520 & 1 & (2.4$\pm$0.3)& (4.0$\pm$0.4)&(3.1$\pm$0.4)& 2.2$\pm$0.4 &
                                                1.6$\pm$0.6 &  2.6$\pm$0.2 & 2.8$\pm$1.3 &11.9$\pm$0.8 \\
\cline{2-13}
          &    avg & \nodata  & \nodata  &   &  5.8$\pm$1.1 & 10.8$\pm$2.2 & 7.0$\pm$1.1 & 2.4$\pm$0.6 &
                                                3.4$\pm$1.1 &  3.3$\pm$1.3 & 5.0$\pm$3.1 &26.4$\pm$2.8 \\
          & simult & \nodata  & \nodata  &   &  4.4$\pm$0.3 &  7.9$\pm$0.4 & 6.2$\pm$0.3 & 2.2$\pm$0.1 &
                                                3.5$\pm$0.2 &  2.8$\pm$0.3 &[5.5]        &21.1$\pm$0.6 \\
\cline{2-13}
          &  e12   & 1323.1   & $-$2.070 & 2 &  4.8$\pm$1.4 & 13.7$\pm$1.6 &11.5$\pm$1.7 & \nodata     &
                                                \nodata     & \nodata      & \nodata     &32.9$\pm$2.7 \\
          &  a'17  & 1366.214 & $-$1.460 & 1 &  4.0$\pm$1.2 & 16.0$\pm$1.9 &10.1$\pm$1.5 & 2.5$\pm$1.0 &
                                                \nodata     & \nodata      & \nodata     &33.0$\pm$2.9 \\
          &  d9    & 1422.28  & $-$1.256 & 2 &  6.9$\pm$0.9 & 13.4$\pm$1.1 & 7.1$\pm$1.1 & 1.8$\pm$0.9 &
                                                \nodata     & \nodata      & \nodata     &29.6$\pm$2.0 \\
          &  d12   & 1366.443 & $-$1.103 & 2 &  7.8$\pm$1.4 & 11.2$\pm$1.0 & 5.2$\pm$1.1 & 2.2$\pm$0.7 &
                                                \nodata     & \nodata      & \nodata     &26.8$\pm$2.2 \\
          &  a'14  & 1419.535 &    0.072 & 2 &  9.0$\pm$3.3 & 10.5$\pm$1.4 & 6.1$\pm$0.5 & 1.8$\pm$0.1 &
                                                5.6$\pm$0.8 & \nodata      & \nodata     &28.0$\pm$3.6 \\
\cline{2-13}
          &   avg  & \nodata  & \nodata  &   &  6.5$\pm$2.1 & 13.0$\pm$2.2 & 8.0$\pm$2.7 & 2.1$\pm$0.3 &
                                                5.6         & \nodata      & \nodata     &30.1$\pm$4.1 \\\cline{2-13}
          &  {\bf adopt} & \nodata  & \nodata  &   &  {\bf 6.0$\pm$1.0} & {\bf 10.5$\pm$2.0} & {\bf 7.0$\pm$1.5} & {\bf 2.5$\pm$0.5} & 
                                                {\bf 3.5$\pm$0.5} &  {\bf 3.2$\pm$0.5} & {\bf 5.0$\pm$2.0} &{\bf 26.4$\pm$2.7} \\
\hline
\multicolumn{5}{c}{ }&
\multicolumn{1}{c}{$N_{13}$(0)}&
\multicolumn{1}{c}{$N_{13}$(1)}&
\multicolumn{1}{c}{$N_{13}$(2)}&
\multicolumn{1}{c}{$N_{13}$(3)}&
\multicolumn{1}{c}{ }&
\multicolumn{1}{c}{ }&
\multicolumn{1}{c}{ }&
\multicolumn{1}{c}{$N_{13}$(tot)} \\
\hline
$^{13}$CO &  (8,0) & 1325.793 &    0.428 & 1 & 12.2$\pm$1.2 & 16.9$\pm$1.5 & 8.4$\pm$1.2 & 4.8$\pm$1.2 & \nodata & \nodata & \nodata & 42.3$\pm$2.6 \\
          &  (7,0) & 1347.575 &    0.747 & 1 & 11.6$\pm$0.9 & 22.1$\pm$1.2 & 8.1$\pm$0.6 & 1.0$\pm$0.5 & \nodata & \nodata & \nodata & 42.8$\pm$1.7 \\
          &  (6,0) & 1370.617 &    0.951 & 1 & 17.0$\pm$2.1 & 21.7$\pm$1.7 & 9.0$\pm$0.7 & 1.1$\pm$0.4 & \nodata & \nodata & \nodata & 48.8$\pm$2.8 \\
          &  (5,0) & 1395.229 &    1.306 & 1 & 10.6$\pm$1.6 & 19.8$\pm$2.1 & 6.4$\pm$0.5 & 1.8$\pm$0.2 & \nodata & \nodata & \nodata & 38.6$\pm$2.7 \\
          &  (4,0) & 1421.340 &    1.537 & 1 & 13.3$\pm$2.8 & 23.4$\pm$3.6 & 7.4$\pm$0.7 & 1.2$\pm$0.2 & \nodata & \nodata & \nodata & 45.3$\pm$4.6 \\
\cline{2-13}
          &   avg  & \nodata  & \nodata  &   & 12.9$\pm$2.5 & 20.8$\pm$2.5 & 7.9$\pm$1.0 & 1.3$\pm$0.4 & \nodata & \nodata & \nodata & 42.9$\pm$3.7 \\
          &  simult& \nodata  & \nodata  &   & 12.8$\pm$0.8 & 21.1$\pm$0.9 & 7.6$\pm$0.3 & 1.5$\pm$0.1 & \nodata & \nodata & \nodata & 43.0$\pm$1.2 \\
\cline{2-13}
          &  d12   & 1370.928 &    0.322 & 1 & 12.7$\pm$1.7 & 21.6$\pm$2.2 &11.1$\pm$2.1 & 4.4$\pm$2.3 & \nodata & \nodata & \nodata & 49.8$\pm$4.2 \\
\cline{2-13}
          &  {\bf adopt} & \nodata  & \nodata  &   & {\bf 12.8$\pm$0.8} & {\bf 21.1$\pm$0.9} & {\bf 7.6$\pm$0.3} & {\bf 1.5$\pm$0.1} & \nodata & \nodata & \nodata & {\bf 43.0$\pm$1.2} \\
\hline
\multicolumn{5}{c}{ }&
\multicolumn{1}{c}{$N_{12}$(0)}&
\multicolumn{1}{c}{$N_{12}$(1)}&
\multicolumn{1}{c}{ }&
\multicolumn{1}{c}{ }&
\multicolumn{1}{c}{ }&
\multicolumn{1}{c}{ }&
\multicolumn{1}{c}{ }&
\multicolumn{1}{c}{$N_{12}$(tot)} \\
\hline
C$^{18}$O &  (5,0) & 1395.455 &    1.306 & 1 &  4.9$\pm$1.3 &  5.0$\pm$1.6 & \nodata & \nodata & \nodata & \nodata & \nodata &  9.9$\pm$2.1 \\
          &  (4,0) & 1421.530 &    1.537 & 1 &  2.8$\pm$0.9 &  2.9$\pm$1.3 & \nodata & \nodata & \nodata & \nodata & \nodata &  5.7$\pm$1.6 \\
\cline{2-13}
          & simult & \nodata  & \nodata  &   &  4.0$\pm$0.8 &  3.4$\pm$1.0 & \nodata & \nodata & \nodata & \nodata & \nodata &  7.4$\pm$1.3 \\
\enddata
\tablecomments{Subscript on $N$($J$) in column headings denotes power of ten for values in that column.
Column densities in parentheses or square braces were fixed in fits to the line profiles.}
\tablenotetext{a}{References for $f$ values:  1 = Morton \& Noreau 1994; 2 = Eidelsberg \& Rostas 2003.}
\end{deluxetable}


\clearpage

\begin{references}
\reference {ang12} Abdel-Naby, Sh. A., Nicoli\'{c}, D., Gorczyca, T. W., Korista, K. T., \& Badnell, N. R. 2012, \aap, 537, A40
\reference {abel16} Abel, N. P., Ferland, G. J., O'Dell, C. R., \& Troland, T. H. 2016, \apj, 819, 136
\reference {ar94} Abgrall, H., Roueff, E., Launay, F., \& Roncin, J.-Y. 1994, CaJPh, 72, 856
\reference {adam03} \'{A}d\'{a}mkovics, M., Blake, G. A., \& McCall, B. J. 2003, \apj, 595, 235
\reference {adam05} \'{A}d\'{a}mkovics, M., Blake, G. A., \& McCall, B. J. 2005, \apj, 625, 857
\reference {agun10} Agundez, M., Goicoechea, J. R., Cernicharo, J., Faure, A., \& Roueff, E. 2010, \apj, 713, 662
\reference {ad71} Allison, A. C., \& Dalgarno, A. 1971, \aap, 13, 331
\reference {and91} Andersson, B.-G., Wannier, P. G., \& Morris, M. 1991, \apj, 366, 464
\reference {ar85} Arnaud, M., \& Rothenflug, R. 1985, \aaps, 60, 425
\reference {bac19} Bacalla, X. L., Linnartz, H., Cox, Nick L. J., et al. 2019, \aap, 622, A31
\reference {bad06} Badnell, N. R. 2006, \apjs, 167, 334
\reference {bad03} Badnell, N. R., O'Mullane, M. G., Summers, H. P. et al. 2003, \aap, 406, 1151
\reference {bw68} Bahcall, J. N., \& Wolf, R. A. 1968, \apj, 152, 701
\reference {bal75} Balona, L. A. 1975, Mem. R. Astron. Soc., 78, 51
\reference {bau98} Bautista, M. A., Romano, P., \& Pradhan, A. K. 1998, \apjs, 118, 259
\reference {ber06} Berengut, J. C., Flambaum, V. V., \& Kozlov, M. G. 2006, \pra, 73, 012504
\reference {bvd91} Black, J. H., \& van Dishoeck, E. F. 1991, \apjl, 369, L9
\reference {bsd78} Bohlin, R. C., Savage, B. D., \& Drake, J. F. 1978, \apj, 224, 132
\reference {bro14} Brooke, J. S. A., Ram, R. S., Western, C. M. et al. 2014, \apjs, 210, 23
\reference {bry09} Bryans, P., Kreckel, H., Roueff, E., Wakelam, V., \& Savin, D. W. 2009, \apj, 694, 286
\reference {bur64} Burgess, A. 1964, \apj, 139, 776
\reference {bfj10} Burgh, E., France, K., \& Jenkins, E. B. 2010, \apj, 708, 334 
\reference {bfm07} Burgh, E., France, K., \& McCandliss, S. 2007, \apj, 658, 446 
\reference {card94} Cardelli, J. A. 1994, Sci, 265, 209
\reference {ccm89} Cardelli, J. A., Clayton, G. C., \& Mathis, J. S. 1989, \apj, 345, 245
\reference {cs88} Cardelli, J. A., \& Savage, B. D. 1988, \apj, 325, 864
\reference {cses90} Cardelli, J. A., Suntzeff, N. B., Edgar, R. J., \& Savage, B. D. 1990, \apj, 362, 551
\reference {cart06} Cartledge, S. I. B., Lauroesch, J. T., Meyer, D. M., \& Sofia, U. J. 2006, \apj, 641, 327
\reference {casu12} Casu, S., \& Cecchi-Pestellini, C. 2012, \apj, 749, A48
\reference {cs01} Chappell, D., \& Scalo, J. 2001, \apj, 551, 712
\reference {ch96} Churchwell, E., Winnberg, A., Cardelli, J., Cooper, G., \& Suntzeff, N. B. 1996, \apj, 469, 209
\reference {cox05} Cox, N. L. J., Kaper, L., Foing, B. H., \& Ehrenfreund, P. 2005, \aap, 438, 187 
\reference {craw90} Crawford, I. A. 1990, Observatory, 110, 145  
\reference {dfl84} Danks, A. C., Federman, S. R., \& Lambert, D. L. 1984, \aap, 130, 62
\reference {dek00} Dekker, H., D'Odorico, S., Kaufer, A., Delabre, B., \& Kotzlowski, H. 2000, Proc. SPIE, 4008, 534
\reference {ds09} Destree, J. D., \& Snow, T. P. 2009, \apj, 697, 684
\reference {dsb09} Destree, J. D., Snow, T. P., \& Black, J. H. 2009, \apj, 693, 804 
\reference {ds94} Diplas, A., \& Savage, B. D. 1994, \apjs, 93, 211
\reference {dm19} Dirks, C, \& Meyer, D. M. 2019, \apj, 872, 140
\reference {drai78} Draine, B, T. 1978, \apjs, 36, 595 (D78) 
\reference {drai90} Draine, B. T. 1990, in Evolution of the Interstellar Medium, ed. L. Blitz, (Astron. Soc. Pacific), p. 193
\reference {dk86} Draine, B. T., \& Katz, N. 1986, \apj, 310, 392
\reference {er03} Eidelsberg, M., \& Rostas, F. 2003, \apjs, 145, 89
\reference {ew78} Elitzur, M., \& Watson, W. D. 1978, \apjl, 222, L41
\reference {ena82} Enard, D. 1982, SPIE, 331, 232
\reference {falg95} Falgarone, E., Pineau des For\^{e}ts, G., \& Roueff, E. 1995, \aap, 300, 870
\reference {fan17} Fan, H., Welty, D. E., York, D. G. et al. 2017, \apj, 850, 194
\reference {ftw55} Feast, M. W., Thackeray, A. D., \& Wesselink, A. J. 1955, MmRAS, 67, 51
\reference {fed95} Federman, S. R., Cardelli, J. A., van Dishoeck, E. F., Lambert, D. L., \& Black, J. H. 1995, \apj, 445, 325
\reference {fed97} Federman, S. R., Knauth, D. C., Lambert, D. L., \& Andersson, B.-G. 1997, \apj, 489, 758
\reference {fed96} Federman, S. R., Rawlings, J. M. C., Taylor, S. D., \& Williams, D. A. 1996, \mnras, 279, L41
\reference {fed03} Federman, S. R., Lambert, D. L., Sheffer, Y., et al. 2003, \apj, 591, 986
\reference {fed94} Federman, S. R., Strom, C. J., Lambert, D. L. et al. 1994, \apj, 424, 772
\reference {fr96} Felenbok, P., \& Roueff, E. 1996, \apjl, 465, L57
\reference {fitz07} Fitzpatrick, E. L. 1997, \apjl, 482, L199 
\reference {fm07} Fitzpatrick, E. L., \& Massa, D. 2007, \apj, 663, 320
\reference {fs97} Fitzpatrick, E. L., \& Spitzer, L. 1997, \apj, 475, 623
\reference {fried11} Friedman, S. D., York, D. G., McCall, B. J. et al. 2011, \apj, 727, 33
\reference {gnac09} Gnaci\'{n}ski, P. 2009, \actaa, 59, 325
\reference {gnac11} Gnaci\'{n}ski, P. 2011, \aap, 532, A122
\reference {gnac13} Gnaci\'{n}ski, P. 2013, \aap, 549, A37
\reference {god09} Godard, B., Falgarone, E., \& Pineau des For\^{e}ts, G. 2009, \aap, 495, 847
\reference {god14} Godard, B., Falgarone, E., \& Pineau des For\^{e}ts, G. 2014, \aap, 570, A27 
\reference {gold13} Goldsmith, P. F. 2013, \apj, 774, 134
\reference {gcc09} Gordon, K. D., Cartledge, S., \& Clayton, G. C. 2009, \apj, 705, 1320
\reference {gvb91} Gredel, R., van Dishoeck, E. F., \& Black, J. H. 1991, \aap, 251, 625  
\reference {gvb93} Gredel, R., van Dishoeck, E. F., \& Black, J. H. 1993, \aap, 269, 477
\reference {gvb94} Gredel, R., van Dishoeck, E. F., \& Black, J. H. 1994, \aap, 285, 300
\reference {hea17} Heays, A. N., Bosman, A. D., \& van Dishoeck, E. F. 2017, \aap, 602, A105
\reference {heid17} Heidarian, N., Irving, R. E., Federman, S. R. et al. 2017, JPhB, 50, 155007
\reference {heid15} Heidarian, N., Irving, R. E., Ritchey, A. M. et al. 2015, \apj, 808, 112
\reference {ht03} Heiles, C., \& Troland, T. H. 2003, \apj, 586, 1067
\reference {her68} Herbig, G. H. 1968, \zap, 68, 243
\reference {her93} Herbig, G. H. 1993, \apj, 407, 142
\reference {hob69} Hobbs, L. M. 1969, \apj, 157, 135
\reference {hbb18} Hodges, J. N., Bittner, D. M., \& Bernath, P. F. 2018, \apj, 855, 21
\reference {hol12} Hollenbach, D., Kaufman, M. J., Neufeld, D., Wolfire, M., \& Goicoechea, J. R. 2012, \apj, 754, 105
\reference {houk78} Houk, N. 1978, Michigan catalogue of two-dimensional spectral types for the HD stars (Ann Arbor: Dept. of Astronomy, Univ. of Michigan)
\reference {hl95} Hubeny, I., \& Lanz, T. 1995, \apj, 439, 875
\reference {hupe12} Hupe, R. C. Sheffer, Y., \& Federman, S. R. 2012, \apj, 761, 38
\reference {ind09} Indriolo, N., Fields, B. D., \& McCall, B. J. 2009, \apj, 694, 257
\reference {ind07} Indriolo, N., Geballe, T. R., Oka, T., \& McCall, B. J. 2007, \apj, 671, 1736
\reference {jenk87} Jenkins, E. B. 1987, in Interstellar Processes, ed. D. J. Hollenbach \& H. A. Thronson, (Kluwer), p. 533
\reference {jenk96} Jenkins, E. B. 1996, \apj, 471, 292
\reference {jenk09} Jenkins, E. B. 2009, \apj, 700, 1299
\reference {js79} Jenkins, E. B., \& Shaya, E. J. 1979, \apj, 231, 55
\reference {jt01} Jenkins, E. B., \& Tripp, T. M. 2001, \apjs, 137, 297
\reference {jt06} Jenkins, E. B., \& Tripp, T. M. 2006, \apj, 637, 548
\reference {jt11} Jenkins, E. B., \& Tripp, T. M. 2011, \apj, 734, 65 (JT11)
\reference {jenk73} Jenkins, E. B., Drake, J. F., Morton, D. C. et al. 1973, \apjl, 181, L122
\reference {jrs05} Jensen, A. G., Rachford, B. L., \& Snow, T. P. 2005, \apj, 619, 891 
\reference {jen10} Jensen, A. G., Snow, T. P., Sonneborn, G., \& Rachford, B. L. 2010, \apj, 711, 1236
\reference {jura74} Jura, M. 1974, \apjl, 190, L33
\reference {kalb16} Kalberla, P. M. W., Kerp, J., Haud, U., et al. 2016, \apj, 821, 117
\reference {kgb18} Kaur, J., Gorczyca, T. W., \& Badnell, N. R. 2018, \aap, 610, A41
\reference {kil78} Kilkenny, D. 1978, \mnras, 182, 629
\reference {kur93} Kurucz, R. 1993, CD-ROM No. 13 (available at http://archive.stsci.edu/hlsps/reference-atlases/cdbs/grid/k93models/)
\reference {lan15} Lan, T.-W., M\'{e}nard, B., \& Zhu, G. 2015, \mnras, 452, 3629
\reference {lh07} Lanz, T., \& Hubeny, I. 2007, \apjs, 169, 83
\reference {ls83a} Larsson \& Siegbahn 1983a, \jcp, 79, 2270; 85, 4208   
\reference {ls83b} Larsson \& Siegbahn 1983b, CP, 76, 175
\reference {lep06} Le Petit, F., Nehm\'{e}, C., Le Bourlot, J., \& Roueff, E. 2006, \apjs, 164, 506
\reference {lien84} Lien, D. J. 1984, \apj, 284, 578
\reference {lis03} Liszt, H. 2003, \aap, 398, 621
\reference {lis07} Liszt, H. S. 2007, \aap, 476, 291
\reference {lod03} Lodders, K. 2003, \apj, 591, 1220
\reference {lyu01} Lyu, C.-H., Smith, A. M., \& Bruhweiler, F. C. 2001, \apj, 560, 865
\reference {math99} Mather, J. C., Fixsen, D. J., Shafer, R. A., Mosier, C., \& Wilkinson, D. T. 1999, \apj, 512, 511
\reference {mmpxx} Mathis, J. S., Mezger, P. G., \& Panagia, N. 1983, \aap, 128, 212 (MMP) 
\reference {mm02} Mazzitelli, G., \& Mattioli, M. 2002, Atomic Data and Nuclear Data Tables, 82, 313
\reference {meyer01} Meyer, D. M., Lauroesch, J. T., Sofia, U. J., Draine, B. T., \& Bertoldi, F. 2001, \apj, 553, 59
\reference {mr90} Meyer, D. M., \& Roth, K. C. 1990, \apj, 349, 91
\reference {mr91} Meyer, D. M., \& Roth, K. C. 1991, \apjl, 376, L49
\reference {mil07} Miller, A., Lauroesch, J. T., Sofia, U. J., Cartledge, S. I. B., \& Meyer, D. M. 2007, \apj, 659, 441
\reference {mort91} Morton, D. C. 1991, \apjs, 77, 119
\reference {mort00} Morton, D. C. 2000, \apjs, 130, 403
\reference {mort03} Morton, D. C. 2003, \apjs, 149, 205
\reference {mn94} Morton, D. C., \& Noreau, L. 1994, \apjs, 95, 301
\reference {myers15} Myers, A. T., McKee, C. F., \& Li, P. S. 2015, \mnras, 453, 2747
\reference {najar08} Najar, F., Ben Abdallah, D., Jaidane, N., \& Ben Lakhdar, Z. 2008, CPL, 460, 31
\reference {najar09} Najar, F., Ben Abdallah, D., Jaidane, N., et al. 2009, \jcp, 130, 204305
\reference {ns83} Nussbaumer, H., \& Storey, P. J. 1983, \aap, 126, 75
\reference {occ92} O'Donnell, J. E., Cardelli, J. A., \& Churchwell, E. 1992, \aj, 104, 2161
\reference {pmc92} Palazzi, E., Mandolesi, N., \& Crane, P. 1992, \apj, 398, 53
\reference {pan05} Pan, K., Federman, S. R., Sheffer, Y., \& Andersson, B.-G. 2005, \apj, 633, 986
\reference {pan01} Pan, K., Federman, S. R., \& Welty, D. E. 2001, \apjl, 558, L105
\reference {pa86} P\'{e}quignot, D., \& Aldrovandi, S. M. V. 1986, \aap, 161, 169
\reference {pm02} Pereyra, A., \& Magalhaes, A. M., 2002, \apjs, 141, 469
\reference {pdf86} Pineau des For\^{e}ts, G., Roueff, E., \& Flower, D. R. 1986, \mnras, 223, 743 
\reference {plan16} Planck Collaboration, Ade, P. A. R., Aghanim, N., et al. 2016, \aap, 594, A28
\reference {por14} Porras, A. J., Federman, S. R., Welty, D. E., \& Ritchey, A. M. 2014, \apjl, 781, L8
\reference {pro06} Prochaska, J. X., Chen, H.-W., \& Bloom, J. S. 2006, \apj, 648, 95
\reference {rach09} Rachford, B. L., Snow, T. P., Destree, J. D. et al. 2009, \apjs, 180, 125
\reference {rach02} Rachford, B. L., Snow, T. P., Tumlinson, J. et al. 2002, \apj, 577, 221
\reference {rr12} Rai, R. K., \& Rastogi, S. 2012, \mnras, 423, 2941
\reference {rfl11} Ritchey, A. M., Federman, S. R., \& Lambert, D. L. 2011, \apj, 728, 36
\reference {rfl18} Ritchey, A. M., Federman, S. R., \& Lambert, D. L. 2018, \apjs, 236 ,36
\reference {rjf19} Ritchey, A., Jenkins, E., \& Federman, S. 2019, AAS meeting 233, 411.07
\reference {rm95} Roth, K. C., \& Meyer, D. M. 1995, \apj, 441, 129
\reference {rou96} Roueff, E. 1996, \mnras, 279, L37
\reference {sahu92} Sahu, M. 1992, Ph.D. Thesis, University of Groningen
\reference {sav77} Savage, B. D., Bohlin, R. C., Drake, J. F., \& Budich, W. 1977, \apj, 216, 291
\reference {sp74} Savage, B. D., \& Panek, R. J. 1974, \apj, 191, 659
\reference {ss96} Savage, B. D., \& Sembach, K. R. 1996, \araa, 34, 279
\reference {ss92} Sembach, K. R., \& Savage, B. D. 1992, \apjs, 83, 147
\reference {sf07} Sheffer, Y., \& Federman, S. R. 2007, \apj, 659, 1352 
\reference {sfl02a} Sheffer, Y., Federman, S. R., \& Lambert, D. L. 2002a, \apjl, 572, L95 
\reference {slf02b} Sheffer, Y., Lambert, D. L., \& Federman, S. R. 2002b, \apjl, 574, L171  
\reference {shef08} Sheffer, Y., Rogers, M., Federman, S. R. et al. 2008, \apj, 687, 1075  
\reference {sb82} Shull, J. M., \& Beckwith, S. 1982, \araa, 20, 163
\reference {svs82} Shull, J. M., \& Van Steenberg, M. 1982, \apjs, 48, 95
\reference {sp78} Smith, P. L., \& Parkinson, W. H. 1978, \apjl, 223, L127
\reference {snow84} Snow, T. P. 1984, \apj, 287, 238
\reference {sm06} Snow, T. P., \& McCall, B. J. 2006, \araa, 44, 367
\reference {srf02a} Snow, T. P., Rachford, B. L., \& Figoski 2002a, \apj, 573, 662
\reference {snow02b} Snow, T. P., Welty, D. E., Thorburn, J. et al. 2002b, \apj, 573, 670
\reference {sof04} Sofia, U. J., Lauroesch, J. T., Meyer, D. M., \& Cartledge, S. I. B. 2004, \apj, 605, 272
\reference {sof99} Sofia, U. J., Meyer, D. M., \& Cardelli, J. A. 1999, \apjl, 522, L137
\reference {son02} Sonnentrucker, P., Friedman, S. D., Welty, D. E., York, D. G., \& Snow, T. P. 2002, \apj, 576, 241
\reference {son03} Sonnentrucker, P., Friedman, S. D., Welty, D. E., York, D. G., \& Snow, T. P. 2003, \apj, 596, 350
\reference {son07} Sonnentrucker, P., Welty, D. E., Thorburn, J. A., \& York, D. G. 2007, \apjs, 168, 58
\reference {sw92} Stahl, O., \& Wilson, T. L. 1992, \aap, 254, 327
\reference {thad72} Thaddeus, P. 1972, \araa, 10, 305
\reference {thor03} Thorburn, J. A., Hobbs, L. M., McCall, B. J. et al. 2003, \apj, 584, 339
\reference {val17} Valdivia, V., Godard, B., Hennebelle, P. et al. 2017, \aap, 600, A114
\reference {val04} Valencic, L. A., Clayton, G. C., \& Gordon, K. D. 2004, \apj, 616, 912
\reference {vd88} van Dishoeck, E. F. 1988, in Rate coefficients in astrochemistry, ed. T. J. Millar \& D. A. Williams, (Dordrecht: Kluwer), p. 49
\reference {vdb82} van Dishoeck, E. F., \& Black, J. H. 1982, \apj, 258, 533
\reference {vdb88} van Dishoeck, E. F., \& Black, J. H. 1988, \apj, 334, 771
\reference {vdb89} van Dishoeck, E. F., \& Black, J. H. 1989, \apj, 340, 273
\reference {vdbpg91} van Dishoeck, E. F., Black, J. H., Phillips, T. G., \& Gredel, R. 1991, \apj, 366, 141
\reference {vddz84} van Dishoeck, E. F., \& de Zeeuw, T. 1984, \mnras, 206, 383
\reference {vel17} Velusamy, T., Langer, W. D., Goldsmith, P. F., \& Pineda, J. L. 2017, \apj, 838, 165
\reference {ver96} Verner, D. A., Ferland, G. J., Korista, K. T., \& Yakovlev, D. G. 1996, \apj, 465, 487
\reference {vvdb09} Visser, van Dishoeck, E. F., \& Black, J. H. 2009, \aap, 503, 323
\reference {vogt94} Vogt, S. S., Allen, S. L., Bigelow, B. C. et al. 1994, Proc. SPIE, 2198, 362
\reference {vh10} Voshchinnikov, N. V., \& Henning, Th. 2010, \aap, 517, A45
\reference {vos12} Voshchinnikov, N. V., Henning, Th., Prokopjeva, M. S., \& Das, H. K. 2012, \aap, 541, A52
\reference {vre07} Vreeswijk, P. M., Ledoux, C., Smette, A. et al. 2007, \aap, 468, 83
\reference {wd85} Walker, D. D., \& Diego, F. 1985, \mnras, 217, 355
\reference {wan99} Wannier, P. G., Andersson, B.-G., Penprase, B. E., \& Federman, S. R. 1999, \apj, 510, 291
\reference {wpa97} Wannier, P., Penprase, B. E., \& Andersson, B.-G. 1997, \apjl, 487, L165
\reference {wbv96} Warin, S., Benayoun, J. J., \& Viala Y. P. 1996, \aap, 308, 535
\reference {wat76} Watson, W. D., Anicich, V. G., \& Huntress, W. T., Jr. 1976, \apj, 205, L165
\reference {wd01} Weingartner, J. C., \& Draine, B. T. 2001, \apj, 563, 842
\reference {welty14} Welty, D. E. 2014, in IAU Symp. 297, The Diffuse Interstellar Bands, ed. J. Cami \& N. L. J. Cox (Cambridge: Cambridge Univ. Press), 153
\reference {wc10} Welty, D. E., \& Crowther, P. A. 2010, \mnras, 404, 1321
\reference {wel06} Welty, D. E., Federman, S. R., Gredel, R., Thorburn, J. A., \& Lambert, D. L. 2006, \apjs, 165, 138
\reference {wh01} Welty, D. E., \& Hobbs, L. M. 2001, \apjs, 133, 345
\reference {whk94} Welty, D. E., Hobbs, L. M., \& Kulkarni, V. P. 1994, \apj, 436, 152
\reference {whm03} Welty, D. E., Hobbs, L. M., \& Morton, D. C. 2003, \apjs, 147, 61
\reference {whlb13} Welty, D. E., Howk, J. C., Lehner, N., \& Black, J. H. 2013, \mnras, 428, 1107
\reference {wlwy15} Welty, D. E., Lauroesch, J. T., Wong, T., \& York, D. G. 2016, \apj, 821, 118
\reference {wmh96} Welty, D. E., Morton, D. C., \& Hobbs, L. M. 1996, \apjs, 106, 533
\reference {wel14} Welty, D. E., Ritchey, A. M., Dahlstrom, J. A., \& York, D. G. 2014, \apj, 792, 106
\reference {wxw12} Welty, D. E., Xue, R., \& Wong, T. 2012, \apj, 745, 173
\reference {welty99} Welty, D. E., Hobbs, L. M., Lauroesch, J. T. et al. 1999, \apjs, 124, 465
\reference {wes11} Weselak, T., Galazutdinov, G., Beletsky, Y., \& Kre{\l}owski, J. 2011, AN, 332, 167
\reference {wmfm93} Whittet, D. C. B., Martin, P. G., Fitzpatrick, E. L., \& Massa, D. 1993, \apj, 408, 573
\reference {wfd96} Wiese, W. L., Fuhr, J. R., \& Deters, T. M. 1996, Atomic Transition Probabilities of Carbon, Nitrogen, and Oxygen: A Critical Data Compilation (Washington: NIST)
\reference {yk78} York, D. G., \& Kinahan, B. F. 1978, \apj, 228, 127
\reference {xu19} Xu, Z., Luo, N., Federman, S. R. et al. 2019, \apj, 882, 86
\reference {zonca11} Zonca, A., Cecchi-Pestellini, C., Mulas, G., \& Malloci, G. 2011, \mnras, 410, 1932

\end{references}
\end{document}